%% file: article.tex
\documentclass[a4paper, 11pt]{article}
\usepackage{graphicx,grffile}
\usepackage{amsmath, amssymb, amsthm}
\usepackage{caption}
\usepackage{subfig}
\usepackage{xcolor}
\usepackage{grffile}
\usepackage{mathrsfs}
\usepackage{multirow}
\usepackage[normalem]{ulem}
\usepackage{bm}

\usepackage{indentfirst} 
\allowdisplaybreaks

\newcommand\s{\boldsymbol}
\newcommand\p{\partial} 
\newtheorem{theorem}{Theorem}

\newcommand\bx{\boldsymbol{x}}

\newcommand\bu{\boldsymbol{u}}
\newcommand\bbf{\boldsymbol{f}}
\newcommand\bbQ{\boldsymbol{Q}}
\newcommand\bv{\boldsymbol{v}}

\newcommand\be{\boldsymbol{e}}

\newcommand\bz{\boldsymbol{0}}
\newcommand\bbR{\mathbb{R}}
\newcommand\bbN{\mathbb{N}}
\newcommand\bbZ{\mathbb{Z}}

\newcommand\bF{\boldsymbol{F}}

\newcommand\dd{\,\mathrm{d}}

\newcommand\mQ{\mathcal{Q}}
\newcommand\mE{\mathcal{E}}
\newcommand\mM{\mathcal{M}}
\newcommand\mH{\mathcal{H}}
\newcommand\mA{\boldsymbol{A}}
\newcommand\mL{\mathcal{L}}
\newcommand\mP{\mathcal{P}}
\newcommand\mO{\mathcal{O}}

\newcommand\bE{\boldsymbol{E}}

\newcommand\pd[2]{\dfrac{\partial {#1}}{\partial {#2}}}

\newcommand{\bi}{\boldsymbol{i}}
\newcommand{\bj}{\boldsymbol{j}}
\newcommand{\bk}{\boldsymbol{k}}
\newcommand{\factori}{i_1!i_2!i_3!}

\numberwithin{equation}{section}

\graphicspath{{images/}}

{\theoremstyle{remark} \newtheorem{remark}{Remark}}

\newcommand\deletei{\bgroup\markoverwith{\textcolor{blue}{\rule[0.5ex]{2pt}{2pt}}}\ULon}

\title{Hermite spectral method for Fokker-Planck-Landau equation modeling
  collisional plasma}

\author{Ruo Li\thanks{CAPT, LMAM \& School of Mathematical Sciences,
    Peking University, Beijing, China, email: {\tt
      rli@math.pku.edu.cn}.},~~Yinuo Ren\thanks{School of Mathematical
    Sciences, Peking University, Beijing, China, 100871, email: {\tt
      renyinuo@pku.edu.cn}.}~~ Yanli Wang\thanks{Beijing Computational
    Science Research Center, email: {\tt ylwang@csrc.ac.cn}.}}

\begin{document}
\maketitle

\input{article_abstract.tex}

\input{article_FPL.tex}
\input{article_ansatz.tex}
\input{article_scheme.tex}
\input{article_experiment.tex}
\input{article_conclusion.tex}

\input{article_appendix.tex}
\bibliographystyle{plain}
\bibliography{article}
\end{document}

%% file: article_abstract.tex
\begin{abstract}
  We propose an Hermite spectral method for the Fokker-Planck-Landau
  (FPL) equation.  Both the distribution functions and the
  collision terms are approximated by series expansions of the
  Hermite functions. To handle the
  complexity of the quadratic FPL collision operator, a 
  reduced collision model is built by adopting the quadratic collision
  operator for the lower-order terms and the diffusive Fokker-Planck
  operator for the higher-order terms in the
  Hermite expansion of the reduced collision operator.
  The numerical scheme is split into three
  steps according to the Strang splitting, where different expansion
  centers are employed for different numerical steps to take
  advantage of the Hermite functions. The standard normalized
  Hermite basis \cite{FPL2018} is adopted
  during the convection and collision steps to
  utilize the precalculated coefficients of the quadratic collision
  terms, while the one constituted by the local macroscopic velocity
  and temperature is utilized for the
  acceleration step, by which the effect of the external force can be
  simplified to an ODE. Projections between different expansion
  centers are achieved by an algorithm proposed in
  \cite{Hu2020Numerical}.  Several numerical examples are studied to
  test and validate our new method.

  \vspace*{4mm}
  \noindent {\bf keyword:} quadratic collision operator, Hermite
  spectral method, Strang splitting
\end{abstract}

\section{Introduction}

The Fokker-Planck-Landau
(FPL) equation is used to describe the evolution of collisional
plasma systems at the kinetic level \cite{Goudon1997, Degond1994}.
It is
a six-dimensional integro-differential equation, which models binary
collisions between charged particles with long-range Coulomb
interactions.  The FPL equation is the limit of the Boltzmann
equation when all binary collisions are grazing \cite{Degond1992}.
It was originally derived by Landau \cite{Landau1936} and was later
derived independently in the Fokker-Planck form
\cite{PhysRev.107.1}. The high-dimensionality of the FPL equation
is a bottleneck for its numerical
simulations. Although several simplified models of the original FPL equation
has been developed, it is still a challenge to solve it both fast
and accurately.


One of the major difficulties in solving FPL equation numerically
is the complexity of the Fokker-Planck
collision operator, which is an integro-differential nonlinear
operator in the microscopic velocity space.  Most of
the statistical methods, such as the DSMC method \cite{bird}, are
limited for the FPL
equation \cite{NANBU1998639} since the FPL collision operator models
the infinite-range potential interactions within the plasma.  Several
deterministic methods are used
to solve the FPL equation or its simplifications. The entropic scheme,
which guarantees a nondecreasing entropy, is well studied in
\cite{Implicit2005, BuetConservative1998, BEREZIN1987163, Degond1994}.
To handle the stiffness of the collision operator, an
asymptotic-preserving (AP) strategy is studied in \cite{dimarco2015}, 
while a conservative spectral
method is adopted in \cite{ZhangGamba2017}.  A positivity-preserving
scheme for the linearized FPL equation is
proposed in \cite{partical1970}, after which it was modified to
preserve energy \cite{Relaxation1985} and then
extended to the two-dimensional FPL equation with cylindrical geometry
\cite{Fokker2014}.  Several other numerical methods, such as the
multipole expansions \cite{Lemou1998} and multigrid techniques
\cite{BUET1997}, have also been proposed. In
\cite{TAITANO2015357}, a fully implicit method is proposed for the
multidimensional Rosenbluth-Fokker-Planck equation. The finite element
methods in \cite{ZAKI1988184, ZAKI1988200} and the semi-Lagrangian
schemes in \cite{CROUSEILLES20101927, sonnendrucker1999semi,
  Qiu2011positivity, Xiong2014High} are also used to solve the
Vlasov equations. Moreover, the FPL equation with
stochasticity is studied in \cite{Jin2010}.


The spectral method has been widely used in numerical methods to solve
the Vlasov equation \cite{GuoTang2000, Holloway1996Spectral}.  In
\cite{PARESCHI20002, Filbet2002numerical}, a spectral method based 
on the Fourier expansion is implemented for the
nonhomogeneous FPL equation.  Moreover, the Hermite spectral method is
utilized in \cite{Bourdiec2006Numerical, Joseph2015,
Gibelli2006Spectral} to discretize the microscopic
velocity space and is utilized  to solve the FPL and
Boltzmann equation \cite{Hu2020Numerical}. One
advantage of the Hermite spectral method is that its first few
moments have explicit physical meanings.  For
example, the density, the macroscopic velocity and the temperature
can be easily derived using the first three expansion
coefficients \cite{PARESCHI20002}, which indicates that they may be
captured more precisely by ingeniously designing the
numerical algorithm.  However, the high-dimensionality and complexity
of the quadratic FPL collision model still impose great
challenges for adopting the Hermite spectral method to
numerically simulate the FPL equation.


In this paper, a numerical method based on the Hermite spectral
method is proposed for the nonhomogeneous FPL equation. The
distribution functions in the FPL equation are approximated by a
series of basis functions derived from Hermite
polynomials. Strang splitting method is then adopted for the FPL equation,
which is split into collision, convection and acceleration
steps. Unlike the general Hermite spectral method
\cite{PARESCHI20002}, we choose different expansion centers in the
Hermite expansion for different numerical steps. For the collision
step, the standard expansion center \cite{FPL2018} is chosen to
utilize the precalculated expansion coefficients of the quadratic
collisional term.  To further reduce
the computational cost of the quadratic FPL collision operator, a
reduced collision model is constructed by
combining the quadratic collision operator and the diffusive FP
operator, which is proved effective by the
numerical examples afterwards.  For the acceleration step, the
expansion center is chosen as the local macroscopic velocity and
temperature, under which the effect of the external force field can be
reduced to an ODE of the macroscopic velocity.  A
projection algorithm introduced in \cite{Hu2020Numerical} is utilized
to handle the projections between distribution functions with
different expansion centers. In the numerical
simulation, the convection, collision and acceleration steps
are solved successively. Both the linear and nonlinear Landau
damping problems are tested, with the decay rate of the
electrostatic energy and the effect of the collisional frequency
studied.  Moreover, the two-stream instability and the bump-on-tail
instability are also simulated to validate this new method.


The rest of this paper is organized as follows: Section \ref{sec:FPL}
introduces the FPL equation, the FPL collision operator and several
related properties. The detailed spectral method used to approximate
the distribution function is introduced in Section \ref{sec:method}.
The series expansion for the FPL collision operator and 
the reduced collision model are explained in Section
\ref{sec:coll}.  The numerical algorithm is proposed in Section
\ref{sec:model}.  Several numerical examples are exhibited in Section
\ref{sec:num}. The conclusions and future work are stated in Section
\ref{sec:conclusion} with some supplementary 
statements given in the Appendix \ref{app}.

%% file: article_FPL.tex
\section{Preliminaries}
\label{sec:FPL}
In this section, we give a brief introduction to the
Fokker-Planck-Landau equation, the Fokker-Planck collision operator
and some related properties.

\subsection{Fokker-Planck-Landau equation}

As in many other kinetic theories in physics, the distribution
function of a specific species $\alpha$ is described by the
distribution function $f_{\alpha}(t, \bx, \bv)$, a seven-dimensional
function of time $t$, the space position
$\bx \in \Omega \subset \bbR^3$ and the microscopic velocity
$\bv \in \bbR^3$. The distribution function $f_{\alpha}$ relates
the density $\rho_{\alpha}$, the macroscopic velocity
$\bu_{\alpha}$ and the temperature $T_{\alpha}$ of species $\alpha$
through
\begin{equation}
\label{eq:rlt}
\begin{aligned}
  \rho_{\alpha}&=\int_{\mathbb{R}^3}f_{\alpha}(t,\bx,\bv) \dd \bv,
  \quad \rho_{\alpha}\bu_{\alpha}&=\int_{\mathbb{R}^3}\bv
  f_{\alpha}(t,\bx,\bv) \dd \bv, \quad \frac{3}{2}
  \rho_{\alpha}T_{\alpha}& =
  \frac{1}{2}\int_{\mathbb{R}^3}|\bv-\bu_{\alpha}|^2f_{\alpha}(t,\bx,
  \bv)\dd \bv.
\end{aligned}
\end{equation}
The Fokker-Planck-Landau (FPL) equation 
describes the time evolution of the distribution functions for
charged particles in a nonequilibrium plasma. The FPL equation with
respect to the species $\alpha$ has the form
\begin{equation}
  \label{eq:FPL}
  \pd{f_{\alpha}}{t} + \bv \cdot \nabla_{\bx} f_{\alpha} + \bF \cdot
  \nabla_{\bv} f_{\alpha} = \mQ[f_{\alpha}], 
\end{equation} 
where the force field $\bF=\bF(t, \bx)$ is produced either externally
or self-consistently. Here, we consider only the case where
$\bF$ is generated by the self-consistent electric field
$ \bE(t, \bx)$, which is coupled to the distribution function
through the Poisson equation \cite{xiong2020}: 
\begin{equation}
  \label{eq:poisson}
  \bE(t, \bx) = -\nabla_{\bx} \psi(t, \bx), \qquad -\Delta_{\bx}\psi =
  \sum_{\eta} q_{\eta} \int_{\bbR^3} f_{\eta}(\bv) \dd \bv,
\end{equation} 
where $q_{\eta}$ is the electric charge of the particle $\eta$.  The
collision terms
\begin{equation}
  \label{eq:col}
  \mQ[f_{\alpha}] =  \sum_{\eta} \nu_{\eta} 
  \mQ_{\eta}[f_{\alpha}, f_{\eta}]  
\end{equation}
describe the collisions between particles of species $\alpha$ and
$\eta$, which are discussed in detail in the next
section. The non-negative parameter $\nu_{\eta}$ is the collision
frequency.  For simplicity, we restrict the study to a plasma
consisting only of electrons and
ions, which are referred to as the species $\alpha$ and $\beta$
respectively, in the following.

\subsection{Collision operator}

\label{sec:FPL operator}

The collision operator $\mQ_{\eta}$, $\eta = \alpha, \beta$ in
\eqref{eq:col} 
is called the Fokker-Planck-Landau (FPL) collision operator, which is
obtained by setting the Boltzmann collision operator concentrating on
grazing collisions \cite{Desvillettes1992}. It has the 
following form:
\begin{equation}
  \label{eq:collision_operator}
  \mQ_{\eta}[f_{\alpha}, f_{\eta}] = \nabla_{\bv} \cdot
  \left[\int_{\bbR^3} {\bf A}(\bv - \bv')\Big(\nabla_{\bv}
    f_{\alpha}(\bv)f_{\eta}(\bv') - \nabla_{\bv'} f_{\eta}(\bv')
    f_{\alpha}(\bv)\Big) \dd \bv' \right], \qquad \eta = \alpha, \beta,
\end{equation}
where the collision kernel $\bf A(\cdot)$, in the form of a
$3\times 3$ negative and symmetric matrix
\begin{equation}
  \label{eq:A}
  {\bf A}(\bv)=\Psi(|\bv|)\Pi(\bv),
\end{equation}
reflects the interaction between particles. Here $\Pi(\bv)$ is the
projection onto the space orthogonal to $\bv$, as
$\Pi_{ij}(\bv)=\delta_{ij}-\frac{v_iv_j}{|\bv|^2}$. For the
inverse-power-law (IPL) model, $\Psi(|\bv|)$ is a non-negative radial
function, i.e.
\begin{equation}
  \Psi(\bv)=\Lambda|\bv|^{\gamma+2},
\label{eq:ipl}
\end{equation}
where $\Lambda$ is a
positive constant and $\gamma$ is the index of
the power of the distance.  Similar to the Boltzmann equation,
we obtain the hard potential model when $\gamma > 0$ and the soft
potential model when $\gamma < 0$. There are two special cases, the
first of which is the model of Maxwell molecules when $\gamma = 0$ and
the other is the model with Coulomb interactions when $\gamma = -3$
\cite{Filbet2002numerical}.

For the different species of electrons and ions, the collision
operator \eqref{eq:collision_operator} may be reduced to different
forms. These two collision operators are discussed in detail below.

\subsubsection{Electron-electron collision}
\label{sec:ee operator}

The quadratic operator $\mQ_{\alpha}[f_{\alpha}, f_\alpha]$ describes
the electron-electron collisions, the form
of which can be obtained by taking $\eta$ as $\alpha$ in the FPL
collision operator \eqref{eq:collision_operator}. With a 
slight abuse of notation, the subscript $\alpha$ referring
electrons is omitted from now on.  Thus, the distribution function
$f_{\alpha}$ is shorten to $f$ and
the collision operator $\mQ_{\alpha}[f_{\alpha}, f_\alpha]$
is shorten to $\mQ[f, f]$ as
\begin{equation}
  \label{eq:collision_ee}
  \mQ[f, f] = \nabla_{\bv} \cdot
  \left[\int_{\bbR^3} {\bf A}(\bv - \bv')\Big(\nabla_{\bv}
    f(\bv)f(\bv') - \nabla_{\bv'} f(\bv')
    f(\bv)\Big) \dd \bv' \right], 
\end{equation}
with the collision kernel ${\bf A}(\cdot)$ defined in
\eqref{eq:A}. For the steady solution of the FPL
equation, we obtain the equilibrium, which has the following
Maxwellian form:
\begin{equation}
  \label{eq:max}
  \mM(\bv) =
  \frac{\rho}{(2 \pi T)^{3/2}} 
  \exp\left(-\frac{|\bv - \bu|^2}{2T}\right), 
\end{equation}
where $\rho$, $\bu$ and $T$ are the density, macroscopic
velocity and temperature of electrons respectively
\eqref{eq:rlt}. Moreover, this operator 
maintains the conservation of mass, momentum and energy as
\begin{equation}
  \label{eq:conserve}
  \int_{\bbR^3} \mQ[f, f] \left(
  \begin{array}{c}
    1 \\ \bv \\ |\bv|^2
  \end{array} \right) \dd\bv
= 0.
\end{equation}
Due to the complicated form of the FPL collision operator, several
simplified operators are introduced to approximate the original
quadratic operator $\mQ[f, f]$, for example the linearized collision
operator
\begin{equation}
  \label{eq:linear_col}
  \mL[f] = \mQ[f, \mM]+    
  \mQ[\mM, f], 
\end{equation}
and the diffusive Fokker-Planck (FP) operator \cite{JinYan2011}
\begin{equation}
  \label{eq:FP}
  \mP_{\rm FP}[f] = \nabla_{\bv} \cdot
  \left[\mM
    \nabla_{\bv} \left(\frac{f}{\mM}\right)\right].
\end{equation}

\subsubsection{Electron-ion collision}

The collisions between electrons and ions are described by the
operator $\mQ_{\beta}[f, f_{\beta}]$, which
  can be obtained by taking $\eta$ as $\beta$ in
\eqref{eq:collision_operator}. 
Since the electrons have a minute mass and high velocity
  compared to the ions, the ions may be collectively treated as a
  stationary positively-charged background. Furthermore,
the temperature of ions $T_{\beta}$ is negligible compared to
  that of electrons $T$. Thus, the distribution function of ions can
be given simply by a Dirac measure in the velocity space
\cite{ZhangGamba2017} as

\begin{equation}
  \label{eq:density_beta}
  f_{\beta}(t, \bx, \bv) = \rho_{\beta}(t, \bx)\delta_{0}(\bv -
  \bu_{\beta}(t, \bx)),
\end{equation}
where $\rho_{\beta}$ and $\bu_{\beta}$ are the density and macroscopic
velocity of ions. Consequently, the collision operator
$\mQ_{\beta}[f, f_\beta]$ can be reduced to
\begin{equation}
  \label{eq:col_beta}
  \mQ_{\beta}[f]  \triangleq \mQ_{\beta}[f, f_\beta] = \rho_{\beta}\nabla_{\bv} \cdot
  \left[{\bf A}(\bv - \bu_{\beta})\nabla_{\bv}f\right].  
\end{equation}
One can also easily check that this reduced
operator still preserves mass and energy as
\begin{equation}
  \label{eq:mass_energy_beta}
  \int_{\bbR^3} \mQ_{\beta}[f] \dd \bv  = 0, \qquad
  \int_{\bbR^3} |\bv - \bu_{\beta}|^2\mQ_{\beta}[f] 
  \dd \bv = 0.
\end{equation}
Refer to \cite{Filbet, dimarco2015} for more details on this reduced
collision operator.


%% file: article_ansatz.tex

\section{Series expansion of the FPL equation}
\label{sec:method}
In this section, we introduce the series expansion to approximate the
distribution function in detail, including the basis functions which
are constructed by Hermite polynomials and several related properties.  
In the
numerical scheme, different expansion centers are utilized in the
Hermite expansion. A fast algorithm for the projections between
distribution functions with different expansion centers is also stated
in this section.

\subsection{Distribution function}

The Hermite expansion has been proved successful in the
numerical method for the Boltzmann equation \cite{Hu2020Numerical},
and the exact expansion coefficients for the quadratic FPL collision
have been computed in \cite{FPL2018}. Thus, the Hermite expansion is
also adopted in the approximation to the FPL equation here.
To be
precise, the distribution function $f$ is discretized as
\begin{equation}
  \label{eq:expansion}
  f(t, \bx, \bv) = \sum_{\bi \in \bbN^3}
  f_{\bi}^{[\tilde{\bu}, 
    \tilde{T}]}(t, \bx) \mH_{\bi}^{[\tilde{\bu},
    \tilde{T}]}(\bv), 
\end{equation}
where the basis functions
$\mH_{\bi}^{[\tilde{\bu}, \tilde{T}]}(\bv)$ are defined as
\begin{equation}
  \label{eq:basis}
  \mH_{\bi}^{[\tilde{\bu}, \tilde{T}]}(\bv) =
  \tilde{T}^{-\frac{|\bi|}{2}} H_{\bi}\left(\frac{\bv -
      \tilde{\bu}}{\sqrt{\tilde{T}}} \right) \frac{1}{(2 \pi
    \tilde{T})^{3/2}} 
  \exp\left(-\frac{|\bv - \tilde{\bu}|^2}{2 \tilde{T}}\right)
\end{equation}
and $\bi$ refers to the multi-index $(i_1,i_2,i_3)$. We also
adopt the following notations for simplicity:
\begin{equation*}
  |\bi|=i_1+i_2+i_3,\qquad \bi!=\factori, 
  \qquad \pd{^{\bi}}{\bv^{\bi}}=\pd{^{i_1+i_2+i_3}}{v_1^{i_1}v_2^{i_2}v_3^{i_3}}.
\end{equation*} 
In \eqref{eq:basis}, $H_{\bi}(\bv)$ represents the Hermite
polynomial
\begin{equation}
  \label{eq:Hermite}
  H_{\bi}(\bv) = (-1)^{|\bi|} \exp\left(\frac{|\bv|^2}{2}\right) 
  \pd{^{\bi}}{\bv^{\bi}}
  \left[\exp\left(-\frac{|\bv|^2}{2}\right) \right],
\end{equation}
where the two parameters $\tilde{\bu} \in \bbR^3$ and
$\tilde{T} \in \bbR_{+}$, namely the expansion center, are of the
same dimension as $\bv$ and $T$.  The coefficients
$f_{\bi}^{[\tilde{\bu}, \tilde{T}]}$ can be explicitly expressed by
\begin{equation}
  \label{eq:coef}
  f_{\bi}^{[\tilde{\bu},
    \tilde{T}]}(t,
  \bx)=\dfrac{\tilde{T}^{\frac{|\bi|}{2}}}{\bi!}\int_{\mathbb{R}^3}
  f(t,\s{x},\s{v})H_{\bi}\left(\dfrac{\bv-\tilde{\bu}}
    {\sqrt{\tilde{T}}}\right)\dd\bv      
\end{equation}
from the orthogonality of Hermite polynomials 
\begin{equation}
  \label{eq:ortho}
  \int_{\bbR^3}H_{\bi}(\bv)H_{\bj}(\bv) 
  \exp\left(-\frac{|\bv|^2}{2}\right) \dd\bv=\begin{cases}
    \bi!,&\text{if } \bi=\bj,\\
    0,&\text {otherwise.}
  \end{cases}
\end{equation}

Several lower-order moments have close relations to macroscopic
variables. For example, relations \eqref{eq:rlt} with
respect to density $\rho$, macroscopic
velocity $\bu$ and temperature $T$ of the electron can be rewritten
by the expansion coefficients as
\begin{equation}
  \label{eq:relation}
  \begin{gathered}
    \rho = f_{\boldsymbol 0}^{[\tilde{\bu}, \tilde{T}]}, \qquad \rho
    \bu = \rho \tilde{\bu} + \left(f_{\be_1}^{[\tilde{\bu},
        \tilde{T}]}, f_{\be_2}^{[\tilde{\bu}, \tilde{T}]},
      f_{\be_3}^{[\tilde{\bu},
        \tilde{T}]}\right)^T, \\
    \frac{1}{2}\rho |\bu|^2 + \frac{3}{2}\rho T = \rho \bu \cdot
    \tilde{\bu} - \frac{1}{2} \rho |\tilde{\bu}|^2 +
    \frac{3}{2}\rho\tilde{T} +\sum_{d=1}^3 f_{2\be_d}^{[\tilde{\bu}.
      \tilde{T}]}.
\end{gathered}
\end{equation} 
  
Some other related macroscopic quantities such as the shear stress and
the heat flux can also be expressed in terms of the expansion
coefficients $f_{\bi}^{[\tilde{\bu}, \tilde{T}]}$, and we refer
\cite{FPL2018} for further details.  Moreover, it is
worth mentioning that if the expansion
parameters chosen are the local
macroscopic velocity and temperature of the particles, i.e.
 $\tilde{\bu} = \bu$ and $\tilde{T} = T$, it
holds that according to \eqref{eq:relation},
\begin{equation}
  \label{eq:relation_12}
  f_{\be_d}^{[\bu, T]} = 0,
  \qquad \sum\limits_{d = 1}^3f_{2\be_d}^{[\bu,
    T]} = 0,  \qquad  d = 1, 2, 3. 
\end{equation}

\subsection{Projections between different expansion centers}

One should be aware that different expansion centers may lead to
different series expansions in \eqref{eq:expansion} and can be selected
to meet different needs. Based on a prior understanding of the
problem, different expansion centers $\tilde{\bu}$ and $\tilde{T}$ are
chosen to accelerate the convergence of the series expansion
\eqref{eq:expansion}.

The selection of expansion centers is discussed in many
studies.  The normalized Hermite basis, or the expansion center
$\tilde{\bu} = 0$ and $\tilde{T} = 1$, is adopted in \cite{UTH,
  Joseph2015, xiong2020}, while the local macroscopic variables
$\tilde{\bu} = \bu(t, \bx)$ and $\tilde{T} = T(t, \bx)$ defined in
\eqref{eq:rlt} are chosen as the expansion center in \cite{Wang} and
for some related problems \cite{Hu2020Numerical}.  Here,
instead of fixing the expansion center throughout, we select an
appropriate one for each numerical step.  Here, instead of fixing the
expansion center throughout, as we mainly focus on the reduction in
the complexity of the series expansions and the feasibility of the
numerical method, we select an appropriate one for each numerical
step, which is further explained in Section \ref{sec:coll} and
\ref{sec:model}.

To achieve efficient projections between distribution functions with
different expansion centers, we adopt the algorithm proposed in
\cite{Hu2020Numerical}, which is described by the following theorem:
\begin{theorem}
\label{thm:project}
Suppose a distribution function $f(\bv)$ in the velocity space
satisfies  
\begin{equation}
\int_{\bbR^3} (1+|\bv|^M) |f(\bv)| \dd \bv < \infty
\end{equation}
for some $M\in \bbZ_+$. 

Define the expansion coefficients of $f(\bv)$ as
\begin{equation}
  \label{eq:psi}
  \begin{aligned}
    & f_{\bi}^{[\tilde{\bu}^{(1)},\tilde{T}^{(1)}]} =
    \frac{(\tilde{T}^{(1)})^{\frac{|\bi|}{2}}}{\bi!} \int_{\bbR^3}
    H_{\bi}\left( \frac{\bv -
        \tilde{\bu}^{(1)}}{\sqrt{\tilde{T}^{(1)}}}\right)
    f(\bv) \dd \bv, \\
    & f_{\bi}^{[\tilde{\bu}^{(2)},\tilde{T}^{(2)}]} =
    \frac{(\tilde{T}^{(2)})^{\frac{|\bi|}{2}}}{\bi!} \int_{\bbR^3}
    H_{\bi}\left( \frac{\bv -
        \tilde{\bu}^{(2)}}{\sqrt{\tilde{T}^{(2)}}}\right) f(\bv) \dd
    \bv.
  \end{aligned}
\end{equation}
with
$\tilde{\bu}^{(s)}=\left(\tilde{u}^{(s)}_1,
  \tilde{u}^{(s)}_2,\tilde{u}^{(s)}_3\right), s = 1, 2 \in \bbR^3$
and $\tilde{T}^{(1)}, \tilde{T}^{(2)} > 0$. Then, for any
$\bi \in \bbN^3$ satisfying $|\bi|\leqslant M$, we have
\begin{equation}
  f_{\bi}^{[\tilde{\bu}^{(2)},\tilde{T}^{(2)}]} = \sum_{l = 0}^{|\bi|}
  \phi_{\bi}^{(l)}.
\end{equation}
Here $\phi_{\bi}^{(l)}$ is defined recursively by
\begin{equation}
  \label{eq:tilde_psi}
  \phi_{\bi}^{(l)} = \begin{cases}
    f_{\bi}^{[\tilde{\bu}^{(1)},\tilde{T}^{(1)}]}, &l = 0, \\
    \dfrac{1}{l} \sum\limits_{d=1}^3 \left(\left(\tilde{u}^{(2)}_{d} - 
        \tilde{u}^{(1)}_{d}\right) 
      \phi_{\bi - \be_d}^{(l-1)} + \frac{1}{2}(\tilde{T}^{(2)} -
      \tilde{T}^{(1)}) \phi_{\bi - 2\be_d}^{(l-1)} \right),
    & 1   \leqslant  l  \leqslant  |\bi|,
  \end{cases}
\end{equation}
where terms with negative indices are treated as zero.
\end{theorem}

We refer to Theorem 3.1 in \cite{Hu2020Numerical} for the proof and
more details of this projection algorithm.

\section{Series expansion of the collision operators and the reduced
  collision model}

\label{sec:coll}

For the moment, we have obtained the series approximation to the
distribution function. In this section, the FPL
collision operator \eqref{eq:collision_operator} is expanded using
the same basis functions. We derive the series expansions and
provide algorithms to compute the expansion coefficients of the
quadratic collision operator $\mQ[f, f]$ \eqref{eq:collision_ee} as
well as its simplified approximation $\mP_{\rm FP}$ and the collision
operator between different species $\mQ_{\beta}[f]$
\eqref{eq:col_beta} in this section.

Moreover, since the computational cost for
the quadratic collision operator \eqref{eq:collision_operator} is
unaffordable, a reduced collision model is built based on the
expansion coefficients to further reduce the computational cost.

\subsection{Quadratic collision operator $\mQ[f, f]$}
\label{sec:quadoperator}

As stated in Section \ref{sec:FPL operator}, the major difficulty of
handling the original FPL collision operator or even solving the FPL
equation is the complexity of the quadratic collision operator
$\mQ[f, f]$. In our method, both precalculation and model reduction
are employed to address this difficulty.

The quadratic collision operator $\mQ[f ,f]$ is to be expanded
similarly to the series expansion of the distribution functions as
\begin{equation}
  \label{eq:expansion_q}
  \mQ[f, f](t, \bx, \bv) =
  \sum_{\bi \in \bbN^3} 
  Q_{\bi}^{[\tilde{\bu},\tilde{T}]}(t, \bx)
  \mH_{\bi}^{[\tilde{\bu},\tilde{T}]}(\bv).
\end{equation}
By the orthogonality \eqref{eq:ortho}, the coefficients are
calculated as
\begin{equation}
\label{eq:qcoef}
Q_{\bi}^{[\tilde{\bu},\tilde{T}]}(t,
\bx)=\frac{\tilde{T}^{\frac{|\bi|}{2}}}{\bi!}
\int_{\mathbb{R}^3}H_{\bi}\left(\frac{\s{v}-\tilde{\bu}}{\sqrt{\tilde{T}}}\right)   
\mQ[f, f](t, \bx, \bv) \dd\bv.
\end{equation}

In our previous work \cite{FPL2018}, an algorithm 
was proposed to evaluate these coefficients in the standard
case of $\rho=1, \tilde{\bu}=\bz$ and $ \tilde{T}=1$, and the
normalized Hermite basis was used to approximate
the distribution function.  To apply the results, the same expansion
center is adopted here. Then, the approximation to the distribution
function \eqref{eq:expansion} and the collision operator
\eqref{eq:expansion_q} are reduced to
\begin{align}
  \label{eq:expansion1}
  f(t,\bx,\bv)&=\sum_{\bi\in\bbN^3}f_{\bi}^{[\bz,1]}(t,\bx)\mH_{\bi}^{[\bz,1]}(\bv),\\
  \label{eq:expansion_q1}
  \mQ[f,f](t,\bx,\bv)&=\sum_{\bi\in\bbN^3}Q_{\bi}^{[\bz,1]}(t,\bx)\mH_{\bi}^{[\bz,1]}(\bv).
\end{align}
The superscript $[\bz, 1]$ is omitted afterwards if the expansion
center $\tilde{\bu} = \bz$ and $\tilde{T} =1$ is used.  Thus, in this
case, the coefficients \eqref{eq:qcoef} are reduced to
\begin{equation}
\label{eq:qcoef1}
Q_{\bi}(t,\bx)=\frac{1}{\bi!}\int_{\mathbb{R}^3}H_{\bi}\left(\bv\right) 
\mQ[f, f](t, \bx, \bv) \dd\bv.
\end{equation}
Substituting \eqref{eq:expansion1} into \eqref{eq:qcoef1}, we can
derive that
\begin{equation}
    \label{eq:final_coe}
    \begin{aligned}
      Q_{\bi}(t, \bx)& =\frac{1}{\bi!}
      \int_{\mathbb{R}^3}H_{\bi}\left(\bv\right)
      \nabla_{\s{v}}\cdot\left[\int_{\mathbb{R}^3}
        {\bf A}(\s{v}-\s{v}')\big(f(\s{v}')
        \nabla_{\s{v}}f(\s{v})-f(\s{v})\nabla_{\s{v}'}f(\s{v}')\big)\dd\s{v}'\right]\dd \bv
      \\
      & =\sum_{\bj \in \bbN^3}\sum_{\bk \in
        \bbN^3}A_{\bi}^{\bj,\bk}f_{\bj}
      f_{\bk},
      \end{aligned}
\end{equation}
where the coefficients $A_{\bi}^{\bj,\bk}$ have the following expression
\cite[Eq.(3.4)]{FPL2018}:
\begin{equation}
  \label{eq:Acomp}
  A_{\bi}^{\bj,\bk}
  =\dfrac{1}{\bi!}\int_{\mathbb{R}^3}H_{\bi}(\bv)\nabla_{\bv}\cdot\left[\int_{\mathbb{R}^3}A(\bv-\bv')
    (\mathcal{H}_{\bj}(\bv')\nabla_{\bv}(\mathcal{H}_{\bk}(\bv))
    -\mathcal{H}_{\bj}(\bv)\nabla_{\bv'}
    (\mathcal{H}_{\bk}(\bv'))\dd\bv'\dd\bv\right].
\end{equation}
Although the complicated form of the coefficients $A_{\bi}^{\bj,\bk}$
results in a formidably high computational cost, we
find that these coefficients are intrinsic to the
collision model and are constant when a specific collision model is
chosen, which in our case means that the index $\gamma$ in the
IPL model \eqref{eq:ipl} is fixed.  This indicates that we can always
precalculate these coefficients completely offline once and then
use them in all cases. In \cite{FPL2018}, an
algorithm was proposed to calculate
accurate values of the coefficients $A_{\bi}^{\bj,\bk}$
(or $A_\alpha^{\lambda,\kappa}$ correspondingly in
\cite[Eq.(3.4)]{FPL2018}) by introducing Burnett polynomials for all
$\gamma>-5$. Due to the lengthy expressions involved in the algorithm,
we do not present the details
here. Readers may refer to Theorem 1, Lemma 2, Proposition 3 and
Theorem 4 in \cite{FPL2018} for the details of this algorithm.

Moreover, utilizing the recurrence relations
of the Hermite polynomial
  \begin{equation}
  \label{eq:Hermite_recur}
  \begin{gathered}
    \dfrac{\p}{\p
    v_d}H_{\bi}(\bv) = i_d H_{\bi-\be_d}(\bv), \quad H_{\bi+\be_d}(\bv) = v_d H_{\bi}(\bv) - i_d
  H_{\bi-\be_d}(\bv), \\
   \dfrac{\p}{\p v_d}\left[H_{\bi}(\bv)
    \exp\left(-\frac{|\bv|^2}{2}\right)\right] =
  -H_{\bi+\be_d}(\bv) \exp\left(-\frac{|\bv|^2}{2}\right), \qquad d =
  1, 2, 3,
\end{gathered}
\end{equation}
the diffusive FP operator $\mP_{\rm FP}[f]$ \eqref{eq:FP}, a
simplified approximation to the quadratic collision operator
$\mQ[f, f]$, can also be expanded as
\begin{equation}
  \label{eq:expan_FP}
  \begin{aligned}
    \mP_{\rm FP}[f] = \sum_{\bi \in \bbN^3} {\rm FP}_{\bi}
    \mH_{\bi}(\bv), \qquad {\rm FP}_{\bi} = \sum_{d=1}^3 \left[\left(1
        - \frac{1}{T}\right) f_{\bi - 2\be_d} + \frac{u_{d}}{T} f_{\bi
        - \be_d}\right] - \frac{|\bi|}{T} f_{\bi},
  \end{aligned}
\end{equation}
where $\bu=(u_{1}, u_{2}, u_{3})^T$ and $T$ are the macroscopic
velocity and temperature of electrons \eqref{eq:rlt}, respectively.

\subsection{Collision operator $\mQ_{\beta}[f]$}
\label{sec:reduce_q}

The collision operator $\mQ_{\beta}[f]$ \eqref{eq:col_beta} between
different species can also be expanded with similar methods.  Without
loss of generality, we set $\rho_{\beta} = 1$.  Therefore, the
collision operator $\mQ_{\beta}[f]$ is expanded as
\begin{equation}
  \label{eq:expan_FPL_beta_ini}
  \mQ_{\beta}[f](t, \bx, \bv) = \sum_{\bi \in
    \bbN^3}  
  \mQ_{\beta, \bi}^{[\tilde{\bu}, \tilde{T}]}(t, \bx)
  \mH_{\bi}^{[\tilde{\bu}, \tilde{T}]}(\bv),
\end{equation}
with the coefficients being 
\begin{equation}
  \label{eq:expan_FPL_beta_coe_ini}
  \begin{aligned}
    \mQ_{\beta, \bi}^{[\tilde{\bu}, \tilde{T}]}(t, \bx) & =
    \frac{1}{\bi!} \int_{\bbR^3} \mQ_{\beta}[f](t, \bx, \bv)
    H_{\bi}\left( \frac{\bv - \tilde{\bu}}{\sqrt{\tilde{T}}}\right)
    \dd \bv \\
    & = \frac{1}{\bi!} \int_{\bbR^3} \nabla_{\bv} \cdot \left[{\bf
        A}(\bv - \bu_{\beta})\nabla_{\bv}f\right] H_{\bi}\left(
      \frac{\bv - \tilde{\bu}}{\sqrt{\tilde{T}}} \right) \dd \bv.
    \end{aligned}
  \end{equation}
  Noting that the macroscopic velocity
  $\bu_\beta$ of ions appears in the expression of the collision
  operator $\mQ_{\beta}[f]$, the expansion center here is chosen as
  $\tilde{\bu}=\bu_{\beta}$ and $\tilde{T}=1$ to
  simplify its series expansion. Thus, \eqref{eq:expan_FPL_beta_ini}
  and \eqref{eq:expan_FPL_beta_coe_ini} are reduced to
  \begin{equation}
  \label{eq:expan_FPL_beta}
  \begin{aligned}
    & \mQ_{\beta}[f](t, \bx, \bv) = \sum_{\bi \in
    \bbN^3}  
  \mQ_{\beta, \bi}^{[\bu_{\beta}, 1]}(t, \bx)
  \mH_{\bi}^{[\bu_{\beta}, 1]}(\bv), \\
  & \mQ_{\beta,
    \bi}^{[\bu_{\beta}, 1]}(t, \bx)  =   \frac{1}{\bi!}
  \int_{\bbR^3}    \nabla_{\bv} \cdot \left[{\bf
        A}(\bv - \bu_{\beta})\nabla_{\bv}f\right]
  H_{\bi}\left(\bv - \bu_{\beta}\right)
  \dd \bv.
  \end{aligned}
\end{equation}
By substituting \eqref{eq:expansion} and \eqref{eq:col_beta} into
\eqref{eq:expan_FPL_beta} and changing variables, the coefficients
can be calculated explicitly as
\begin{equation}
\label{eq:detail_expan_FPL_beta_coe}
  Q_{\beta, \bi}^{[\bu_{\beta}, 1]}=
  \frac{\Lambda}{\bi!}  \sum_{\bj \in \bbN^3}
  f_{\bj}^{[\bu_{\beta},
    1]} 
  \sum_{m,n=1}^{3}i_m \left[\delta_{mn} \sum_{s=1}^3
    G_{ss}(\gamma, \bi- \be_m,\bj + \be_n)
    - G_{mn}(\gamma, \bi - \be_m, \bj+
    \be_n) \right],
\end{equation}
where $f_{\bi}^{[\bu_{\beta}, 1]}$ denotes the
expansion coefficients of the distribution function $f(t, \bx, \bv)$
under the expansion center $\tilde{\bu} = \bu_{\beta}$ and
$\tilde{T} = 1$ as
\begin{equation}
  \label{eq:dis_f_alpha_beta}
  f(t,\bx,\bv)=\sum_{\bi\in\bbN^3}f_{\bi}^{[\bu_{\beta},1]}(t,\bx)\mH_{\bi}^{[\bu_{\beta},1]}(\bv). 
\end{equation}
Here, $ G_{mn}(\gamma, \bi, \bj)$ is defined in
\cite[Eq.(3.14)]{FPL2018} and also precalculated. The detailed
calculation of \eqref{eq:detail_expan_FPL_beta_coe} is given in the
Appendix \ref{app:linear_coe} and we refer readers to Proposition 3
and Theorem 4 in \cite{FPL2018} for the calculation of
$ G_{mn}(\gamma, \bi, \bj)$.

\subsection{The reduced collision model}
\label{sec:reduced_col}

For the series expansion of the quadratic collision operator
\eqref{eq:expansion_q1}, although the coefficients
$A_{\bi}^{\bj,\bk}$ can be precalculated and 
kept for later use, both the storage cost and the
computational cost for one single collision are too expensive for
spatially nonhomogeneous problems. To be precise, the cost is
$\mathcal{O}(M^9)$, with $M$ being the
expansion order, which is introduced in Section
\ref{sec:model}. To cope with these
issues, we build a reduced quadratic operator $\mQ^{\rm new}[f]$ as
an approximation to $\mQ[f, f]$ in \eqref{eq:collision_ee}, which
consists of two parts:
\begin{equation}
  \label{eq:col_new}
  \mQ^{\rm new}[f] = \nu\mQ^{\rm new}[f, f] + \nu_{\beta}\mQ^{\rm new}_{
    \beta}[f].
\end{equation}
The expansion center here is set as
$\tilde{\bu} = \bz$ and $\tilde{T} = 1$ for the reduced model,
following the choice in Section \ref{sec:quadoperator}, which is
omitted below for simplicity. 
The collision model $\mQ^{\rm new}[f, f]$ is
expanded similarly to \eqref{eq:expansion_q1} as
\begin{equation}
  \label{eq:new_collision}
  \mQ^{\rm new}[f, f](t, \bx, \bv) =  
  \sum_{\bi \in \bbN^3} 
  Q_{\bi}^{{\rm new}}(t, \bx)
  \mH_{\bi}(\bv). 
\end{equation}
To build the reduced collision model, we assume that the lower-order
terms in the expansion are much more important than the
higher-order ones, especially for capturing
macroscopic variables such as the density $\rho$, macroscopic velocity
$\bu$ and temperature $T$. Thus, the expansion coefficients from the
more precise model \eqref{eq:expansion_q1} are adopted for the
lower-order terms, and those from the diffusive FP operator
\eqref{eq:expan_FP} are utilized to make up for the higher-order
terms. Precisely, the expansion coefficients of
$\mQ^{\rm new}[f, f]$ \eqref{eq:new_collision} are determined as
\begin{equation}
  \label{eq:new_collision_coe}
  Q_{\bi}^{{\rm new}}(t, \bx) = 
  \begin{cases}
    Q_{\bi}(t, \bx), & |\bi|
    \leqslant M_0,\\
    \mu_0 {\rm FP}_{\bi}(t, \bx) ,& |\bi| > M_0,
  \end{cases}
\end{equation}
where $Q_{\bi}(t, \bx)$ are calculated as \eqref{eq:expansion_q1}
using the precalculated coefficients and ${\rm FP}_{\bi}(t, \bx)$ are
the expanding coefficients of the diffusive FP operator
\eqref{eq:expan_FP}. The expansion order of the quadratic collision
term $M_0$, to which we also refer as the quadratic length, and the
decay rate of higher-order coefficients $\mu_0$ are the parameters of
this model. In the numerical experiment, the damping rate $\mu_0$ is
chosen as $\mu_0 = {\rm DIM} -1$ according to the isotropic model
derived for the Fokker-Planck equation in \cite{Villani1998on}, where
${\rm DIM}$ is the number of dimensions of the microscopic velocity space.

For the collision operator between different species $\mQ_{\beta}[f]$
\eqref{eq:expan_FPL_beta}, its computational cost is much less
than that of the quadratic collision
operator \eqref{eq:expansion_q} and hence we do not reduce it further
in the reduced collision model. For the
convenience of computation, the same expansion center
$\tilde{\bu} = \bz$ and $\tilde{T} = 1$ is chosen for
$\mQ_{\beta}^{\rm new}[f]$, where Theorem \ref{thm:project} is
utilized to build the new collision operator for $\mQ_{\beta}[f]$
as
\begin{equation}
  \label{eq:col_beta_new}
  \mQ^{\rm new}_{\beta}[f] = \sum_{\bi \in \bbN^3} 
  \mQ_{\beta, \bi}^{{\rm new}}(t, \bx)
  \mH_{\bi}(\bv),
\end{equation}
where $\mQ_{\beta, \bi}^{{\rm new}}(t, \bx)$ is projected from
$\mQ_{\beta, \bi}^{[\bu_{\beta}, 1]}(t, \bx)$
\eqref{eq:expan_FPL_beta} by Theorem \ref{thm:project}.

Consequently, together with \eqref{eq:col_new},
\eqref{eq:new_collision_coe} and \eqref{eq:col_beta_new}, the reduced
collision operator $\mQ^{\rm new}[f]$ is expanded as
\begin{equation}
  \label{eq:col_new_coe}
  \mQ^{\rm new}[f] = \sum_{\bi \in \bbN^3} 
  Q_{\bi}^{{\rm new}}(t, \bx)
  \mH_{\bi}(\bv)
\end{equation}
with
\begin{equation}
  \label{eq:col_new_coe_1}
  \begin{aligned}
    Q_{\bi}^{{\rm new}}(t, \bx) & = \nu Q_{\bi}^{{\rm new}}(t, \bx) +
    \nu_{\beta}
    Q_{\beta, \bi}^{{\rm new}}(t, \bx) \\
    & =
  \begin{cases}
    \nu Q_{\bi}(t, \bx) + \nu_{\beta}Q_{\beta, \bi}^{{\rm new}}(t,
    \bx)
    ,& |\bi| \leqslant M_0,\\[4mm]
    \nu\mu_0 {\rm FP}_{\bi}(t, \bx) + \nu_{\beta}Q_{\beta, \bi}^{{\rm
        new}}(t, \bx), & |\bi| \geqslant M_0.
  \end{cases}
    \end{aligned}
\end{equation}

In the numerical scheme, which is further discussed in Section
\ref{sec:col step}, the computational cost to obtain expansion
coefficients for the quadratic collision term is $\mO(M_0^9)$,
and those for the linear part and the projection are $\mO(M^3)$
and $\mO(M^4)$ \cite{Hu2020Numerical}, respectively. Therefore, the
total computational cost to obtain the collision
term is $\mO(M_0^9 + M^4)$.  Since $M_0$ is always much smaller than
$M$ in the numerical computation, the reduced collision model can
tremendously reduce the computational cost compared with the original
computational cost of $\mO(M^9)$.

\begin{remark}
  As aforementioned, a larger $M_0$ produces a more
  accurate model, but there is no fixed principle 
  regarding how to choose $M_0$, which may be 
  determined on a case-by-case basis, restrained
    by the storage cost. The numerical results show that even a small
  $M_0$ can capture several expected physical phenomena successfully,
  which is further demonstrated in Section \ref{sec:num}.
\end{remark}


%% file: article_scheme.tex
\section{Numerical algorithm for the FPL equation}
\label{sec:model}
In the previous section, the expansion of
the distribution functions and the collision terms
was discussed.  In this section, we
introduce the specific numerical method for
solving the FPL equation, which is an extension of the method in
\cite{VPFP2016}.

Due to the complex form of the FPL equation, the Strang splitting
method \cite{FVM} is adopted here to split the FPL equation into three
parts:

\begin{itemize}
\item the convection step:
  \begin{equation}
    \label{eq:convection}
    \pd{f(t, \bx, \bv)}{t} + \bv \cdot \nabla_{\bx} f(t, \bx, \bv) = 0, 
  \end{equation}
\item the collision step:
  \begin{equation}
    \label{eq:collision_step}
    \pd{f(t, \bx, \bv)}{t} = \mQ[f(t, \bx, \bv)],
  \end{equation}
  \item  the acceleration step:
  \begin{gather}
    \label{eq:force}
      \pd{f(t, \bx, \bv)}{t} + \bE(t, \bx) \cdot \nabla_{\bv} f(t, \bx, \bv) = 0,  \\
      \label{eq:force1}
      \bE(t, \bx) = -\nabla_{\bx} \psi(t, \bx), \qquad
      -\Delta_{\bx}\psi = \sum_{\eta} q_{\eta}\int_{\bbR^3} f_{\eta}(\bv)
      \dd \bv.
    \end{gather}
\end{itemize}

To obtain a finite system for computation, we make an approximation to the
distribution function as
\begin{equation}
  \label{eq:discretization}
  f(t, \bx, \bv) \approx\sum_{ \bi \in I_M}
  f_{\bi}^{[\tilde{\bu}, 
    \tilde{T}]}(t, \bx) \mH_{\bi}^{[\tilde{\bu}, \tilde{T}]},
\end{equation}
where $I_M$ is the set of indices, with
\begin{equation}
  I_M=\{\bi = (i_1,i_2,i_3): 0\leqslant |\bi|\leqslant M, i_1, i_2, i_3\in \bbN\},
\end{equation}
and $M\in\mathbb{Z}_+$ is the expansion order. The distribution
function $f(t, \bx, \bv)$ is determined by the coefficients
$\{f_{\bi}^{[\tilde{\bu}, \tilde{T}]}(t, \bx), |\bi|\leqslant M\}$,
which are stored as a vector in the implementation as
\begin{equation}
  \label{eq:fvec}
  \bbf^{[\tilde{\bu}, \tilde{T}]} = \left(f_{\bz}^{
      [\tilde{\bu}, \tilde{T}]}, f_{\be_1}^{[\tilde{\bu},
      \tilde{T}]},
    f_{\be_2}^{[\tilde{\bu},\tilde{T}]}
    f_{\be_3}^{[\tilde{\bu}, \tilde{T}]}, \cdots,
    f_{\bi}^{[\tilde{\bu}, \tilde{T}]},
    \cdots\right)_{|\bi|\leqslant M}^T.
\end{equation}
Thus, the reduced collision term \eqref{eq:col_new_coe} is also
approximated as
\begin{equation}
  \label{eq:discretizationq}
  \mQ^{\rm new}[f]  \approx\sum_{ \bi\in I_M}
  Q_{\bi}^{{\rm new}}(t, \bx) \mH_{\bi},
\end{equation}
which is determined by the coefficients
$\{Q_{\bi}^{{\rm new}}(t, \bx), |\bi|\leqslant M\}$ and stored as
\begin{equation}
  \label{eq:qvec}
  \bbQ^{\rm new} = \left(Q_{\bz}^{\rm new}, Q_{\be_1}^{\rm
      new}, Q_{\be_2}^{\rm new}, Q_{\be_3}^{\rm new}, \cdots, Q_{\bi}^{\rm new},
    \cdots\right)_{|\bi|\leqslant M}^T.
\end{equation}
Here, 
the length of these vectors is
\begin{equation}
  \label{eq:N}
  N = \frac{(M+1)(M+2)(M+3)}{6}.
\end{equation} 
For the spatial space, we restrict our study to
the one-dimensional spatial space, and the standard finite volume
discretization is adopted along that direction. Let
 $\Gamma_h$ be a uniform mesh in $\Omega \in \bbR$,
with an index $s$ as the identifier of each cell and $x_0$
as the left point. The mesh $\Gamma_h$ can be
expressed by
\begin{equation}
  \label{eq:Gamma}
  \Gamma^h = \{\Gamma_s  = x_0 + (s h , (s+1)h): s \in
  \bbN\}, 
\end{equation}
and the volume average value of
$\bbf^{[\tilde{\bu}, \tilde{T}]}$ and $\bbQ^{\rm new}$ at the cell
$\Gamma_s$ are $\bbf_s^{[\tilde{\bu}, \tilde{T}]}$ and
$\bbQ_{s}^{\rm new}$, respectively. In the following sections,
a numerical scheme is
proposed to update the distribution function or, more precisely,
the coefficients $\bbf_s^{[\tilde{\bu}, \tilde{T}]}$ at each step.

We should point out that the expansion center $\tilde{\bu}$ and
$\tilde{T}$ in \eqref{eq:fvec} is chosen differently at each step for
different purposes.  The expansion center is set as the standard
expansion center $\tilde{\bu} = \bz$ and $\tilde{T} =1$ at the
collision step to utilize the reduced collision model
\eqref{eq:new_collision}. The same expansion center is utilized at the
convection step to reduce the computational cost of the projection.
Furthermore, the local velocity and
temperature are adopted as the expansion center at the acceleration
step. Therefore, the governing equation \eqref{eq:force} could be
reduced to an ODE. The selection of the
expansion centers at each step is further explained in the following
sections.

\subsection{Convection step}
We begin the explanation of the numerical scheme from the convection
step.  To reduce the computational cost, we
choose $\tilde{\bu} = \bz$ and $\tilde{T} = 1$, the same expansion
center as that in the collision step here.  Then, the approximation to
the distribution function \eqref{eq:discretization} is reduced to
\begin{equation}
  \label{eq:discretization_con}
  f(t, \bx, \bv) \approx\sum_{ \bi \in I_M}
  f_{\bi}(t, \bx) \mH_{\bi}.
\end{equation}
In this case, one projection is saved, which reduces
 the computational cost of $O(M^4)$ due to the same
expansion center being used at the convection and collision
steps.

By substituting \eqref{eq:discretization_con} into
\eqref{eq:convection} and matching the corresponding coefficients, we
derive the equations for coefficients $f_{\bi}$ as
\begin{equation}
  \label{eq:f_alpha}
  \pd{}{t} f_{\bi} + \pd{}{x} \left( (i_1+1)f_{\bi +
      \be_1} + f_{\bi - 
      \be_1}   \right) = 0, \qquad |\bi| \leqslant M,
\end{equation}
where terms with negative indices are regarded as zero.  With the
coefficient vector introduced in \eqref{eq:fvec}, \eqref{eq:f_alpha}
can be rewritten as
\begin{equation}
  \label{eq:f_eq}
  \pd{\bbf}{t} + \mA \pd{\bbf}{x} = 0,
\end{equation}
where $\mA$ is an $N\times N$ matrix, 
the entries of which are determined by \eqref{eq:f_alpha}.

Supposing that $\bbf_s^n$ is the numerical solution to $\bbf$ at time
$t^n$ and cell $s$, the convection equation \eqref{eq:f_eq} is solved
by the forward Euler scheme as
\begin{equation}
  \label{eq:scheme}
  \bbf_s^{n+1, \ast} =\bbf_s^n -
  \frac{\Delta t}{\Delta x}[F_{s+1/2}^n - 
  F_{s-1/2}^n], 
\end{equation}
where $\bbf_s^{n+1, \ast}$ denotes the numerical solution after the
convection step at time $t^{n+1}$ and $F_{s+1/2}^n$ is the numerical
flux through the boundary of the cells $\Gamma_s$ and $\Gamma_{s+1}$.
In our method, the HLL flux \cite{VPFP2016} is utilized, which
has the following form:
\begin{equation}
  \label{eq:HLL}
  F_{s+1/2}^n = \left\{
    \begin{array}{ll}
      \mA \bbf_{s}^{n} & \lambda^L
                                \geqslant 0,
      \\[2mm] 
      \dfrac{  \lambda^R \mA \bbf_{s}^{n} - \lambda^L \mA 
      \bbf_{s+1}^{n} + 
      \lambda^R\lambda^L\left(\bbf_{s+1}^{n} -
      \bbf_{s}^{n}\right) }{\lambda^R - 
      \lambda^L}, & \lambda^L < 0 < \lambda^R, \\[2mm]
      \mA \bbf_{s+1}^{n}, & \lambda^R \leqslant 0,
    \end{array}
  \right.
\end{equation}
where $\lambda^L $ and $\lambda^R$ are the smallest and largest
characteristic velocities, with
$\lambda^L = - C_{M+1}$ and $\lambda^R = C_{M+1} $. Here,
$C_{M+1}$ is the maximum root of the Hermite polynomial of degree
$M+1$. To obtain a
high-order numerical scheme, the linear reconstruction
\cite{Hu2020Numerical} is adopted for the distribution function. In
addition, the time step is decided by the CFL condition
\begin{equation}
  \label{eq:CFL}
  \frac{ \Delta t C_{M+1} }{\Delta x} <
  {\rm CFL}. 
\end{equation}

\subsection{Collision step}
\label{sec:col step}

The reduced collision model \eqref{eq:col_new_coe} is utilized here
for the collision step. Substituting \eqref{eq:discretization_con} and
\eqref{eq:discretizationq} into \eqref{eq:collision_step}, we 
obtain the governing equations of $f$ as
\begin{equation}
  \label{eq:collision_eq}
  \pd{\bbf}{t} = \bbQ^{\rm new}.
\end{equation}
Here, \eqref{eq:collision_eq} is solved again by the forward Euler
scheme as
  \begin{equation}
  \label{eq:Euler_col}
  \bbf_s^{n+1, \ast\ast} =
  \bbf_s^{n+1, \ast } + \Delta t \bbQ_s^{{\rm new}, n+1, \ast},
\end{equation}
where $ \bbQ_s^{{\rm new}, n+1,\ast}$ is the numerical solution of the
reduced collision operator $\mQ^{\rm new}$ after the convection step at
time $t^{n+1}$ and cell $s$, with $\bbf_s^{n+1, \ast}$ and
$\bbf_s^{n+1, \ast\ast}$ being the numerical solution after the
convection step and after the collision step at time $t^{n+1}$,
respectively.  High-order Runge-Kutta numerical schemes can also be
adopted to update the collision term in \eqref{eq:Euler_col}.

\subsection{Acceleration step}

At the acceleration step, the expansion center is chosen as the local
macroscopic velocity and temperature or, more precisely,
$\tilde{\bu} = \bu(t, x)$ and $\tilde{T} = T(t, x)$, both defined in
\eqref{eq:rlt}.  Consequently, the governing equation \eqref{eq:force}
is reduced to an ODE system \cite{Wang}, which greatly reduces the
computational cost.

For the one-dimensional spatial problem, where the macroscopic
velocity $\bu(t, x)$ is reduced to $\bu = (u_1, 0, 0)$, the numerical system for the
acceleration step is simply reduced to solving an ODE system of the
macroscopic velocity $u_1$ as
\begin{equation}
  \label{eq:force_1}
  \begin{gathered}
    \pd{u_1}{t} - E_1 = 0, \qquad E_1(t, x) = -\pd{\psi(t,x)}{x},
    \qquad -\partial_{xx}\psi = \sum_{\eta} q_{\eta} \int_{\bbR^3}
    f_{\eta}(\bv) \dd \bv.
\end{gathered}
\end{equation}
A detailed deduction of \eqref{eq:force_1} can be
found in the Appendix \ref{app:acc}.

Since the expansion center at the acceleration step is different from
that at the collision step, Theorem \ref{thm:project} is
utilized to carry out the projections.  We organize the procedure as
below to perform the acceleration step:
\begin{enumerate}
\item Find $\left(\bbf_s^{[\bu, T]}\right)^{n+1, \ast\ast}$ from
  $\bbf_s^{n+1, \ast\ast}$ based on Theorem \ref{thm:project}, where
  $\bbf_s^{n+1, \ast\ast}$ are the numerical solutions after the
  collision step at $t = t^{n+1}$.
\item Solve \eqref{eq:force1} to obtain
  $(F_1)_{s}^{n+1, \ast\ast}$ with the finite difference scheme
  \cite{VPFP2016}.
\item Solve \eqref{eq:force} by the forward Euler scheme
  \begin{equation}
    \label{eq:forwardEuler}
    (u_1)_{s}^{n+1} =   (u_1)_{s}^{n+1,\ast\ast} + \Delta t (F_1)_{ s}^{n+1, \ast\ast},
  \end{equation}
  where $(u_1)_{s}^{n+1,\ast\ast}$ is the macroscopic velocity at cell
  $s$ after the collision step at time $t=t^{n+1}$.
\item Obtain $\left(\bbf_{s}^{[\bu, T]}\right)^{n+1}$ by updating the
  expansion center to $(u_1)_{s}^{n+1}$.
\item Find $\bbf_{s}^{n+1}$ from
  $\left(\bbf_{s}^{[\bu, T]}\right)^{n+1}$ based on Theorem
  \ref{thm:project}.
\end{enumerate}


%% file: article_experiment.tex
\section{Numerical experiments}
\label{sec:num}
In this section, several numerical examples are presented to test the
new algorithm. In all the tests, the CFL is set as $0.45$.  The Landau
damping problems are studied first to show the capability of the new
algorithm to simulate the FPL equations quantitatively.  Two-stream
instability and bump-on-tail instability are also tested to show that
the numerical method can detect the evolution in the microscopic
velocity space with the reduced collision model.

\subsection{Linear Landau damping problem}
 \label{sec:lld} 

 The Landau damping problem is one of the most popular problems in
 plasma physics. It is caused by the strong interactions between
 the electromagnetic wave and particles with velocities comparable to
 the phase velocity, which tend to synchronize with the wave
 \cite{Chen1984}. Particles with velocities
 slightly lower than the phase velocity are accelerated and thus gain
 energy from the wave, while those with slightly higher
 velocities are decelerated and thus lose energy to the wave, which
 results in an exponential decrease in the
 electrostatic energy of the wave. The linear Landau damping problem
 has been studied in \cite{Filbet2002numerical}, where several
 specific settings of the problems are proposed and simulated. The
 numerical results in \cite{Filbet2002numerical} can be used for
 comparison here.

 The setting of the linear Landau damping problem is adopted from
 \cite{ZhangGamba2017} with $\rho_{\beta} = 1$, $\bu_{\beta} = 0$ and
 the initial data being
\begin{equation}
  \label{eq:ini_ex1}
  f(x,\bv) = \frac{1}{(2\pi)^{3/2}} \exp\left( -\dfrac{|\bv|^2}{2}\right) [1 + A \cos(k       
  x)], \qquad (x, \bv) \in [0, 2\pi/k] \times \bbR^3,
\end{equation}
where $A$ is the amplitude of the perturbation. The periodic
boundary condition is implemented in this example. In the Landau
damping problem, our interest lies in the evolution of the square root
of the electrostatic energy which is defined as
\begin{equation}
  \label{eq:energy}
  \mathcal{E}(t) = \left(\sum_{j} \Delta x E_{1,j}(t)^2\right)^{1/2}.
\end{equation}
According to Landau's theory, $\mE(t)$ should decrease
exponentially with a fixed rate $\omega_i$, which can be regarded as
the imaginary part of the frequency $\omega$. The theoretical damping
rate is often estimated as \cite{ZhangGamba2017,
  sonnendrucker2013numerical}
\begin{equation}
  \label{eq:damping_rate}
  \gamma = \gamma_{L} + \gamma_{C}, 
\end{equation}
where the damping rate of collisionless plasma $\gamma_L$ is
\begin{equation}
  \label{eq:damping_rate_collisionless}
  \gamma_L = 
  \begin{cases}
    -\sqrt{\dfrac{\pi}{8}} \dfrac{1}{k^3} \exp\left(-\dfrac{1}{2k^2} -
      \dfrac{3}{2}\right), & k \textrm { is large}, \\[5mm]
    -\sqrt{\dfrac{\pi}{8}} \left(\dfrac{1}{k^3} - 6k\right)
    \exp\left(-\dfrac{1}{2k^2} - \dfrac{3}{2} - 3k^2 - 12 k^4\right),
    & k \textrm{ is small},
    \end{cases}
\end{equation}
and $\gamma_C$ is the collisional ``correction'' to the
collisionless damping rate:
\begin{equation}
  \label{eq:damping_rate_collision}
  \gamma_{C} = -\frac{1}{3} \nu \sqrt{2/\pi},
\end{equation}
which depends only on the collisional
frequency and reflects the effect of the collision, where $\nu$ refers
to the collisional frequency. In this test, the amplitude of the
perturbation $A$ is set as $10^{-5}$. 
In addition, the expansion order $M$ is set as
$M = 20$, and the grid size as $N = 800$.

\begin{figure}[!htb]
  \centering \subfloat[$k = 0.3, \nu = \nu_{\beta} = 0$]
  {\includegraphics[width=0.49\textwidth, height=0.35\textwidth,
    clip]{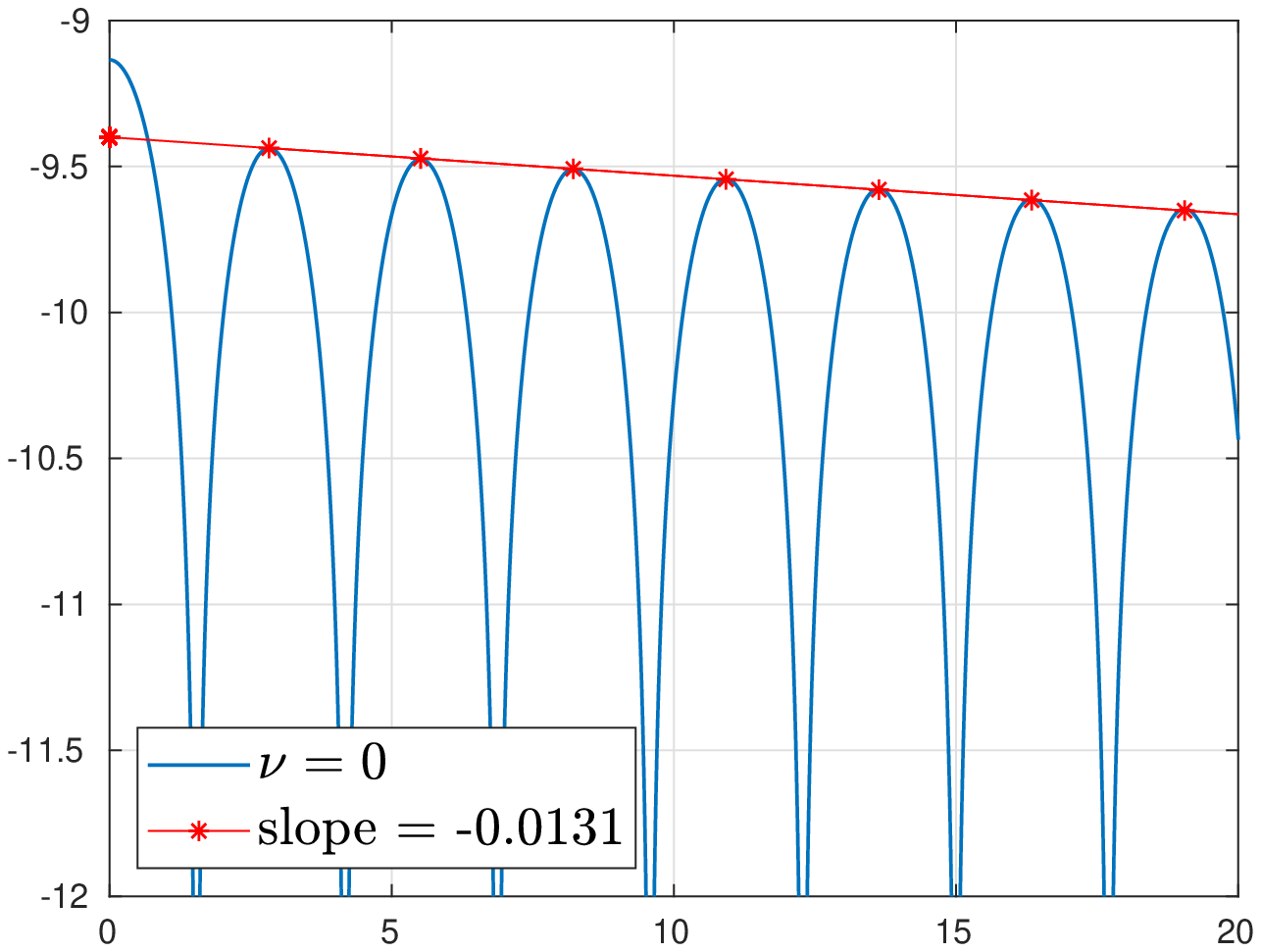}}\hfill
  \subfloat[$k = 0.3, \nu = \nu_{\beta}= 0.01$]
  {\includegraphics[width=0.49\textwidth,
    height=0.35\textwidth,clip]{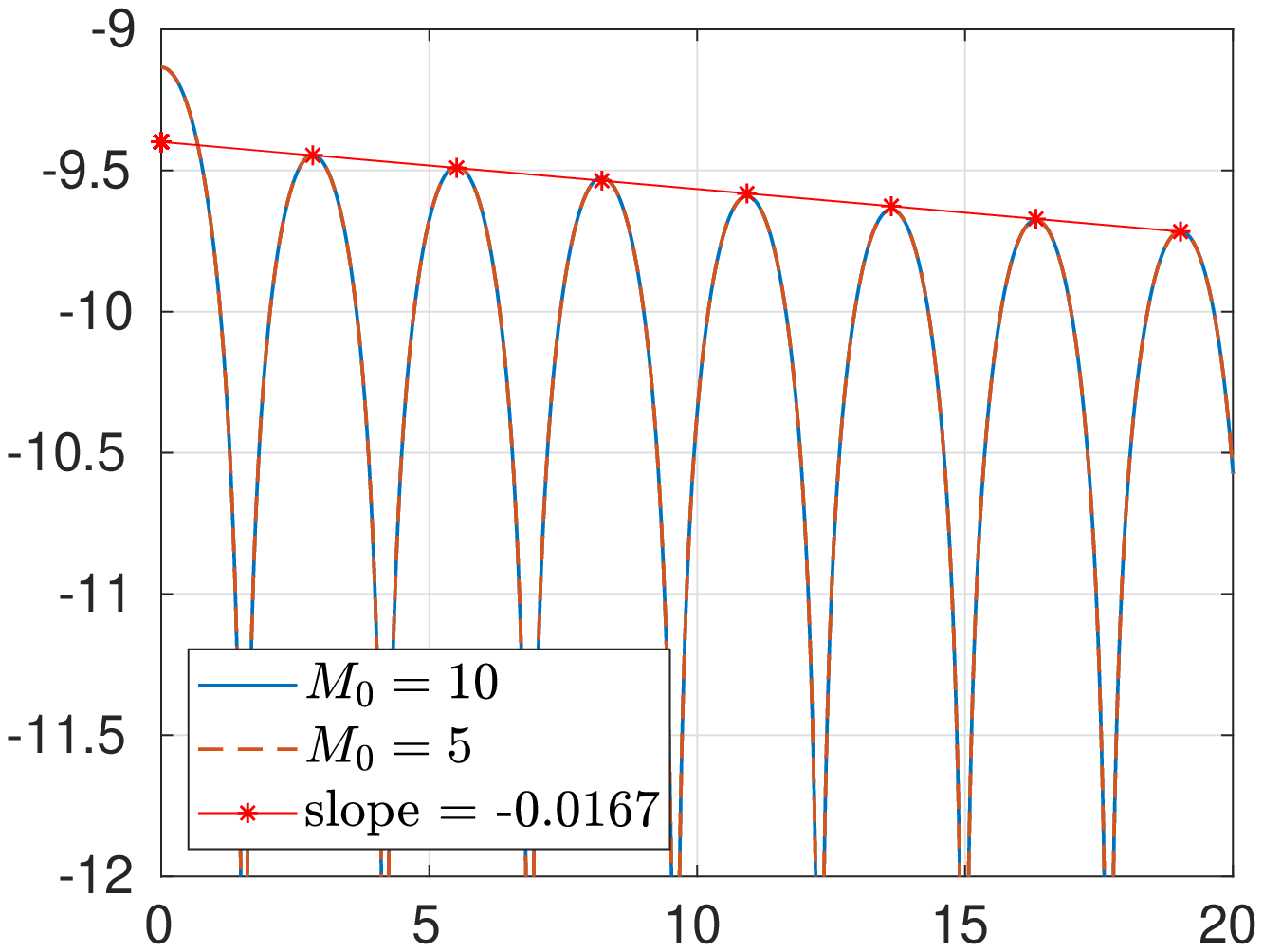}}
  \caption{Time evolution of $\ln(\mE(t))$ with $N=800$ and
      $M=20$ for different $\nu$ in the linear Landau damping problem.
       The wave number $k = 0.3$. For the
      collisional case, the red dashed line corresponds to $M_0 =5$
      while the blue line corresponds to $M_0 =10$. }
\label{fig:ex1_k03_M0}
\end{figure}

\begin{figure}[!htb]
  \centering \subfloat[$k = 0.5, \nu = \nu_{\beta}= 0$]
  {\includegraphics[width=0.49\textwidth, height=0.35\textwidth,
    clip]{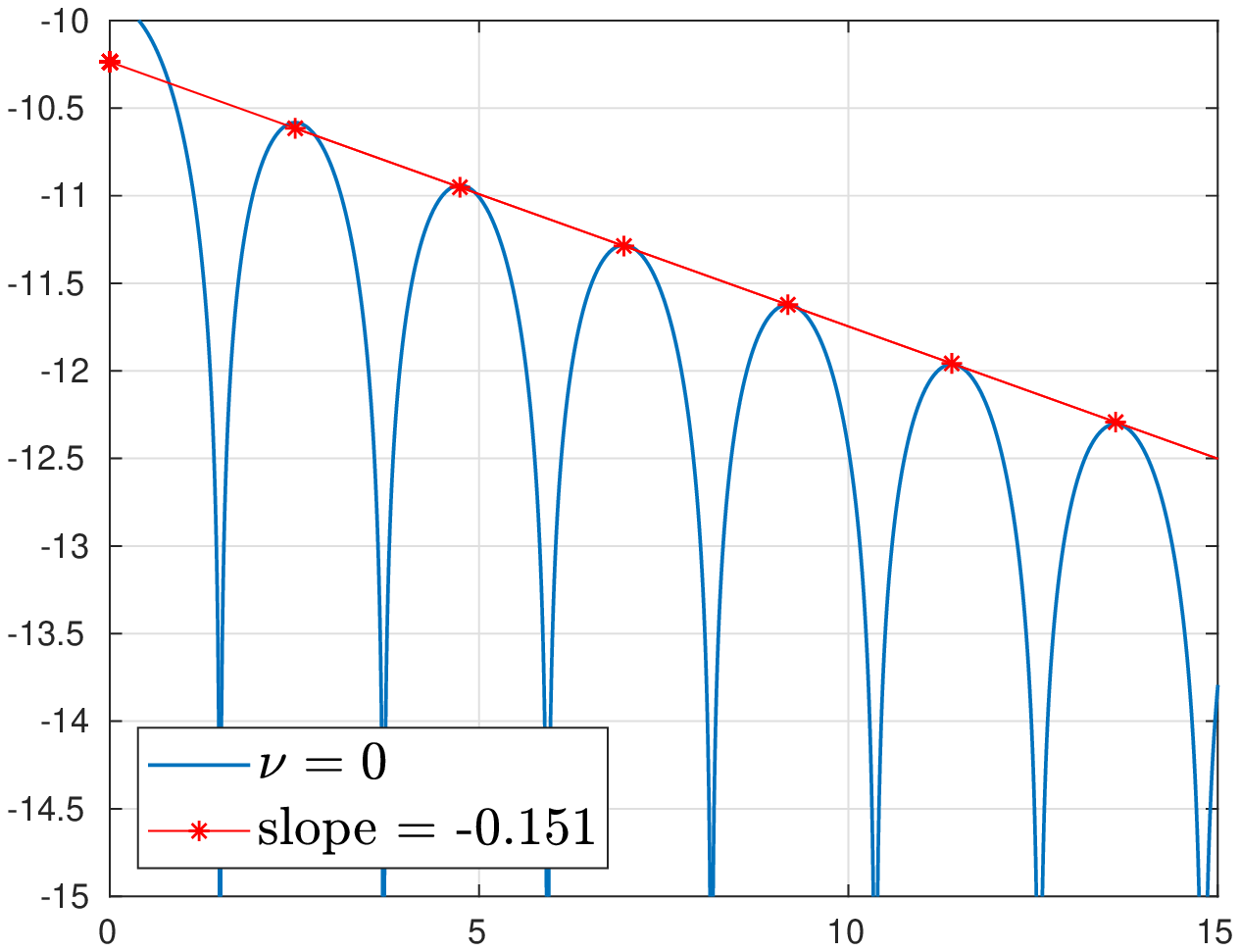}}\hfill
  \subfloat[$k = 0.5, \nu = \nu_{\beta}= 0.01$]
  {\includegraphics[width=0.49\textwidth,
    height=0.35\textwidth,clip]{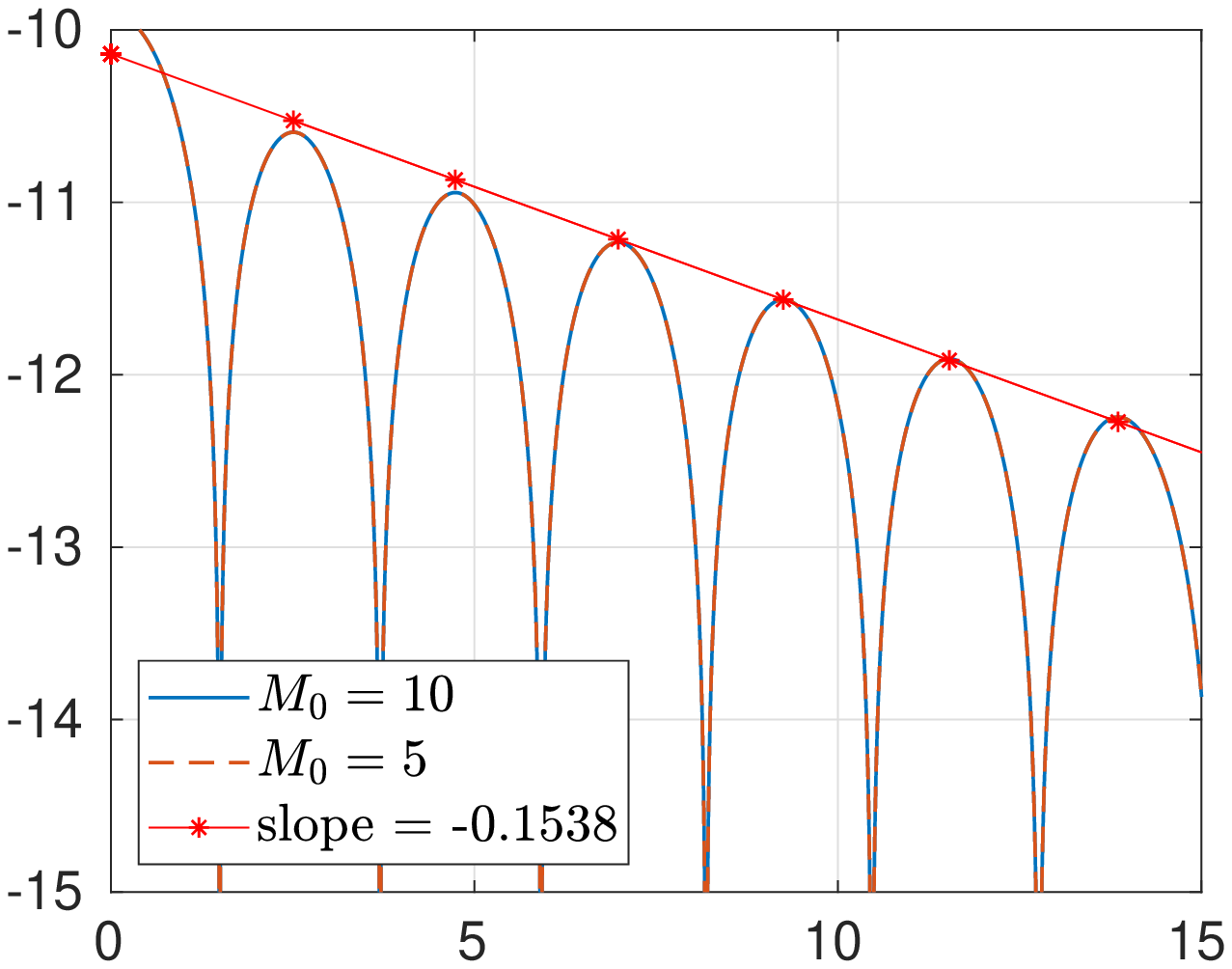}}
  \caption{Time evolution of $\ln(\mE(t))$ with $N=800$ and $M=20$ for
    different $\nu$ in the linear Landau damping problem. 
    The wave number $k = 0.5$. For the collisional
    case, the red dashed line corresponds to $M_0 =5$ while the blue
    line corresponds to $M_0 =10$. }
  \label{fig:ex1_k05_M0}
\end{figure}

Figures \ref{fig:ex1_k03_M0} and \ref{fig:ex1_k05_M0} show the
time evolution of the electrostatic energy $\mE(t)$ with the
wave number $k$ set as $0.3$ and $0.5$, respectively. For both wave
numbers, the Coulomb case $\gamma = -3$ is studied, and the collision
frequency is set as $\nu = \nu_{\beta} = 0$ and $0.01$ to demonstrate
the effect of collision.  The quadratic length $M_0$ is chosen as $5$
and $10$, respectively.  The results show that this method
successfully simulates the linear Landau damping problem and the
numerical damping rate of the electrostatic energy is almost identical
to the theoretical result in \eqref{eq:damping_rate_collisionless} for
both wave numbers.  When the collision is added, the electrostatic
energy shows a faster decay due to the effect of the collision.  This
is reflected in the larger damping rates compared to that of the
collisionless case, and the increase in
damping rates exactly matches the theoretical results in
\eqref{eq:damping_rate_collision}.  This proves the accuracy of both
our Hermite spectral method and our reduced collision model.

Most importantly, the numerical solution with the quadratic length
$M_0 =5$ is almost the same as that with $M_0=10$.  
This indicates that for the linear
Landau damping problem, even with a small quadratic length $M_0=5$, 
our collisional model can capture
the linear Landau damping phenomenon satisfactorily.  For this reason,
the quadratic length is set as $M_0 = 5$ in the linear Landau damping
experiments.

\begin{figure}[!htb]
  \centering \subfloat[$k = 0.3, \nu = \nu_{\beta}= 0.01$]
  {\includegraphics[width=0.49\textwidth, height=0.35\textwidth,
    clip]{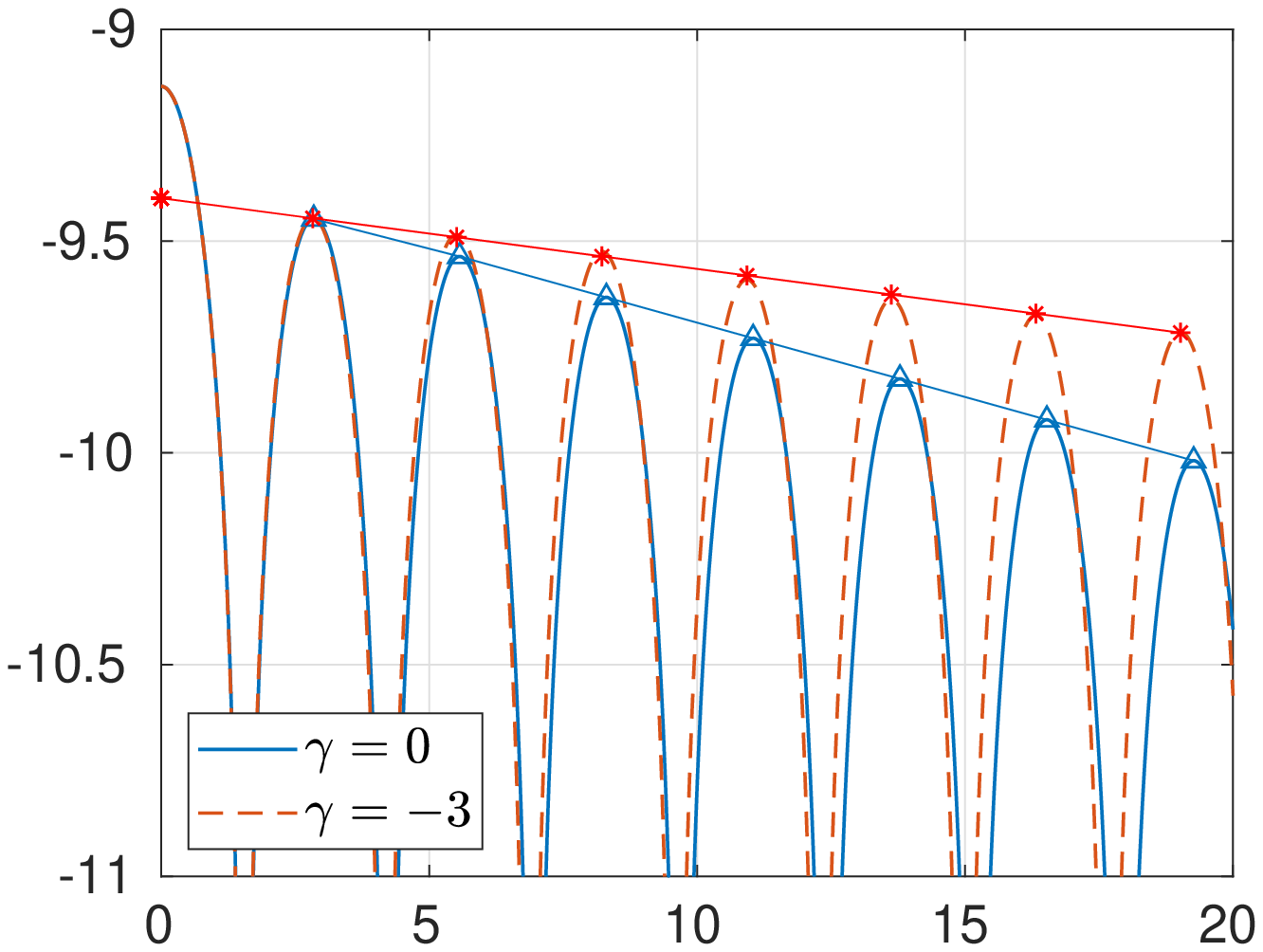}}\hfill
  \subfloat[$k = 0.5, \nu = \nu_{\beta}= 0.01$]
  {\includegraphics[width=0.49\textwidth,
    height=0.35\textwidth,clip]{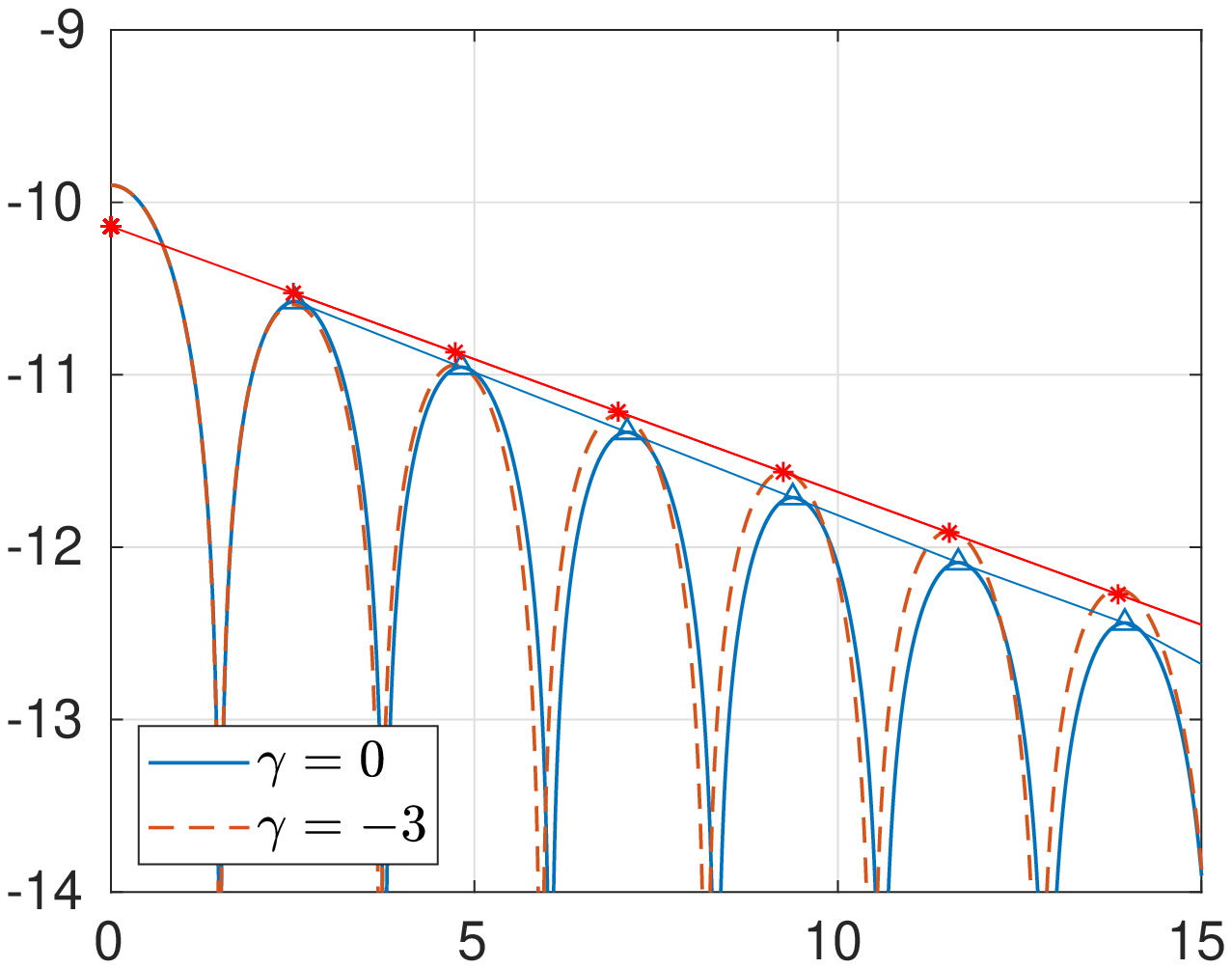}}
  \caption{Time evolution of $\ln(\mE(t))$ with $N=800$ and $M_0=5$ for
    different $\gamma$ in the linear Landau damping problem. 
    The blue line corresponds to $\gamma = 0$
    while the red dashed line corresponds to $\gamma = -3$.}
  \label{fig:ex1_nu001_gamma}
\end{figure}

Then, we test the effect of different IPL models on our numerical
method. The time evolution of the electrostatic energy $\mE(t)$
for different potential indices $\gamma$, the
index in the IPL model \eqref{eq:ipl} aforementioned, 
is tested. Specifically, the model of
Maxwell molecules $\gamma = 0$ and the model with Coulomb interactions
$\gamma = -3$ are tested and compared in Figure
\ref{fig:ex1_nu001_gamma}.  Here, we also set the collisional
frequency $\nu $ as $0.01$ and the wave number $k$ as $0.3$ and $0.5$,
respectively.  The numerical result illustrates that our method is
capable of simulating the linear Landau damping for different
$\gamma$, and we can conclude that the collision model with
softer potential imposes a smaller damping rate.

\subsection{Nonlinear Landau damping}

As shown in the last section, when the wave amplitude $A$ is
sufficiently small, the linear regime is valid, which yields
exponentially decreasing electrostatic energy.  However, the Landau
damping problem with larger amplitude, which 
diverges from the linear theory and hence is also known
as nonlinear Landau damping, is quite a different case.
Typically, one finds that the amplitude decays, grows and
oscillates before settling down to a relatively steady state
\cite{Chen1984}.

In this section, we study the nonlinear Landau damping problem
numerically. The nonlinear Landau damping is primarily attributed to
the ``trapping" phenomenon, 
where a particle is caught in the potential well of a wave,
shuttles back and forth, and ends up gaining and losing energy to the
wave \cite{Chen1984}.

In this numerical experiment, the form of the initial data is the same
as that in the last section, with $A$ augmented to $0.2$ and
the electrostatic energy is again studied.  The nonlinear Landau
damping problem with this particular initial data 
was also studied in \cite{Filbet, ZhangGamba2017}, to
which we refer readers for a
  comparison of the numerical results.  The case of Maxwell molecules
$\gamma=0$ is studied, and the spatial grid size, expansion
order and quadratic length are set as $N=800$,
$M = 20$ and $M_0 = 5$, respectively.  Moreover, to avoid recurrence
\cite{Filter2017}, the expansion order is chosen as $M = 200$ for the
collisionless case.

\begin{figure}[!htb]
  \centering \subfloat[$k = 0.3, \nu = \nu_{\beta}= 0$]
  {\includegraphics[width=0.45\textwidth, height=0.35\textwidth,
    clip]{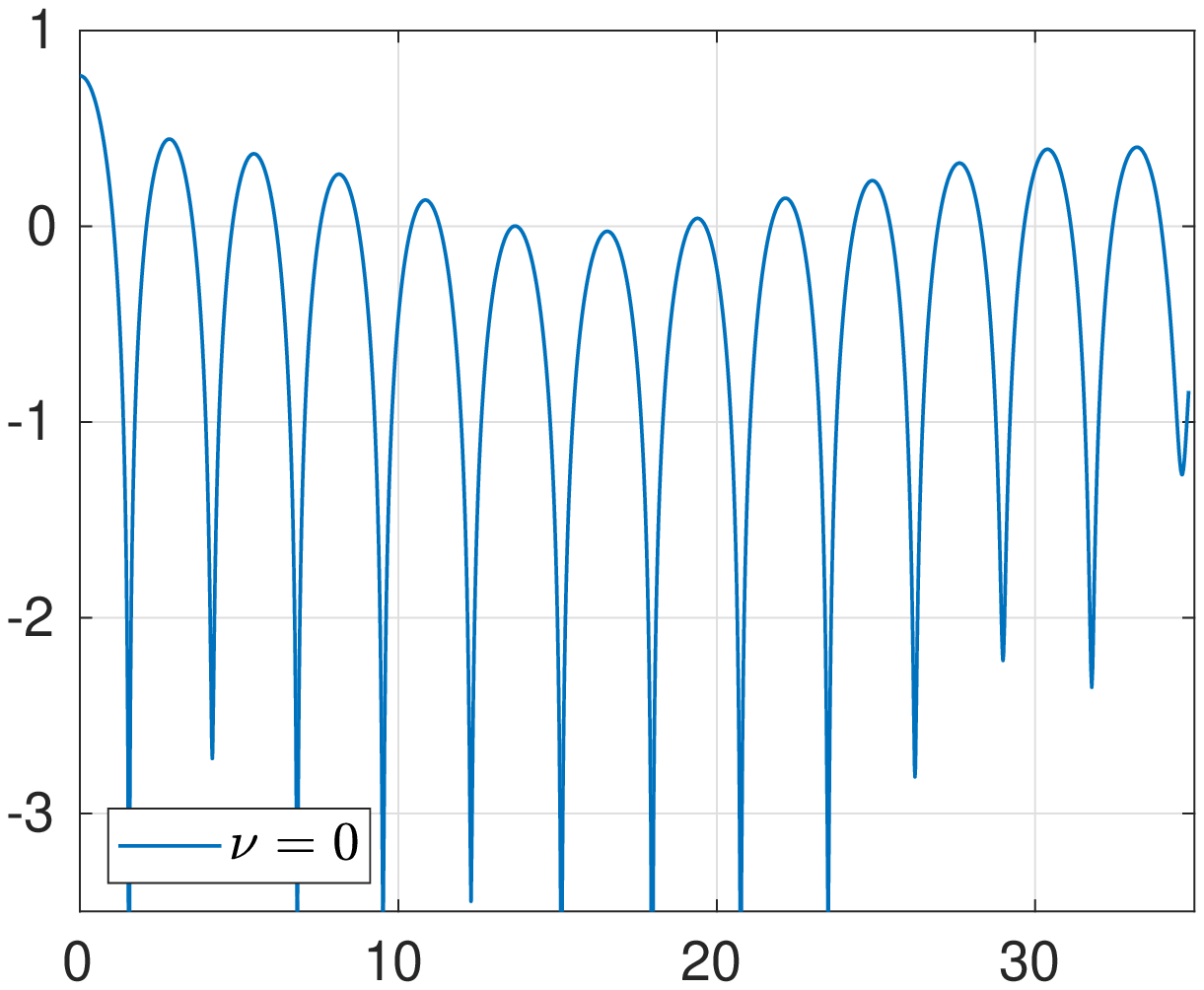}}\hfill
  \subfloat[$k = 0.3, \nu = \nu_{\beta}= 0.01$]
  {\includegraphics[width=0.45\textwidth, height=0.35\textwidth,
    clip]{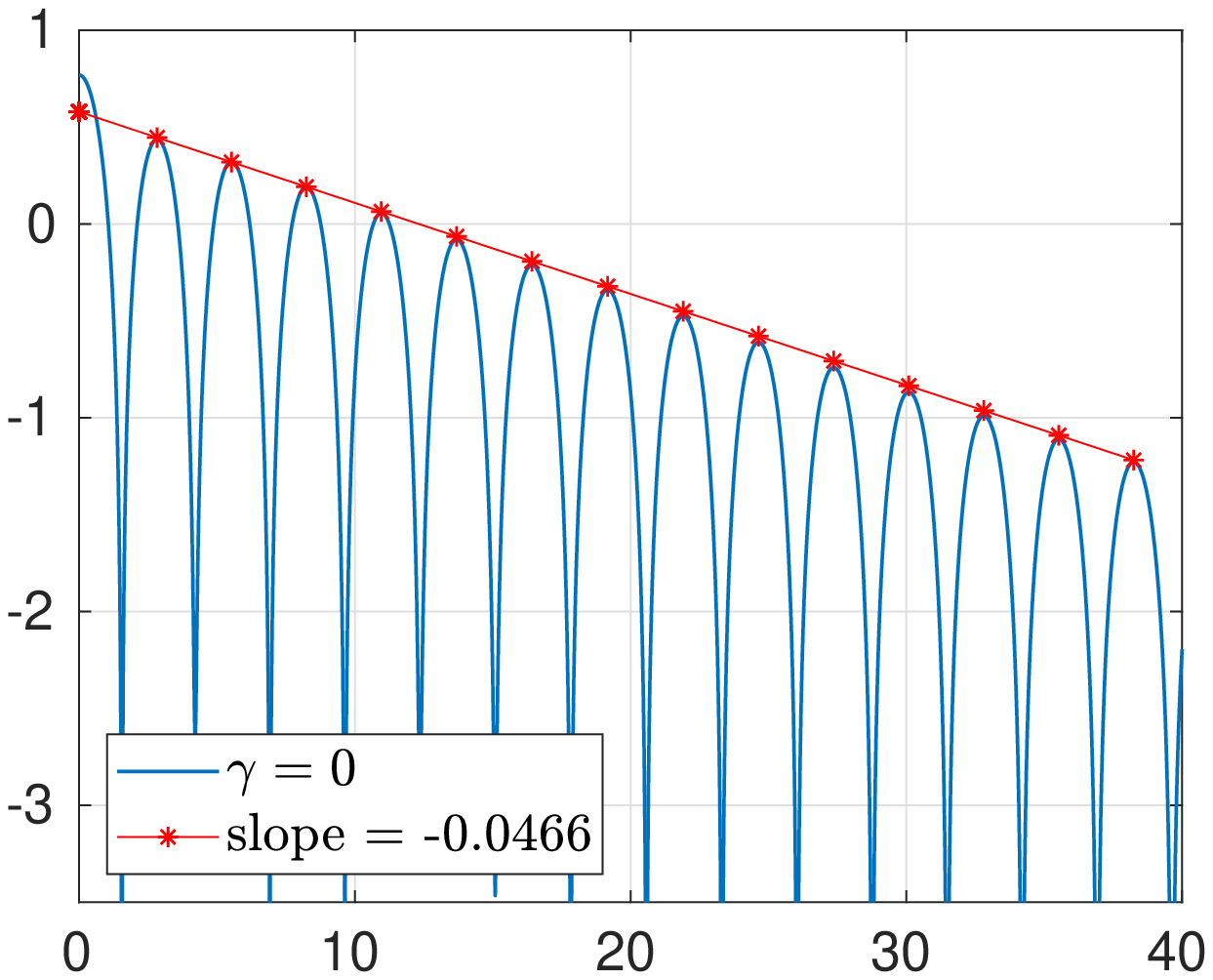}}\hfill
  \subfloat[$k = 0.3, \nu= \nu_{\beta} = 0.05$]
  {\includegraphics[width=0.45\textwidth,
    height=0.35\textwidth,clip]{ex2_nu001_k03_gamma0_c.eps}} \hfill
  \subfloat[$k = 0.3, \nu = \nu_{\beta}= 0.1$]
  {\includegraphics[width=0.45\textwidth,
    height=0.35\textwidth,clip]{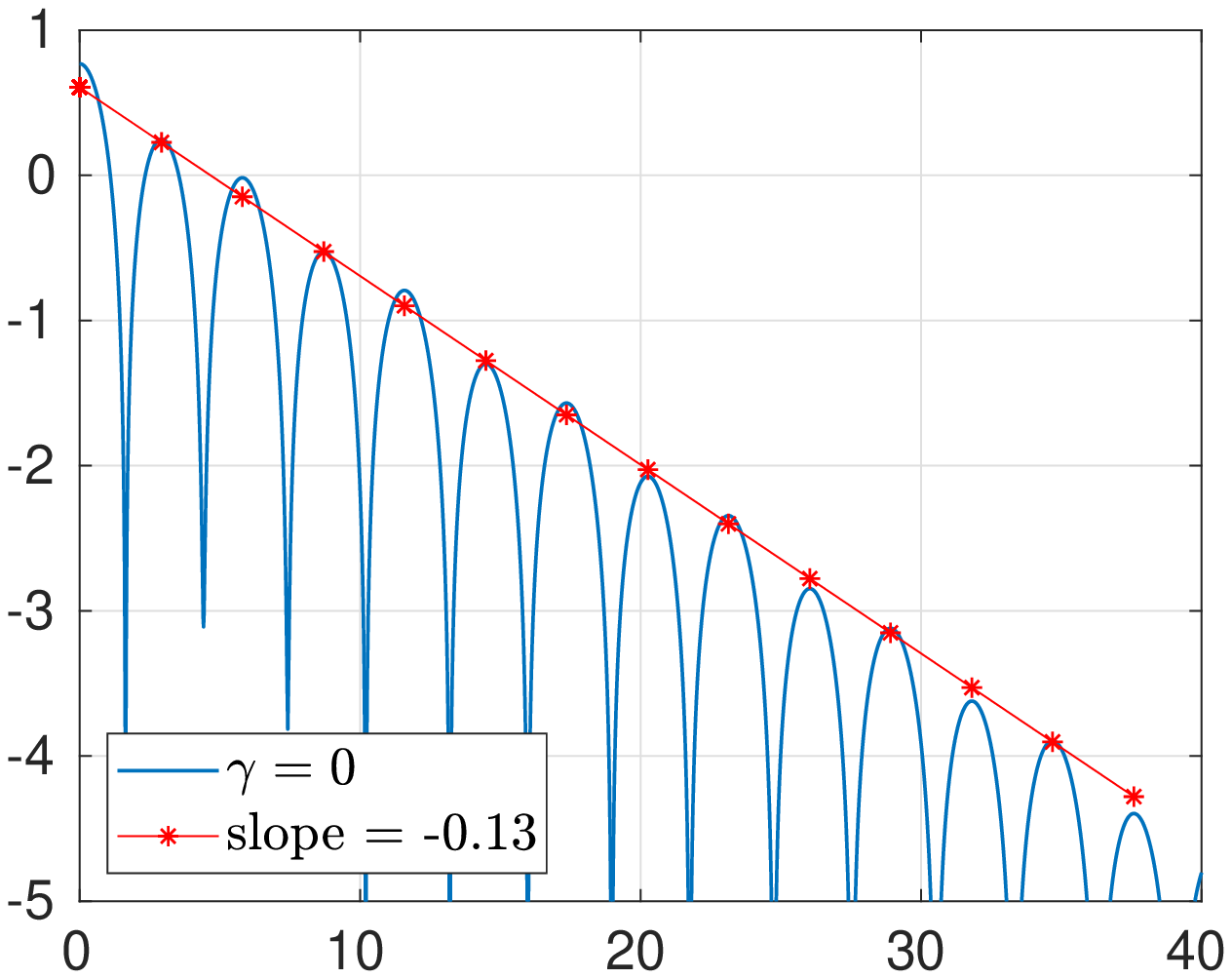}}
  \caption{Time evolution of $\ln(\mE(t))$ with $N=800$ and $M_0=5$
    for different collisional frequencies
    $\nu = \nu_{\beta}= 0, 0.01, 0.05$ and $0.1$
    in the nonlinear Landau damping problem. The wave number
    $k = 0.3$.}
  \label{fig:ex2_frequency_03}
\end{figure}

\begin{figure}[!htb]
  \centering \subfloat[$k = 0.5, \nu= \nu_{\beta} = 0$]
  {\includegraphics[width=0.45\textwidth, height=0.35\textwidth,
    clip]{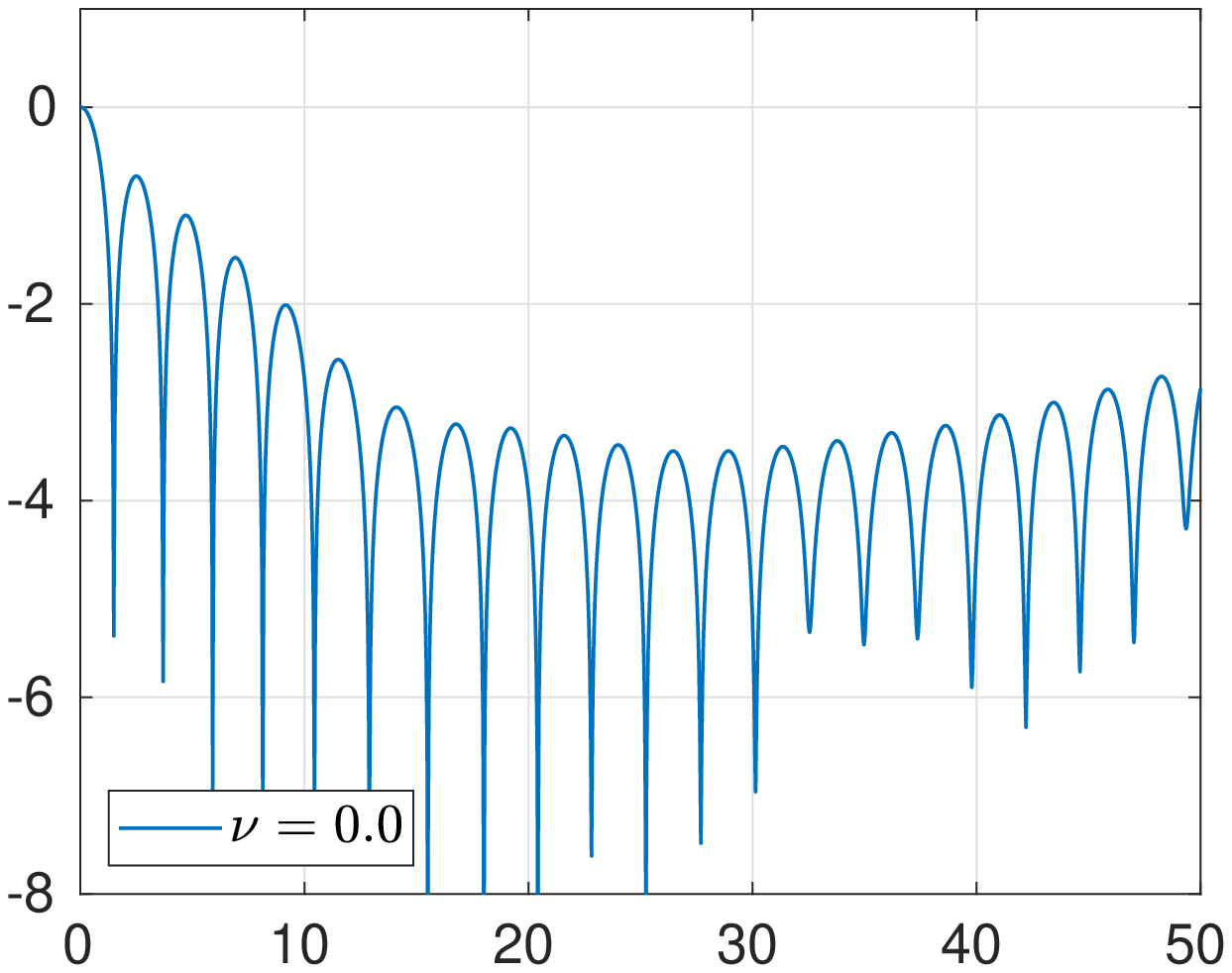}}\hfill
  \subfloat[$k = 0.5, \nu = \nu_{\beta}= 0.01$]
  {\includegraphics[width=0.45\textwidth, height=0.35\textwidth,
    clip]{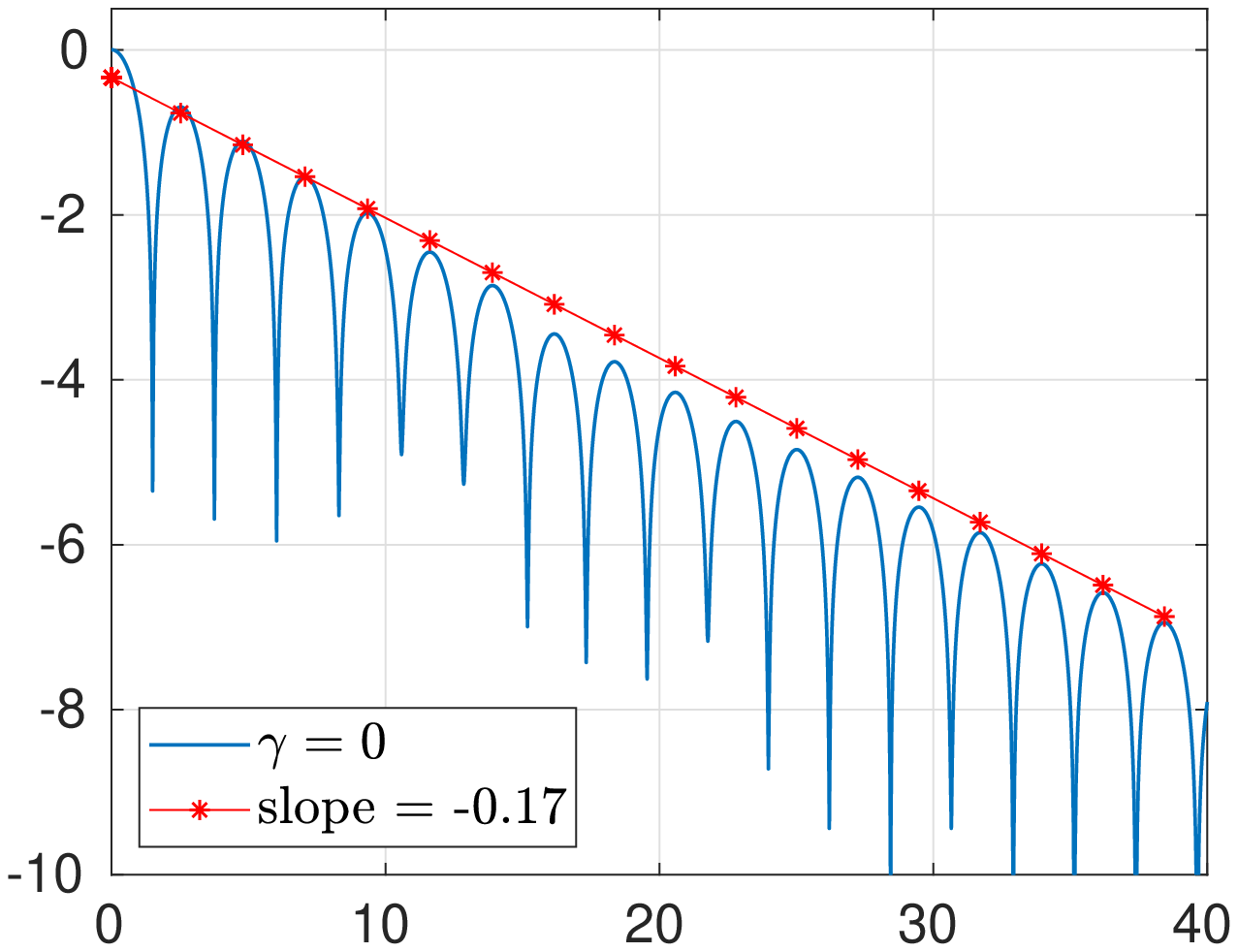}}\hfill
  \subfloat[$k = 0.5, \nu = \nu_{\beta}= 0.05$]
  {\includegraphics[width=0.45\textwidth,
    height=0.35\textwidth,clip]{ex2_nu001_k05_gamma0_c.eps}} \hfill
  \subfloat[$k = 0.5, \nu = \nu_{\beta}= 0.1$]
  {\includegraphics[width=0.45\textwidth,
    height=0.35\textwidth,clip]{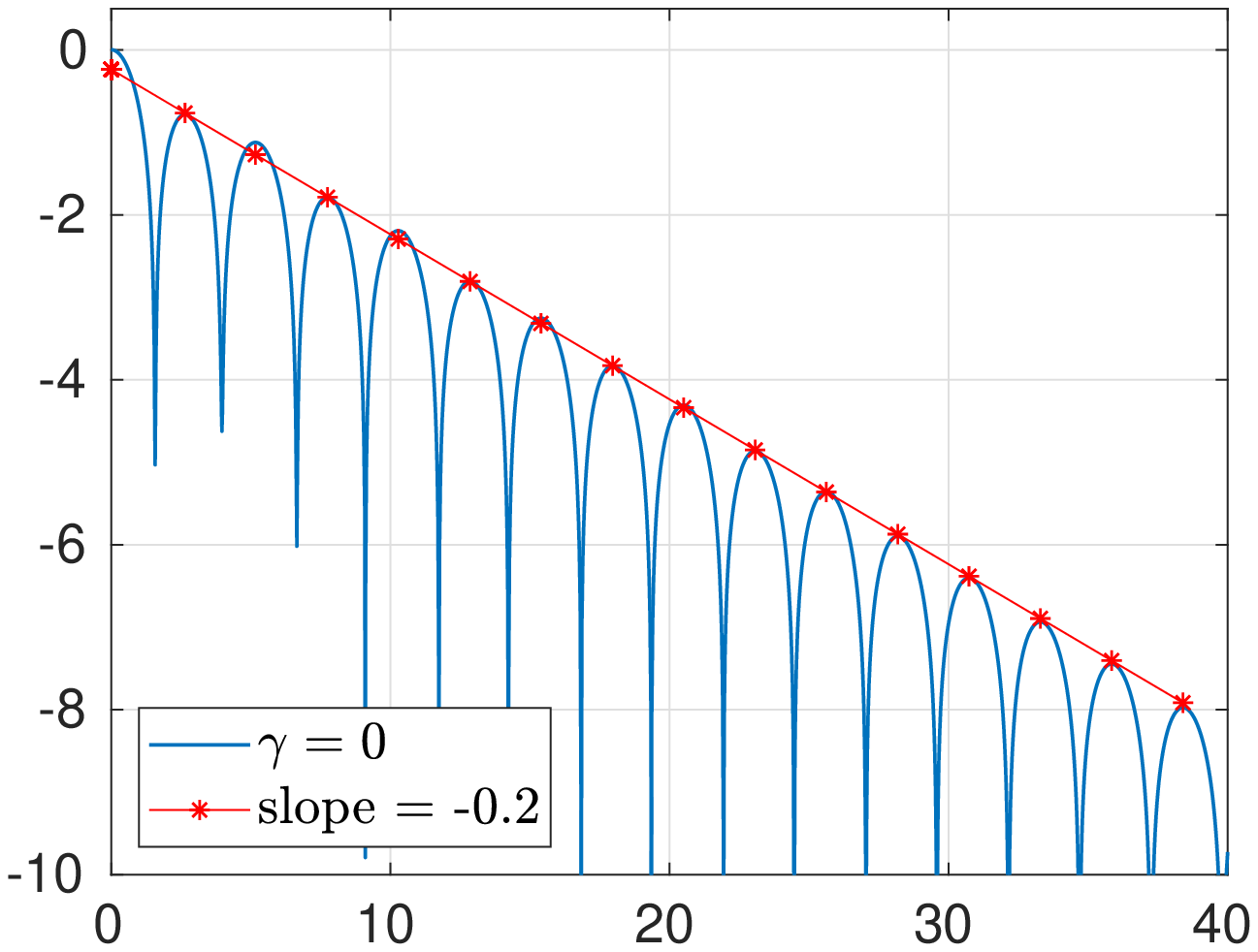}}
  \caption{Time evolution of $\ln(\mE(t))$ with $N=800$ and $M_0=5$
    for different collisional frequencies
    $\nu = \nu_{\beta} = 0, 0.01, 0.05$ and $0.1$
    in the nonlinear Landau damping problem. The wave number
    $k = 0.5$.}
  \label{fig:ex2_frequency_05}
\end{figure}

Figures \ref{fig:ex2_frequency_03} and
\ref{fig:ex2_frequency_05} show the time evolution of electrostatic
energy for $k = 0.3$ and $k = 0.5$ with collisional frequency
$\nu = \nu_{\beta}=0, 0.01,0.05$ and $0.1$.  We can conclude
that for the nonlinear collisionless problem, instead of exponential
damping as in the linear case, the electrostatic energy decreases
exponentially at the beginning and then grows exponentially at
 a smaller rate, which is
consistent with the results achieved by \cite{Cheng2014,
  ZhangGamba2017}. For the collisional case, we find that the
electrostatic energy exhibits an exponential-like damping for both
wave numbers $k = 0.3$ and $0.5$ and the damping rate increases with
the collisional frequency. These results are reasonable because
stronger collision implies more frequent energy exchange between
particles and results in less ``trapping'' phenomena
and faster damping rates. This numerical result
also accords with that in \cite{ZhangGamba2017, Filbet}.

\begin{figure}[!htb]
    \centering \subfloat[$k = 0.3, \nu = \nu_{\beta}= 0.05$]
    {\includegraphics[width=0.45\textwidth, height=0.35\textwidth,
      clip]{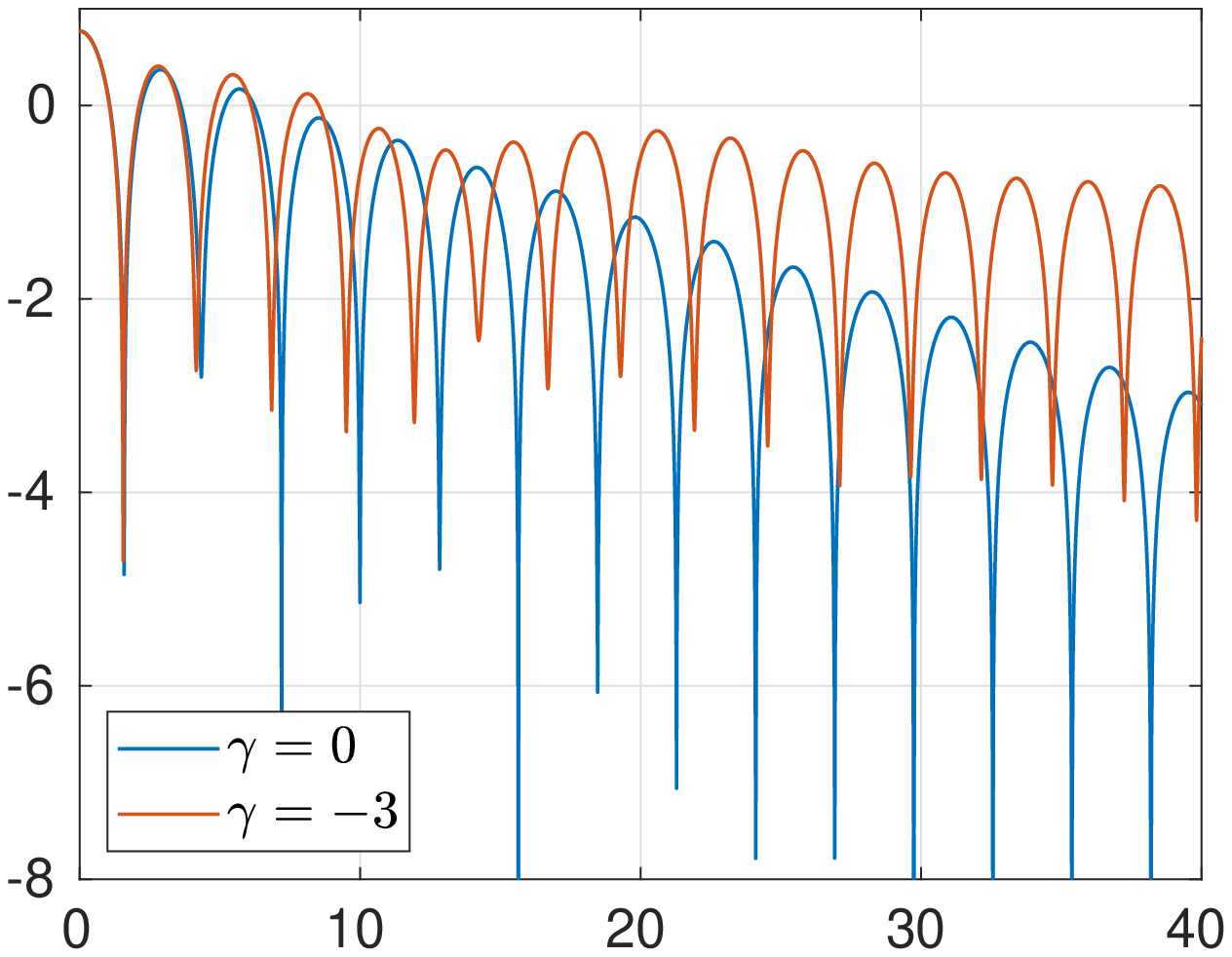}}\hfill
    \subfloat[$k = 0.3, \nu = \nu_{\beta}= 0.1$]
    {\includegraphics[width=0.45\textwidth,
      height=0.35\textwidth,clip]{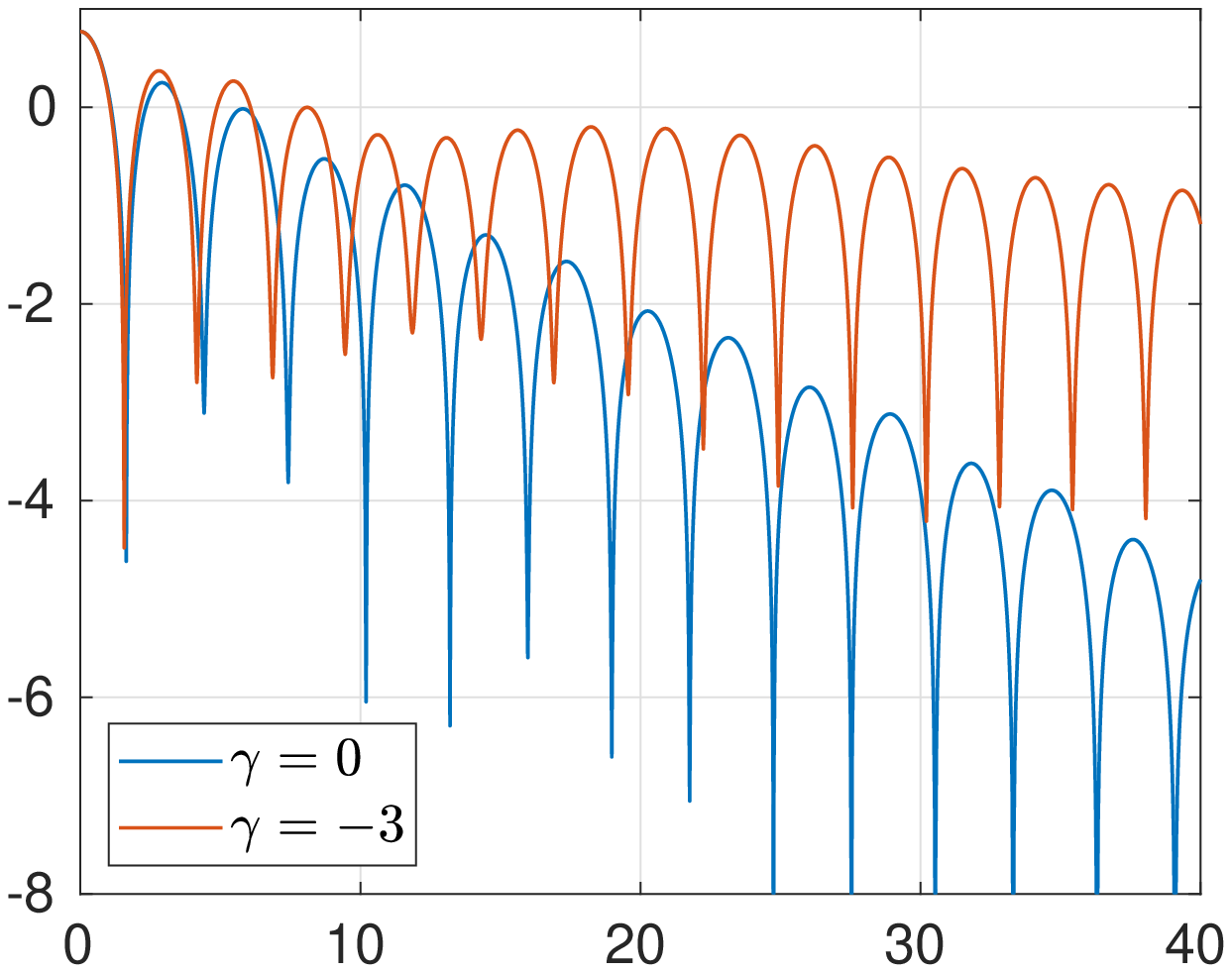}} \\
    \subfloat[$k = 0.5, \nu = \nu_{\beta}= 0.05$]
    {\includegraphics[width=0.45\textwidth, height=0.35\textwidth,
      clip]{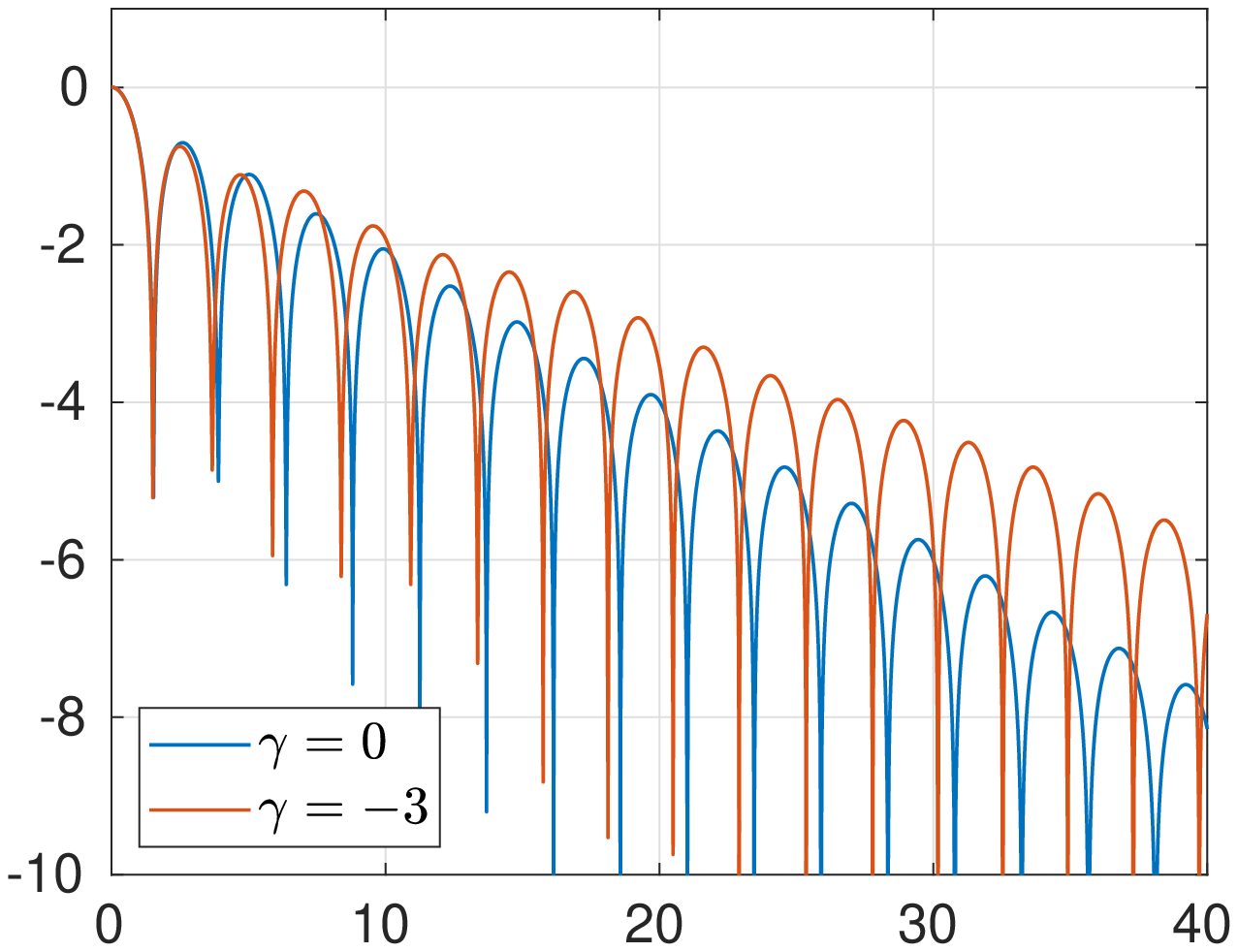}}\hfill
    \subfloat[$k = 0.5, \nu = \nu_{\beta}= 0.1$]
    {\includegraphics[width=0.45\textwidth,
      height=0.35\textwidth,clip]{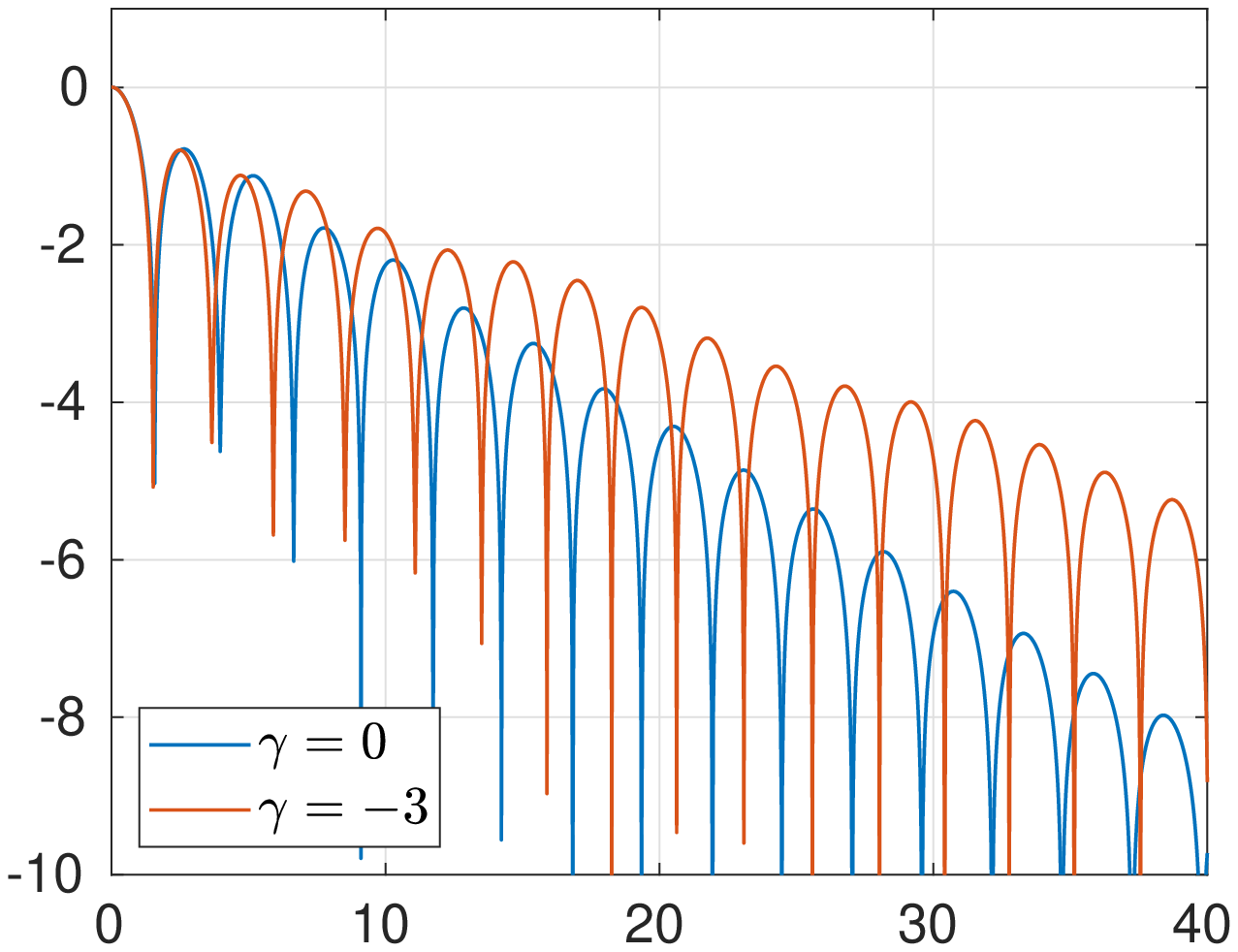}}
    \caption{Time evolution of $\ln(\mE(t))$ with $N=800$ and $M_0=5$
      for different potential indices $\gamma$ in the 
      nonlinear Landau damping problem, where the blue line
      represents $\gamma = 0$ and red line represents $\gamma = -3$.
      The first row corresponds to the wave number $k = 0.3$ and the
      bottom row corresponds to the wave number $k =0.5$.}
    \label{fig:ex2_gamma}
  \end{figure}

  The cases of different potential indices in the IPL model are also
  studied, where the model of Maxwell molecules $\gamma = 0$ and the
  model with Coulomb interactions $\gamma = -3$ are tested.  Figure
  \ref{fig:ex2_gamma} shows the time evolution of the electrostatic
  energy for wave number $k = 0.3$ and $0.5$ 
  under different collisional frequencies
  $\nu = \nu_{\beta}= 0.05$ and $0.1$, from which we find that the
  damping rate for the Maxwell case $\gamma = 0$ is much larger than
  that for the Coulomb case $\gamma = -3$. This
  result is compatible with a similar conclusion in the linear case.


\subsection{Two-stream instability}

\begin{figure}[!htb]
  \centering \subfloat[Initial MDF $g(0, x, v_1)$]
  {\includegraphics[width=0.32\textwidth, height=0.21\textwidth,
    clip]{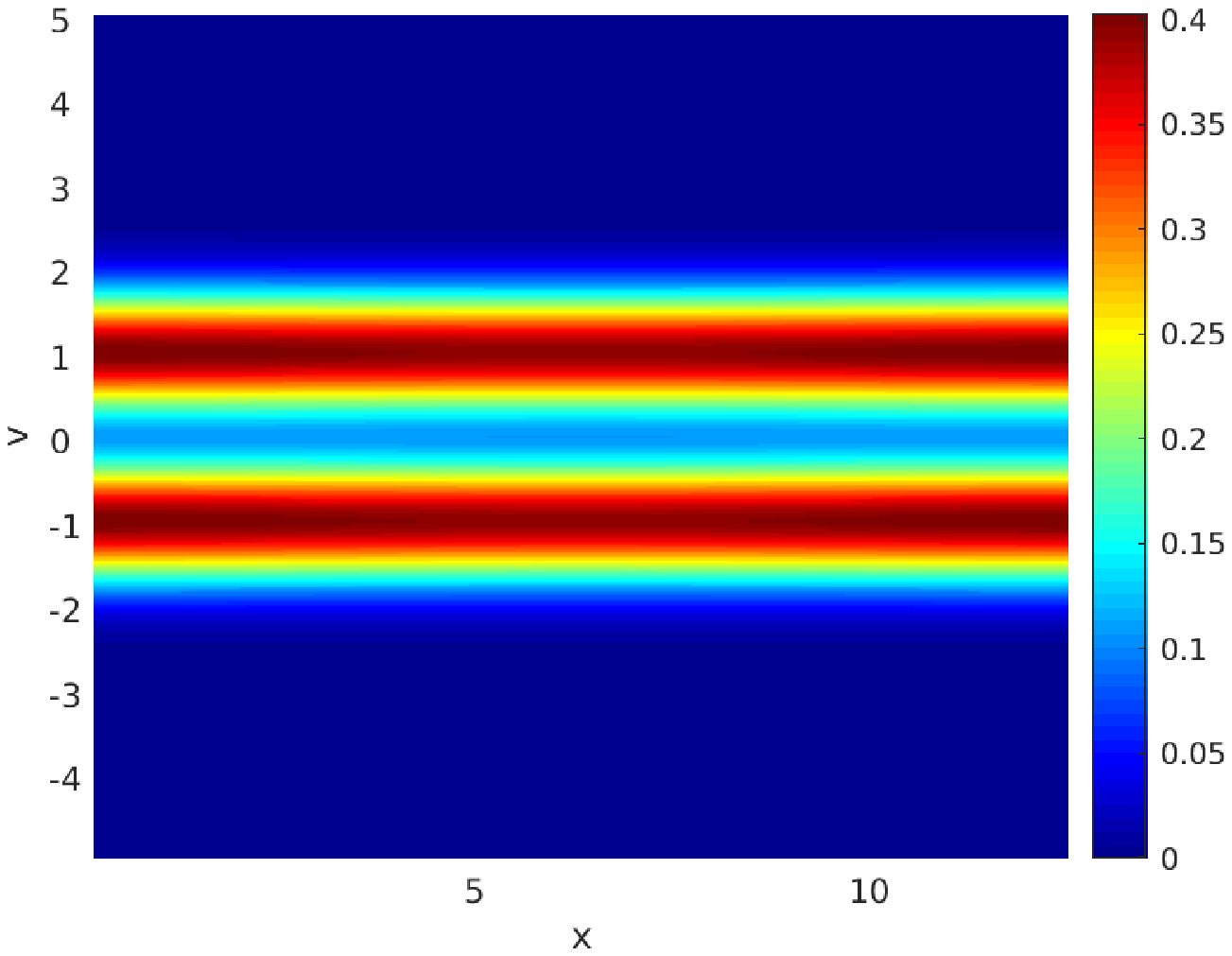}}\hfill
  \subfloat[Contours of $g(0, x, v_1)$]
  {\includegraphics[width=0.32\textwidth,
    height=0.21\textwidth,clip]{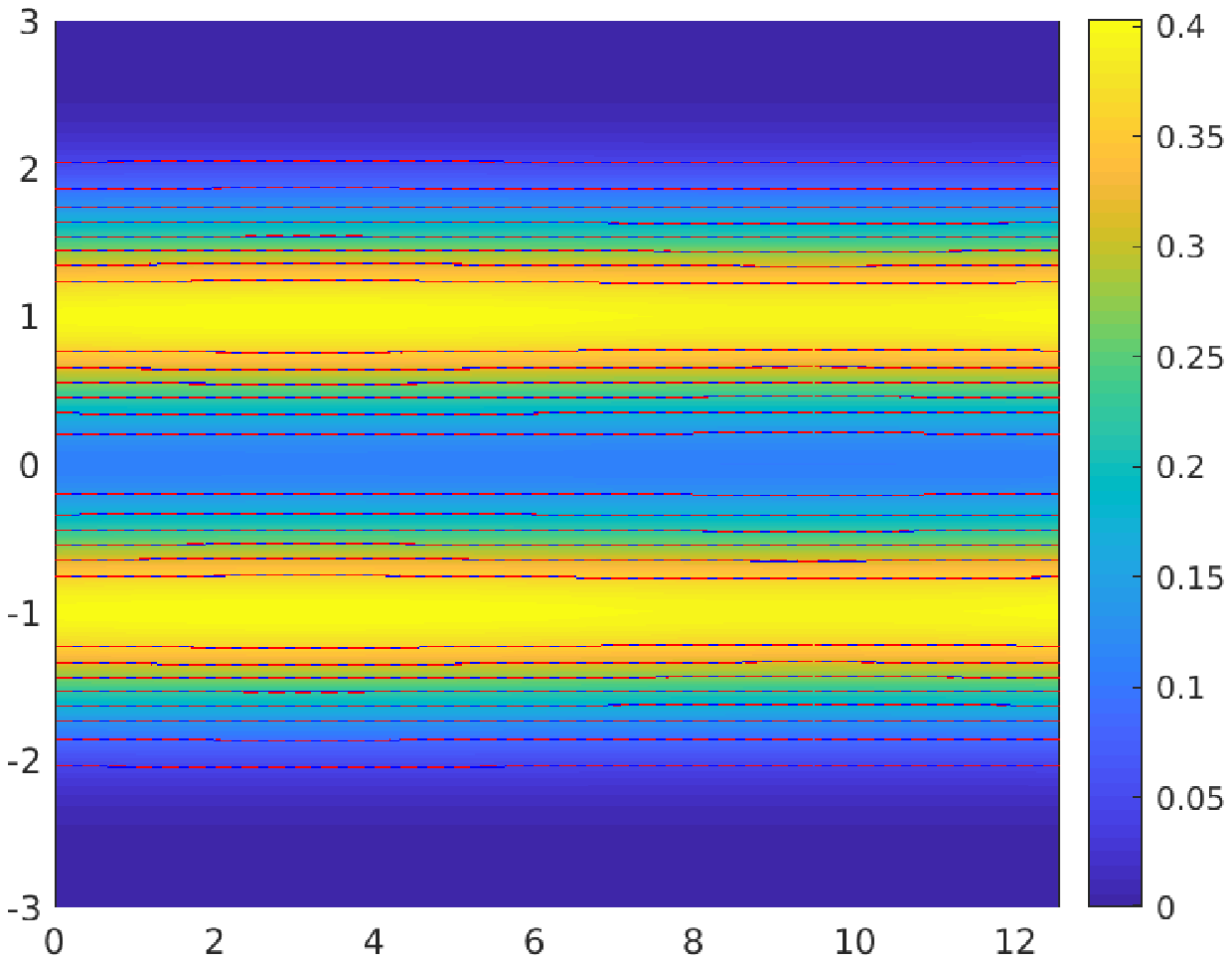}} \hfill
  \subfloat[Initial MDF $g(0, \frac{\pi}{4}, v_1)$]
  {\includegraphics[width=0.32\textwidth,
    height=0.21\textwidth,clip]{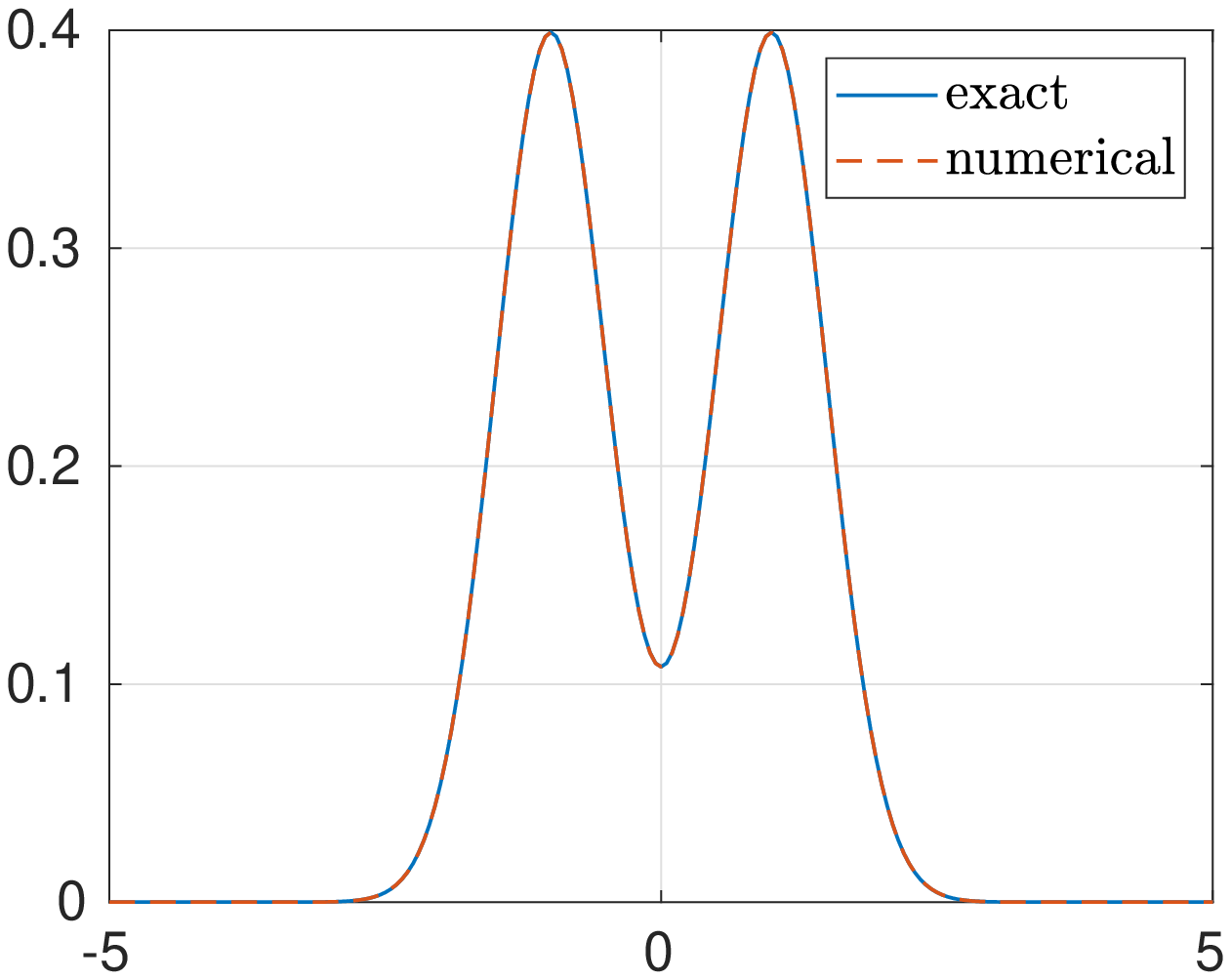}}
  \caption{Initial marginal distribution functions of the two-stream
  instability problem. In (b) and (c),
    the blue solid lines correspond to the exact solution, and the red
    dashed lines correspond to the numerical approximation.  Figure
    (a) shows only the numerical approximation. Figure (c) shows the
    numerical approximation and the exact solution at the position
    $x = \frac{\pi}{4}$. }
  \label{fig:ex3_ini}
\end{figure}

Two-stream instability is a common instability in plasma physics and
of primary concern for studying the nonlinear
effect of plasma in the future. It occurs when the fluid consists of
two electron streams with different velocities. The mechanism of
two-stream instability is similar to that of Landau damping, 
where particles at different velocities
transfer energy to each other \cite{Bittencourt}.

\begin{figure}[!htb]
  \centering \subfloat[$t = 20, \nu = 0$]
  {\includegraphics[width=0.32\textwidth, height=0.21\textwidth,
    clip]{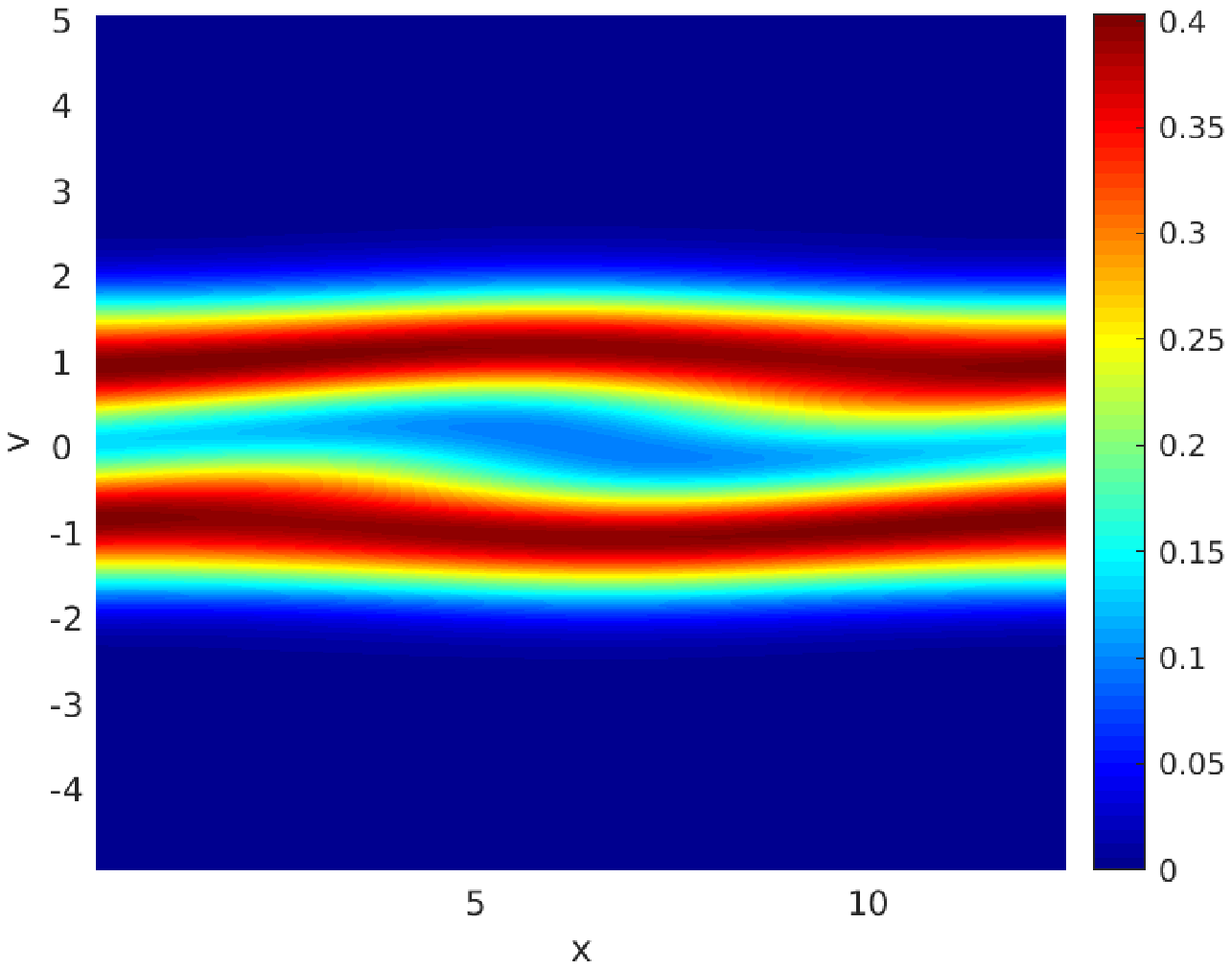}}\hfill
  \subfloat[$t = 20, \nu = 0.001$]
  {\includegraphics[width=0.32\textwidth,
    height=0.21\textwidth,clip]{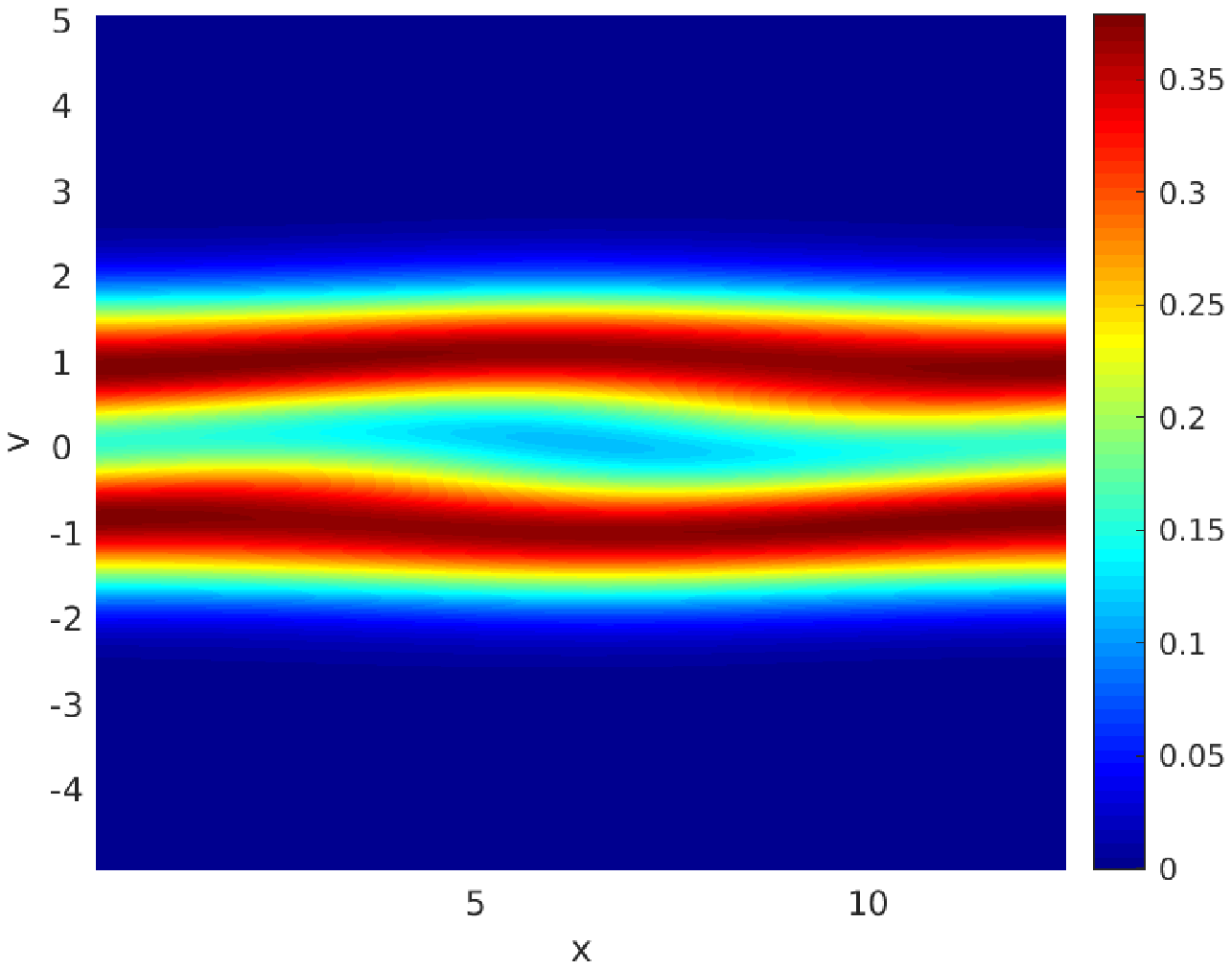}}
  \hfill \subfloat[$t = 20, \nu = 0.01$]
  {\includegraphics[width=0.32\textwidth,
    height=0.21\textwidth,clip]{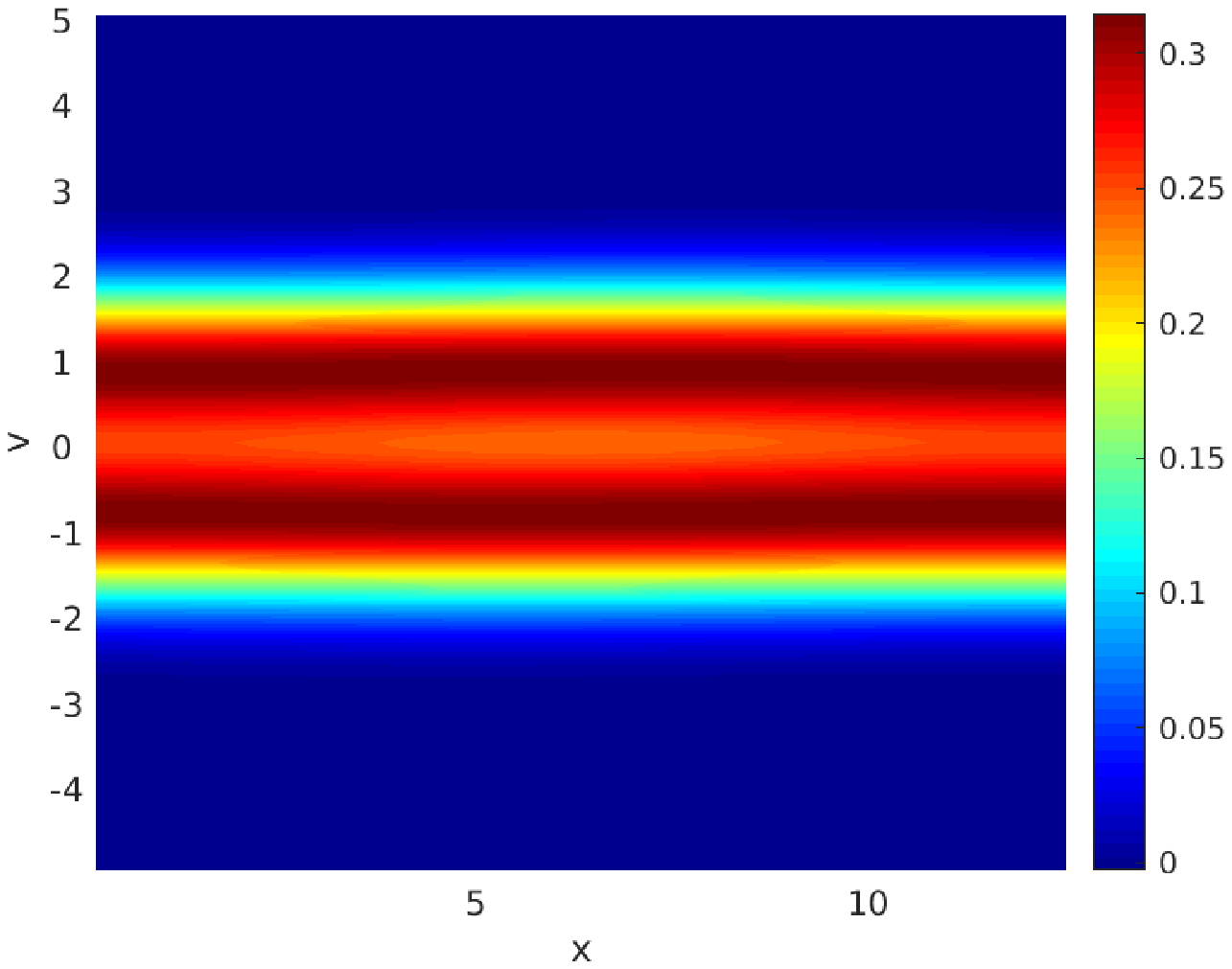}}   \\
  \subfloat[$t = 30, \nu = 0$]
  {\includegraphics[width=0.32\textwidth, height=0.21\textwidth,
    clip]{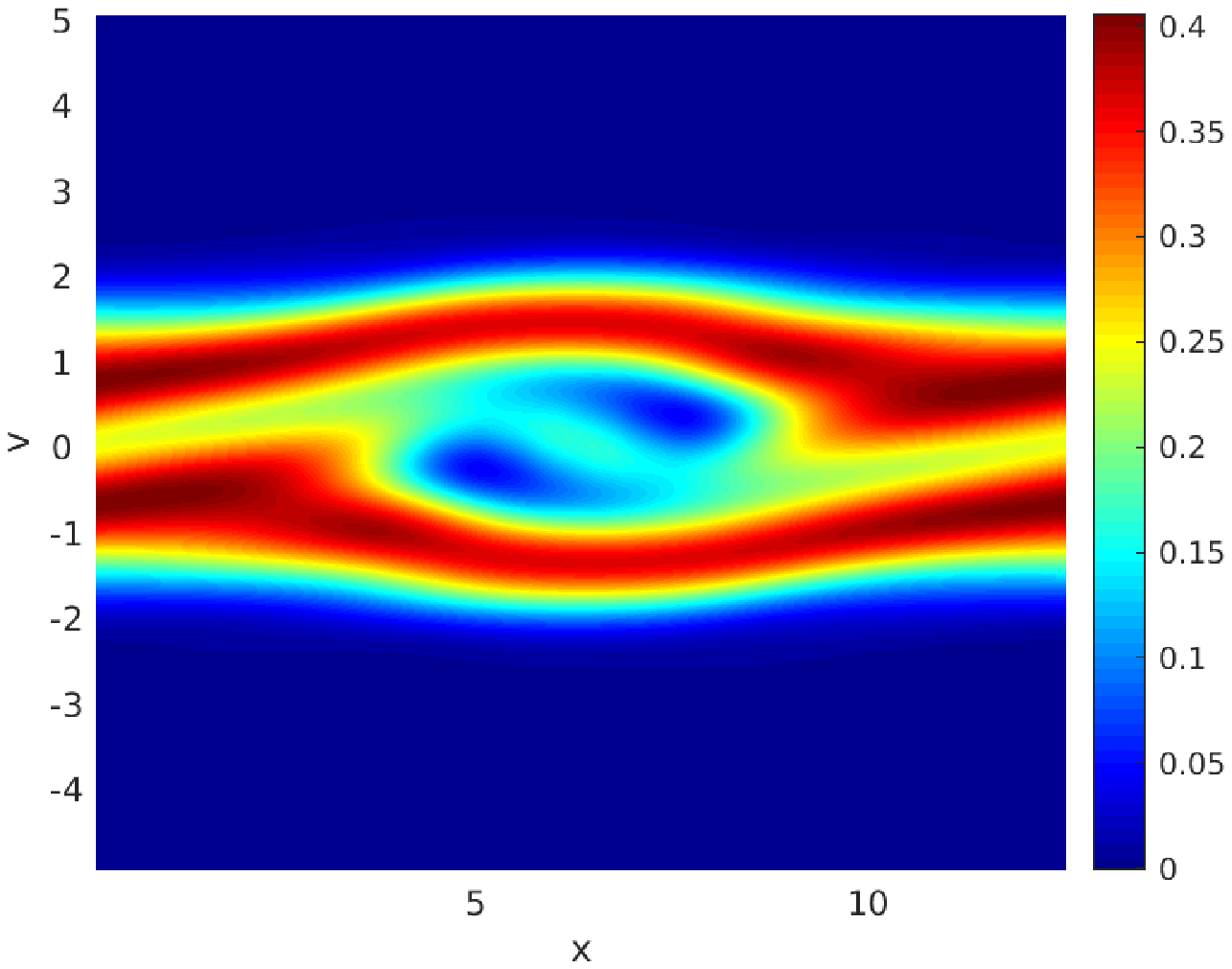}}\hfill
  \subfloat[$t = 30, \nu = 0.001$]
  {\includegraphics[width=0.32\textwidth,
    height=0.21\textwidth,clip]{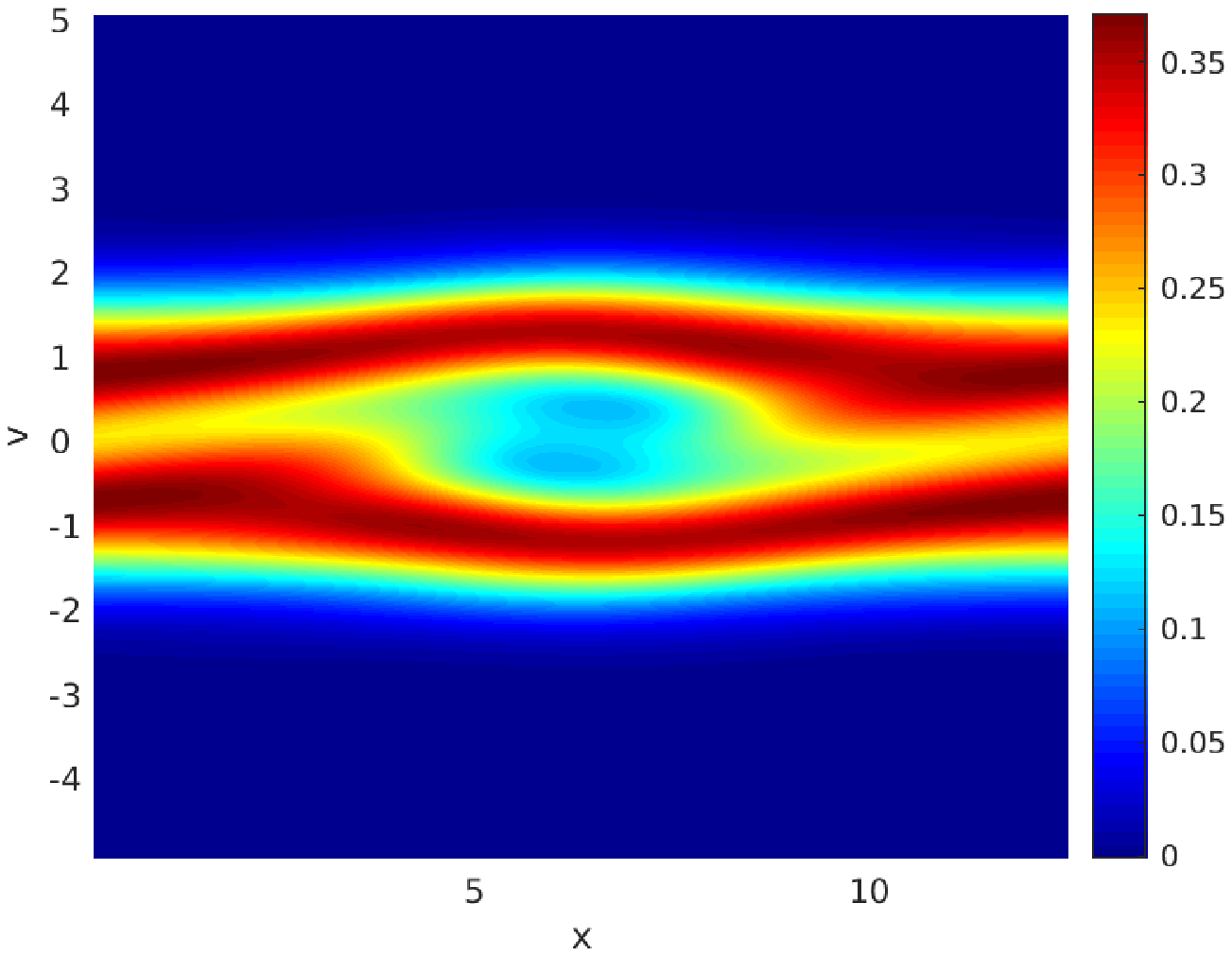}}
  \hfill \subfloat[$t = 30, \nu = 0.01$]
  {\includegraphics[width=0.32\textwidth,
    height=0.21\textwidth,clip]{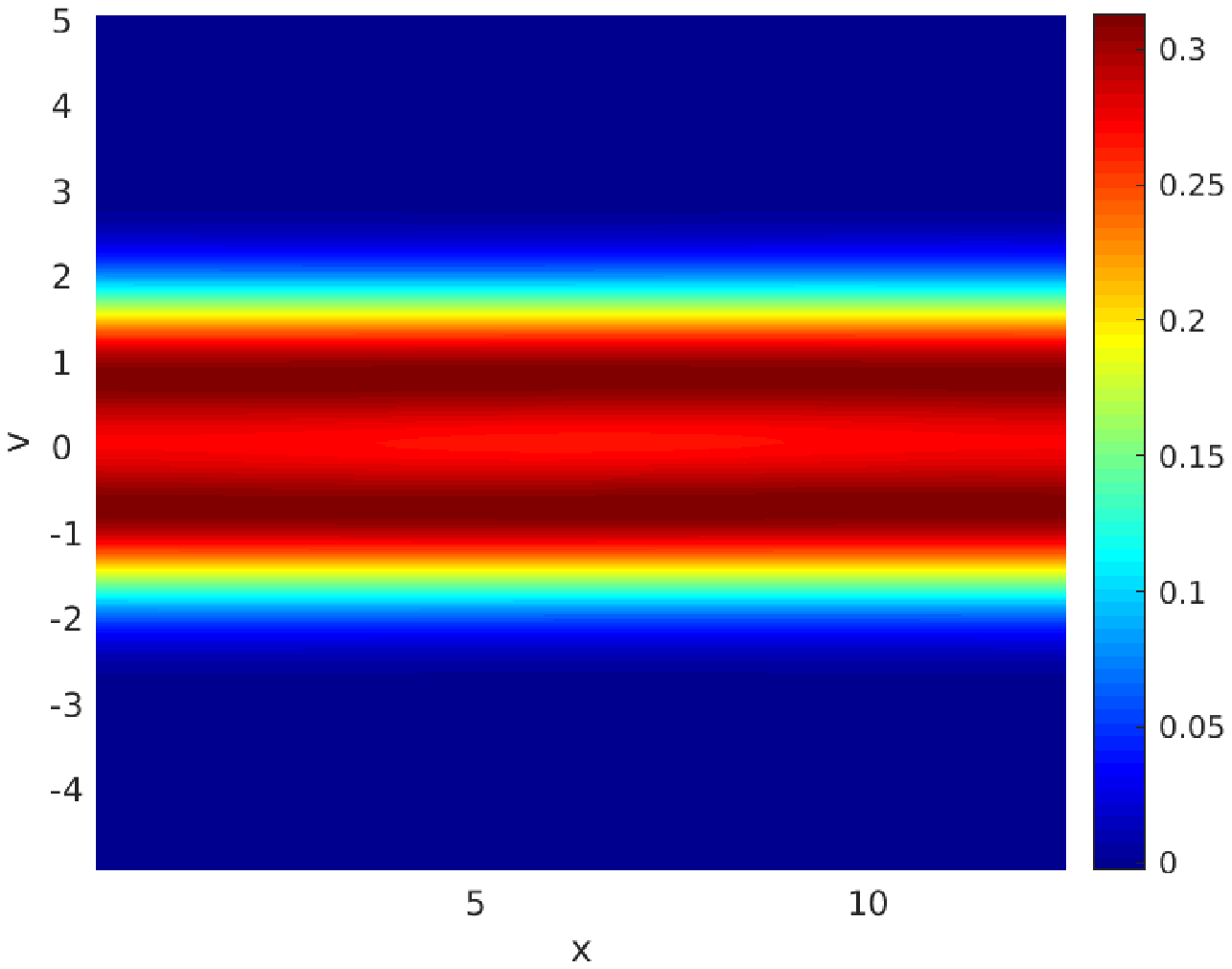}} \\
  \subfloat[$t = 50, \nu = 0$]
  {\includegraphics[width=0.32\textwidth, height=0.21\textwidth,
    clip]{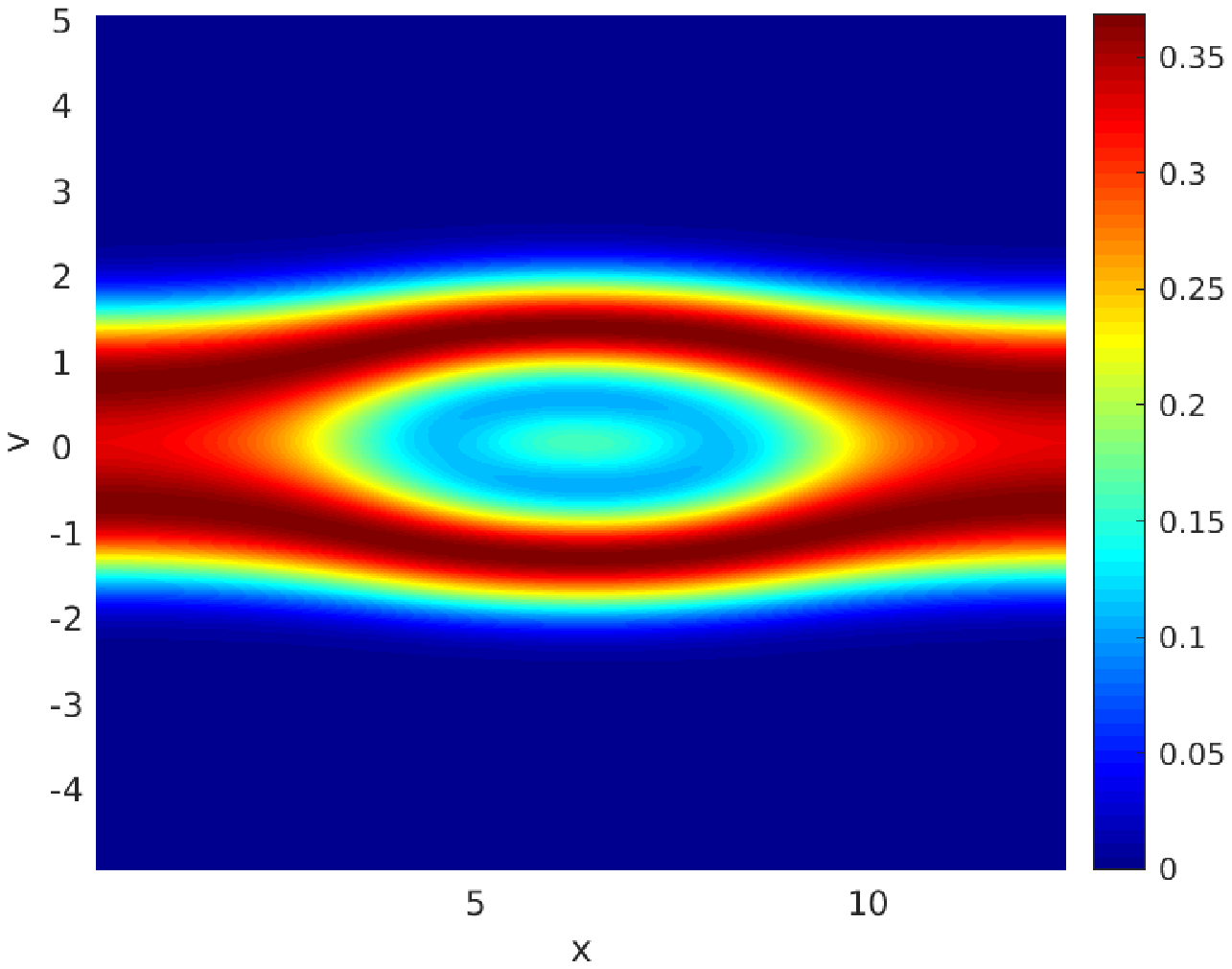}}\hfill
  \subfloat[$t = 50, \nu = 0.001$]
  {\includegraphics[width=0.32\textwidth,
    height=0.21\textwidth,clip]{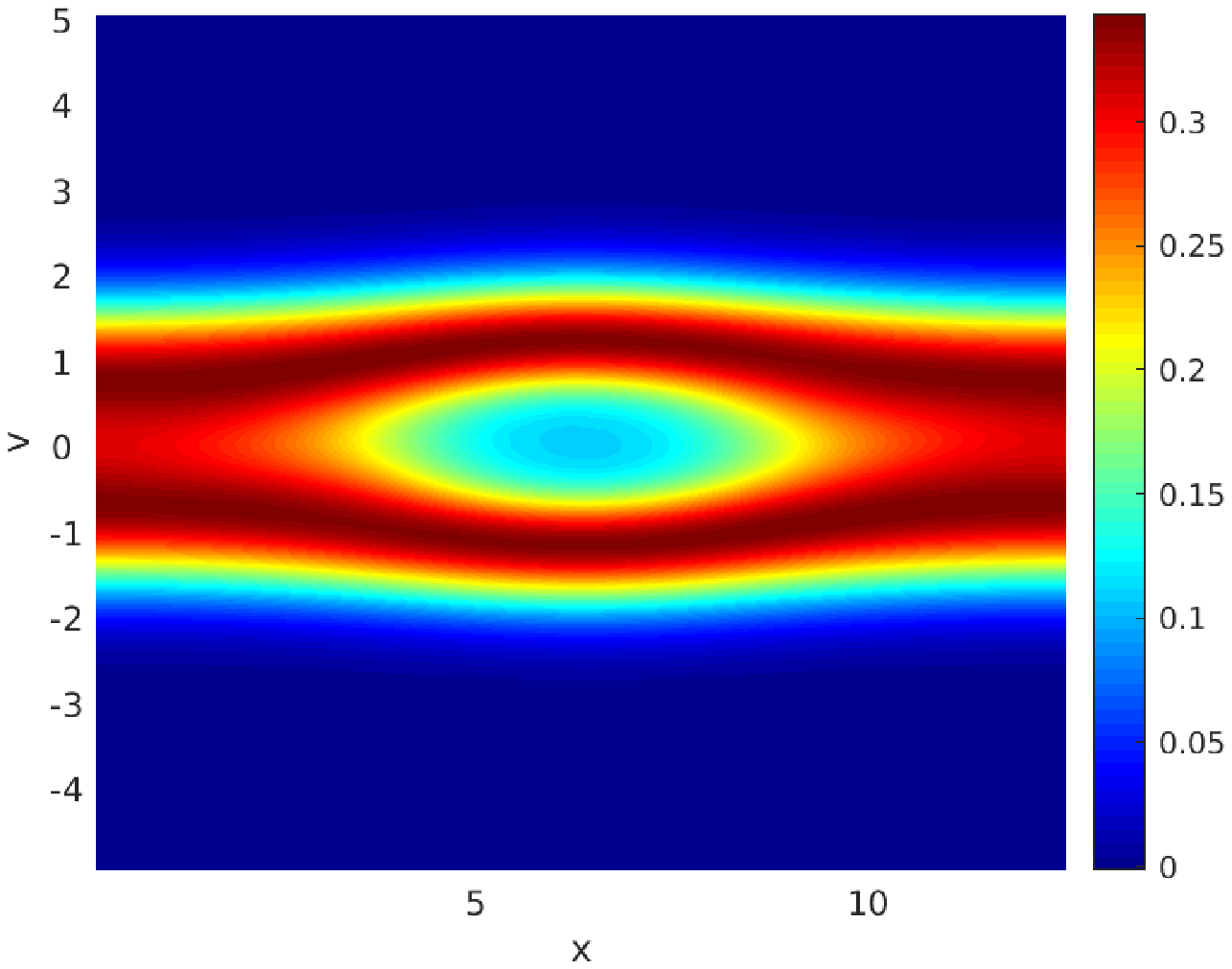}}
  \hfill \subfloat[$t = 50, \nu = 0.01$]
  {\includegraphics[width=0.32\textwidth,
    height=0.21\textwidth,clip]{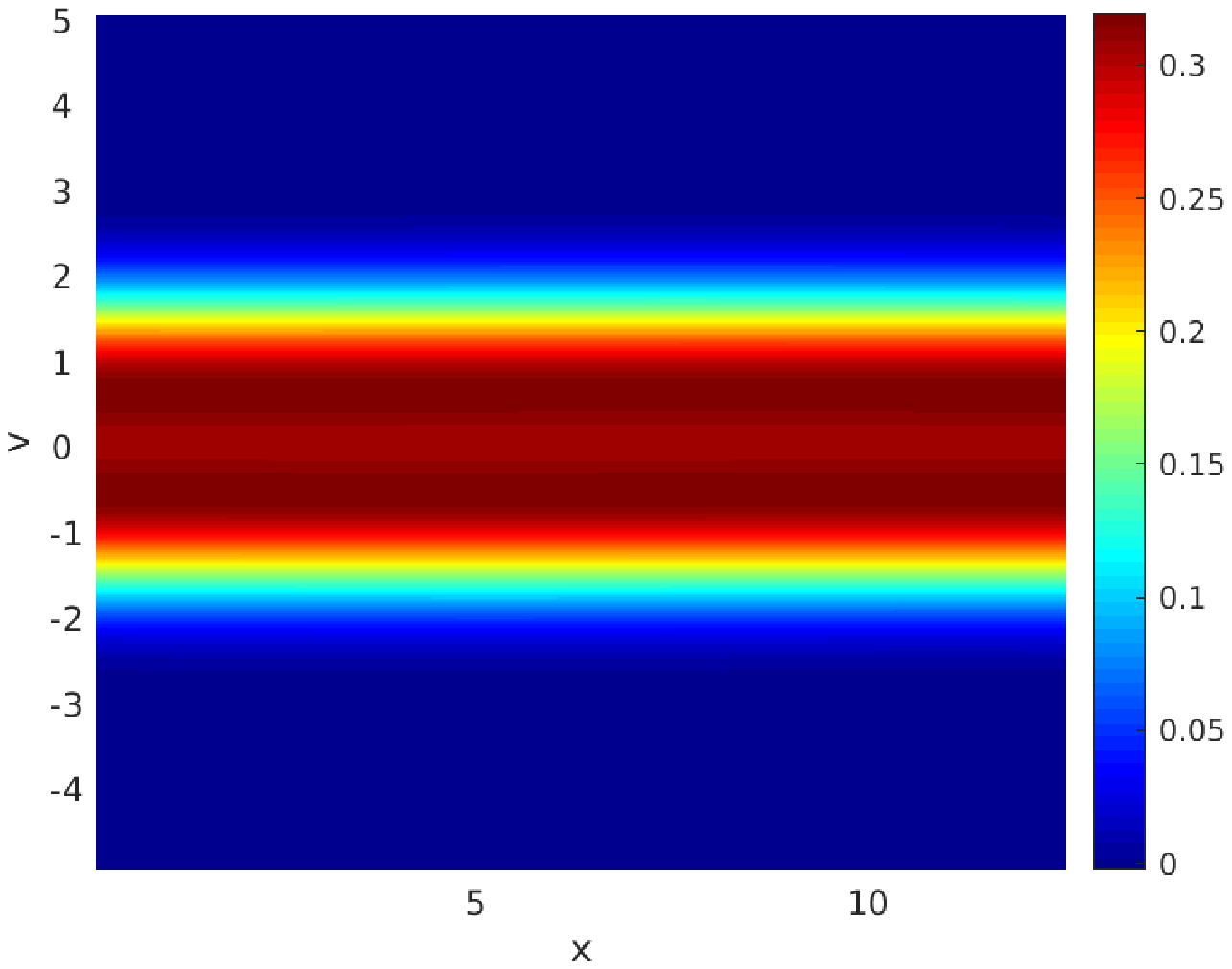}} 
  \caption{Evolution of the marginal distribution function
    $g(t, x, v_1)$ under different collisional frequencies $\nu$
    in the two-stream instability problem. The
    left column corresponds to $\nu =0$, the middle column corresponds
    to $\nu = 0.001$, and the right column corresponds to
    $\nu = 0.01$.}
  \label{fig:ex3_twostream_nu}
\end{figure}

In this numerical experiment, the initial data is given with a
nonisotropic two-stream flow
\begin{equation}
  \label{eq:two_stream}
  f = \frac{(1+A \cos (k x))}{\sqrt{2 \pi T}}\left[0.5\exp
    \left(-\frac{\left|\bv-(u_1, 0, 0)^T\right|^{2}}
      {2 T}\right)+0.5\exp \left(-\frac{\left|\bv+(u_1,
          0,0)^T\right|^{2}}{2 T}\right)\right],  
\end{equation}
with $A=0.01$, $T = 0.25$ and $u_1=1$.  Here, only the
electron-electron collsion is considered and thus the electron-ion
collision frequency $\nu_{\beta}$ is set as $0$. Similar initial data
and assumptions can be found in \cite{ZhangGamba2017}. The time
evolution of the particles with the collisional model of Coulomb
interactions $\gamma = -3$ is studied, and the wave number $k$ is
chosen as $k = 0.5$. The grid size and expansion order
are chosen as $N = 400$ and $M =40$, respectively. Moreover, the
quadratic length is
set as $M_0 = 5$. Here, the collisional frequency is set as $\nu=0$,
$0.001$ and $0.01$ to present the effect of the collision. The
marginal distribution function
\begin{equation}
  \label{eq:marginal}
  g(t, x, v_1) = \int_{\bbR^2} f(t, x, v_1, v_2, v_3) \dd v_2 \dd v_3 
\end{equation}
is also plotted to show the electron ``trapping''
phenomenon. Clearly, our chosen parameters can approximate the
initial distribution function satisfactorily (see Figure
\ref{fig:ex3_ini}).  To suppress the recurrence and the nonphysical
oscillations, the filter developed in \cite{hou2007computing,
  Filter2017} is applied here.

Figure \ref{fig:ex3_twostream_nu} shows the time evolution of the
marginal distribution function \eqref{eq:marginal} in the $x-v_1$
plane. From these, we can find that for the collisionless case, the
linear two-stream instability grows exponentially at first,
and then the nonlinearity becomes dominant and
``trapping'' emerges. At the same time, the original distribution
begins to twist and curve until an electron hole-like structure
finally forms, which is consistent with the results in
\cite{heath2012discontinuous}. For the collisional case, a
smaller electron hole-like structure forms with the increase
in the collisional frequency $\nu$, and no visible hole-like structure
occurs in the case of 
collisional frequency $\nu = 0.1$. This again substantiates the effect
of collision to reduce the ``trapping'' phenomenon.

\begin{figure}[!htb]
  \centering \includegraphics[width=0.49\textwidth,
  height=0.35\textwidth, clip]{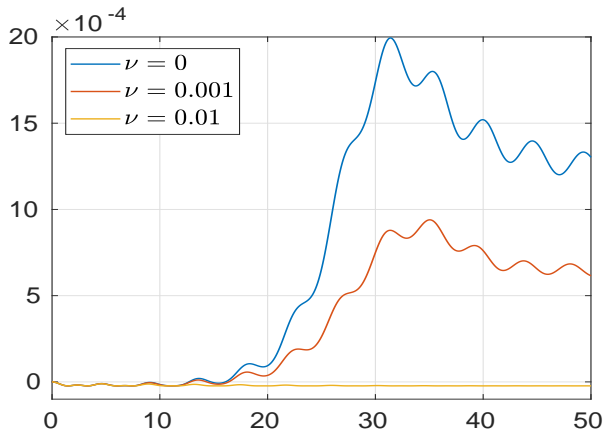}
  \caption{Time evolution of the variation in the total energy
    $ \mE_t(t)$ for different collisional frequencies
    in the two-stream instability problem. The variation
    is defined as $(\mE_{t}(t) - \mE_{t}(0)) / \mE_t(0)$. }
  \label{fig:ex3_energy}
\end{figure}

The time evolution of the total energy is also studied to test the
conservation property of this numerical scheme. The total energy
$\mE_t(t)$ is defined as
\begin{equation}
  \label{eq:total_energy}
  \mE_t(t) = \frac{1}{2} \Delta x \sum_j \int_{\bbR^3} f(t, x_j, \bv) |\bv|^2
  \dd \bv  + \frac{1}{2}\mE(t)^2. 
\end{equation}
The evolution of the total energy $\mE_t(t)$ for different collisional
frequencies is plotted in Figure \ref{fig:ex3_energy}, from which we
can see that although the numerical scheme cannot exactly preserve the
total energy, the variation of the total energy is minute, especially
in the linear instability stage, where the
variation is almost negligible.

\subsection{Bump-on-tail instability}

Bump-on-tail instability is another important micro-instability
which is a special case
of two-stream instability when the two electron streams have
different densities~\cite{Cheng2014}.  The distribution function is
unstable, which leads
to growth in the initial perturbation followed by saturation and
oscillation of the particles trapped in the potential through the wave
\cite{Magdi1979, NAKAMURA1999122}.
  
In this numerical experiment, we also begin with a nonisotropic distribution
function as
\begin{equation}
  \label{eq:bumpontail}
  f =\frac{(1+A \sin (k x))}{\sqrt{2 \pi T}}\left[n_m\exp
    \left(-\frac{\left|\bv-(u_1, 0, 0)^T\right|^{2}}
      {2 T}\right)+n_b\exp \left(-\frac{\left|\bv+(u_1,
          0,0)^T\right|^{2}}{2 T}\right)\right],  
\end{equation}
where $A=0.01$, $T=0.25$ and $u_1=1$. $n_m=0.7$, which
represents the magnitude of the ``mainstream'', and
$n_b=0.3$, representing the magnitude of the ``bump'' on the
tail of the ``mainstream''.

\begin{figure}[!htb]
  \centering \subfloat[Initial MDF $g(0, x, v_1)$]
  {\includegraphics[width=0.32\textwidth, height=0.21\textwidth,
    clip]{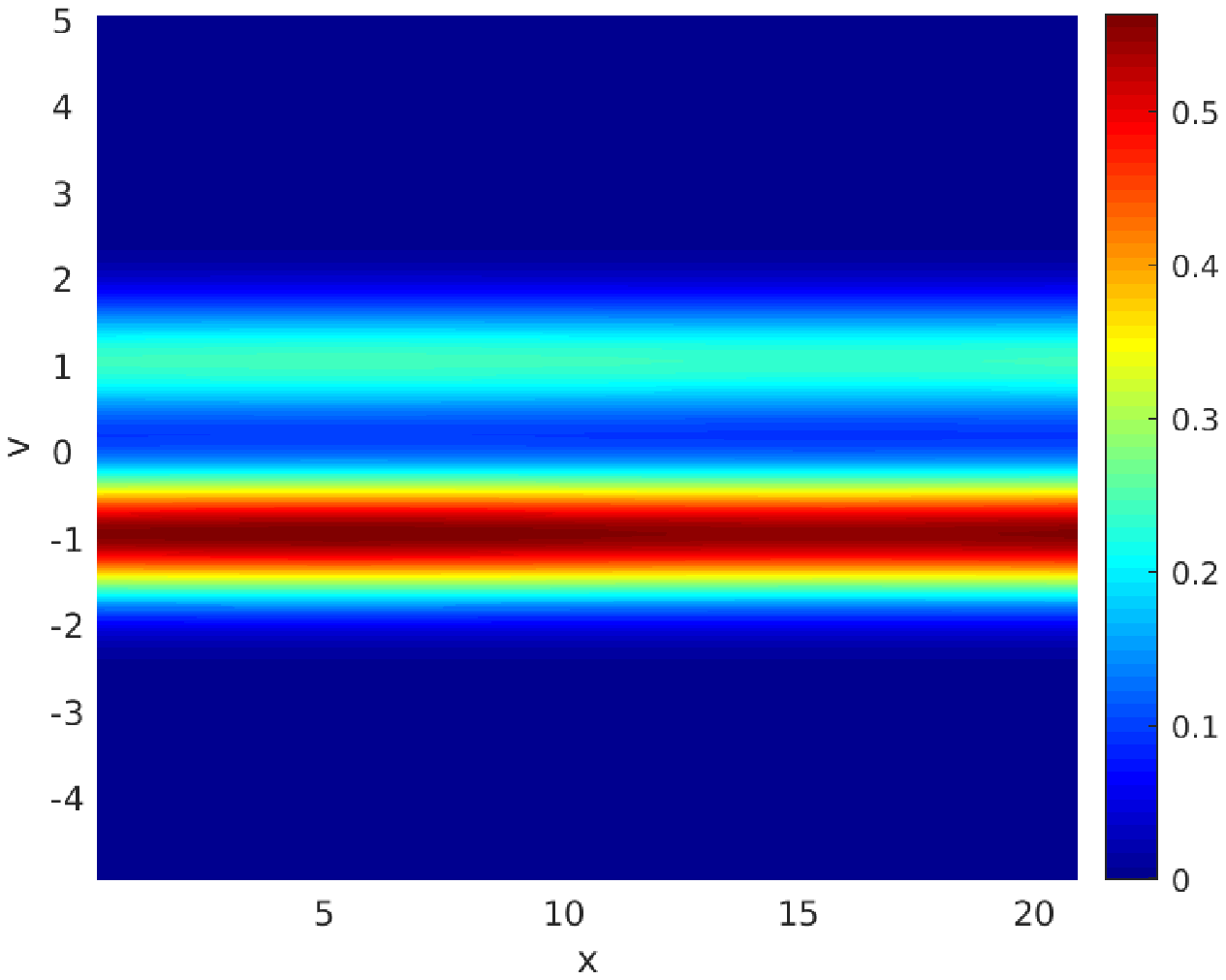}}\hfill
  \subfloat[Contours of $g(0, x, v_1)$]
  {\includegraphics[width=0.32\textwidth,
    height=0.21\textwidth,clip]{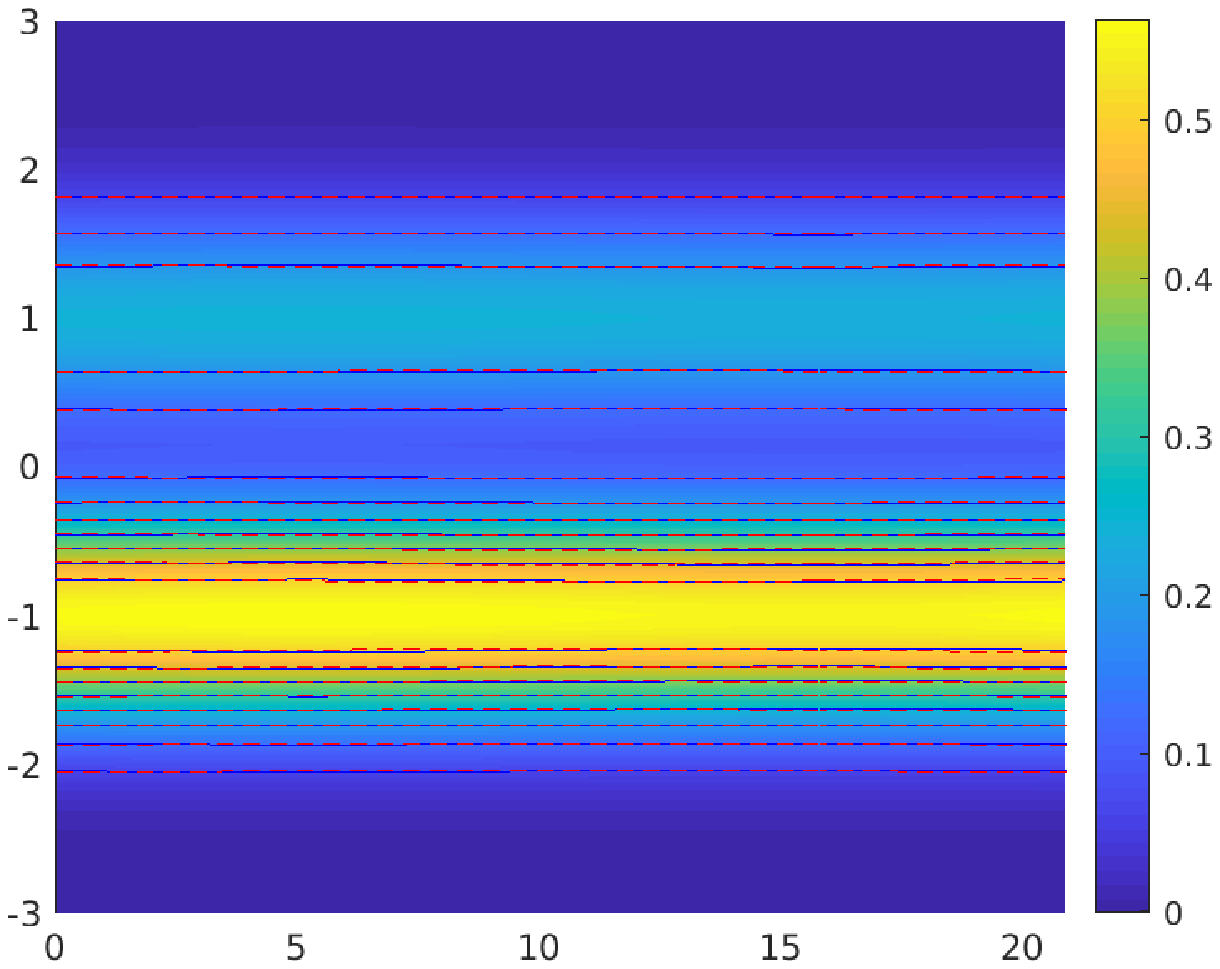}} \hfill
  \subfloat[Initial MDF $g(0, 0, v_1)$]
  {\includegraphics[width=0.32\textwidth,
    height=0.21\textwidth,clip]{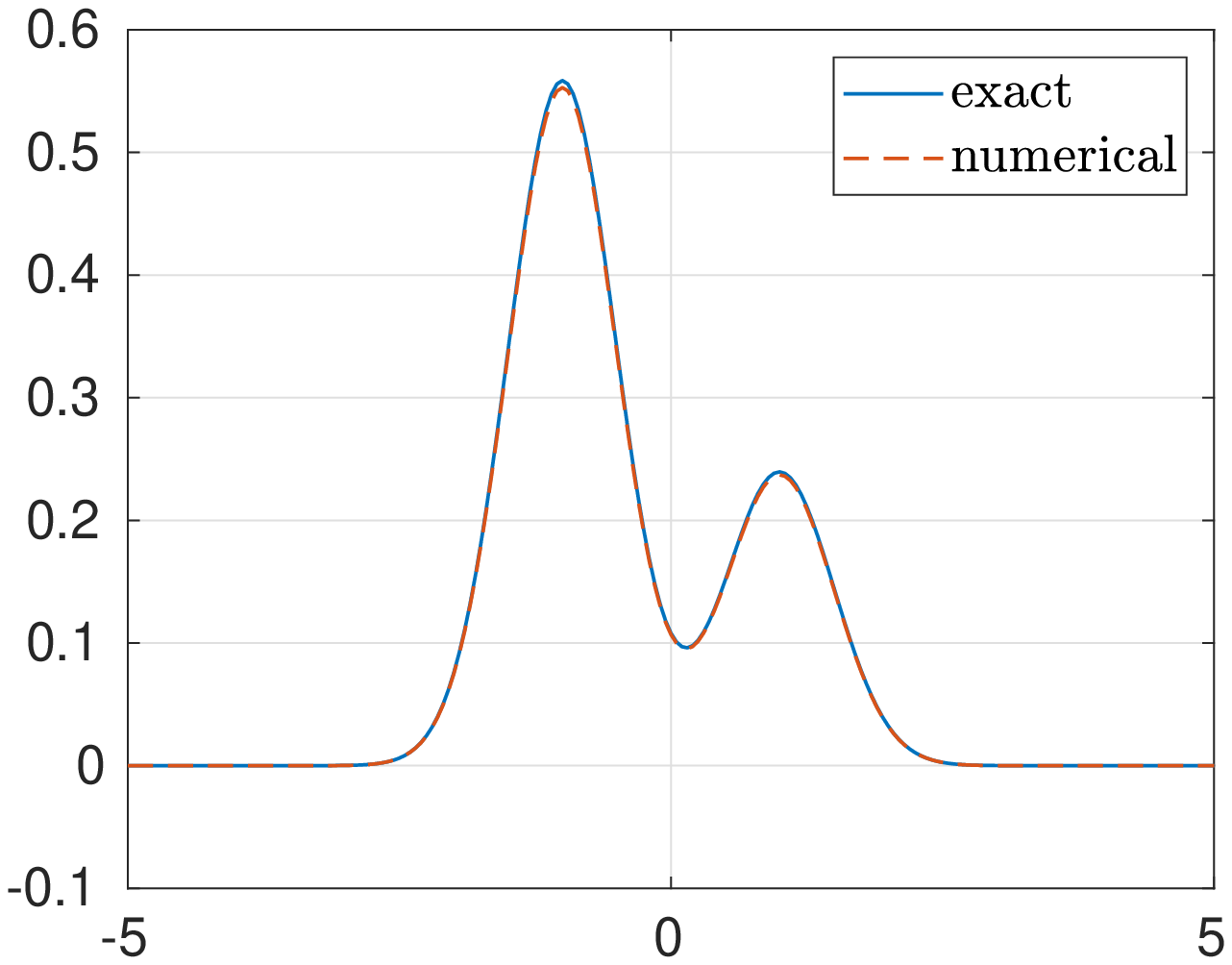}}
  \caption{Initial marginal distribution functions
  of the bump-on-tail instability problem. In (b) and (c),
    the blue solid lines correspond to the exact solution, and the red
    dashed lines correspond to the numerical approximation.  Figure
    (a) shows only the numerical approximation. Figure (c) shows the
    numerical approximation and the exact solution at the position
    $x = 0$. }
  \label{fig:ex4_ini}
\end{figure}

The wave number $k$ is chosen as $k = 0.3$.  The grid size is chosen
as $N = 400$, and the expansion order is set as $M =40$, which gives a
satisfying approximation to the initial distribution function (see
Figure \ref{fig:ex4_ini}). Moreover, the quadratic length is set as $M_0 = 5$.
Similar to the previous numerical experiment, we focus on the model
with Coulomb interactions, and the time evolution of 
particles with collision frequencies
$\nu=0$, $0.001$ and $0.01$ is studied.

\begin{figure}[!htb]
  \centering \subfloat[$t = 20, \nu = 0.0$]
  {\includegraphics[width=0.32\textwidth, height=0.21\textwidth,
    clip]{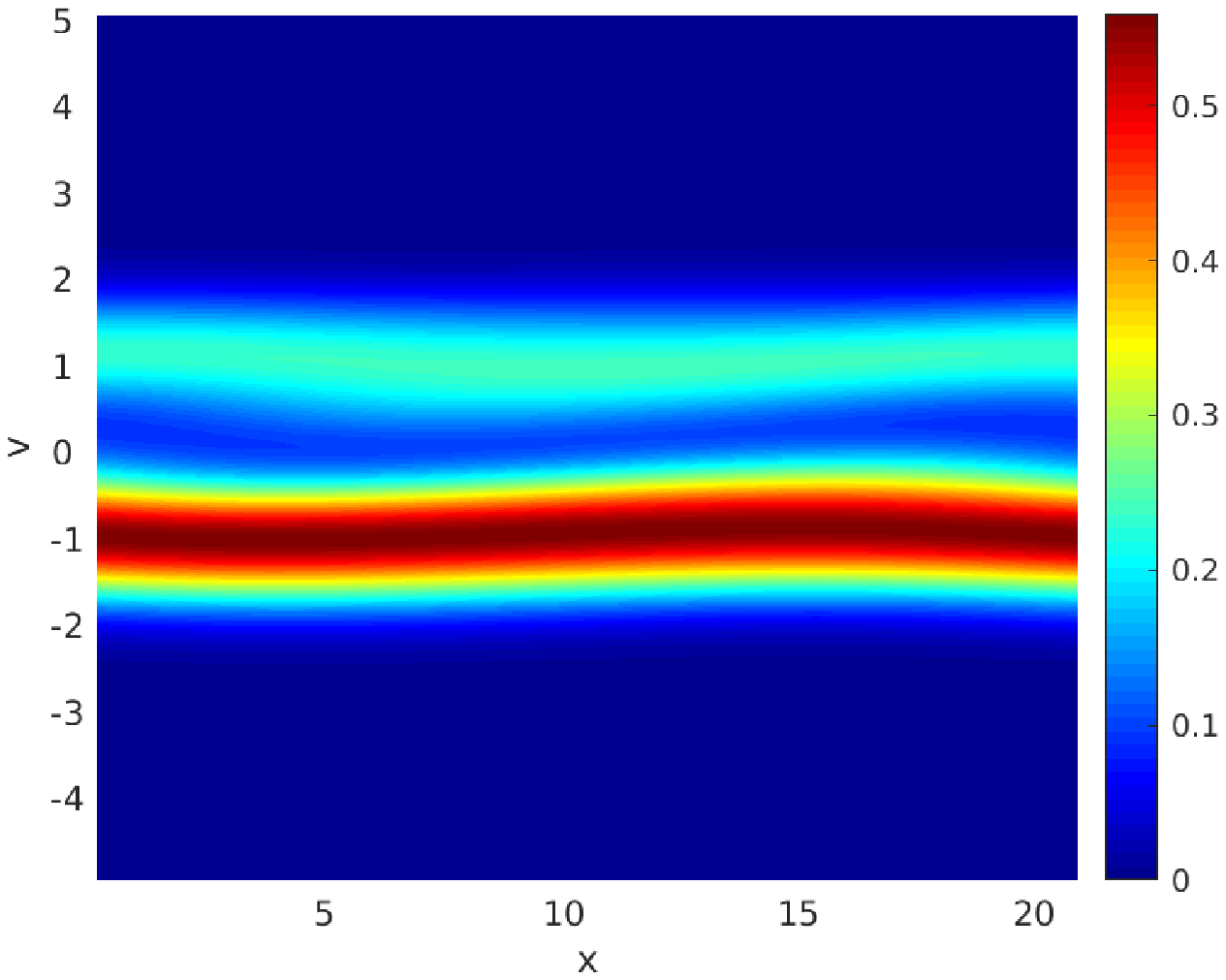}}\hfill
  \subfloat[$t = 20, \nu = 0.001$]
  {\includegraphics[width=0.32\textwidth,
    height=0.21\textwidth,clip]{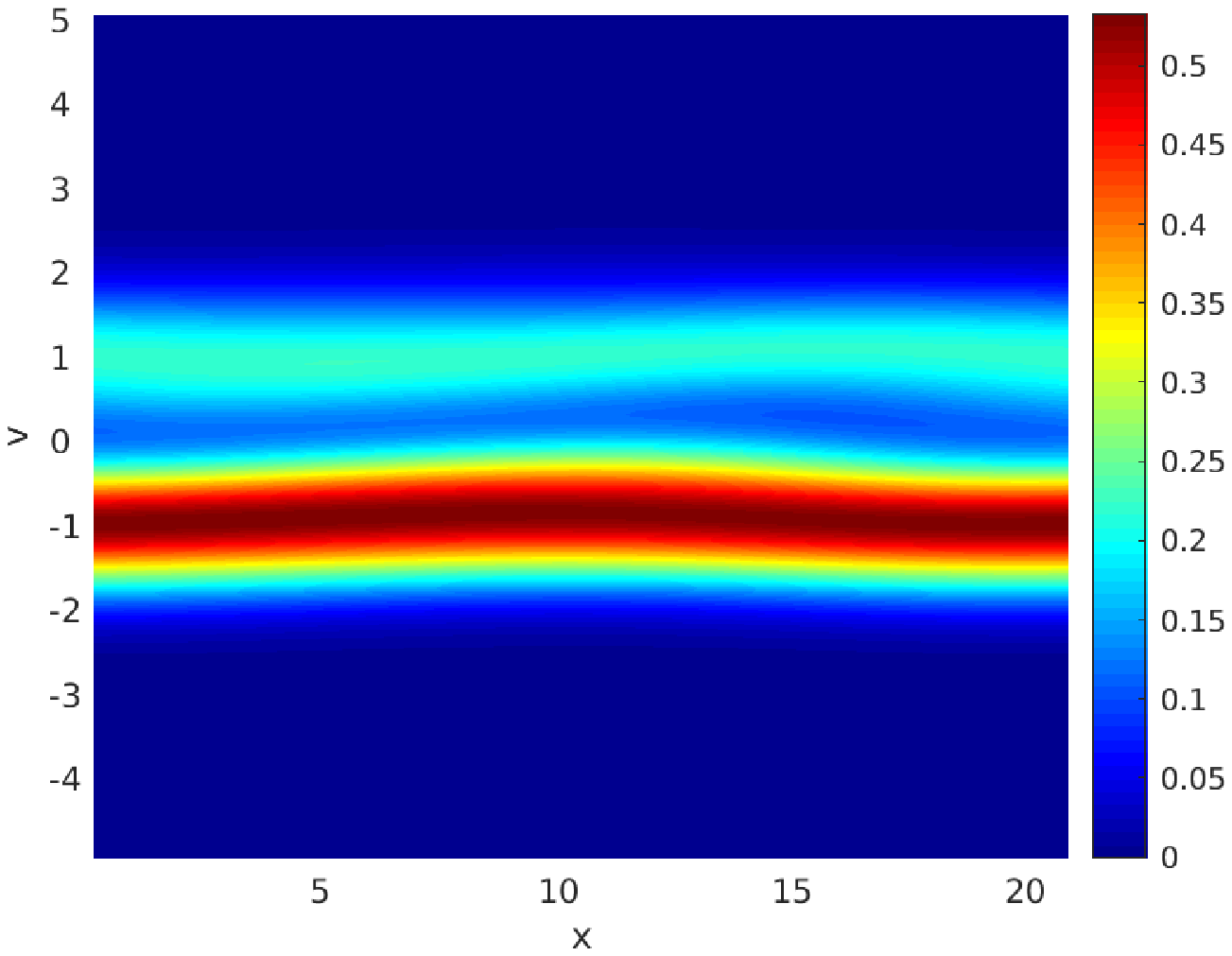}} \hfill
  \subfloat[$t = 20, \nu = 0.01$]
  {\includegraphics[width=0.32\textwidth,
    height=0.21\textwidth,clip]{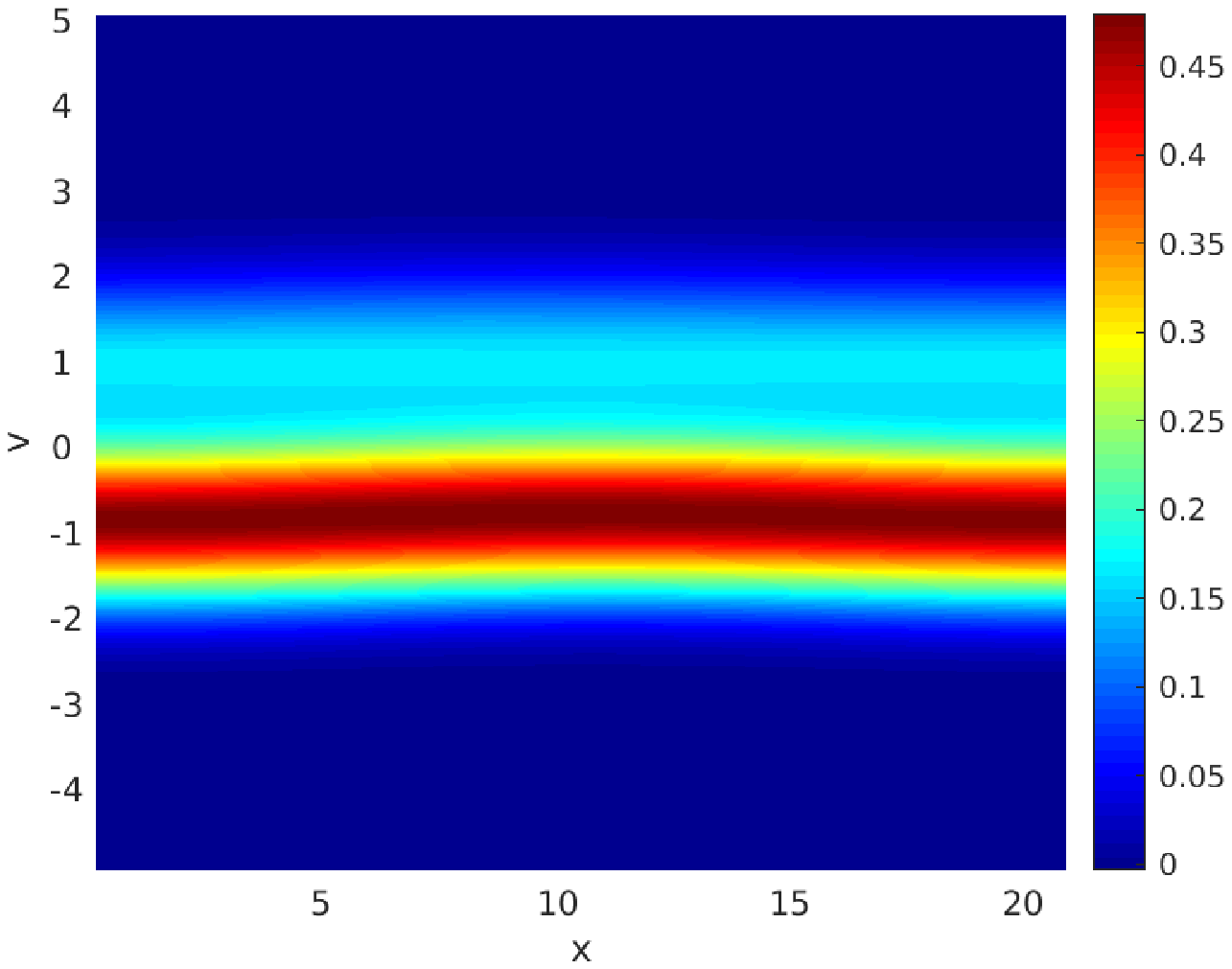}}   \\
  \subfloat[$t = 30, \nu = 0.0$]
  {\includegraphics[width=0.32\textwidth, height=0.21\textwidth,
    clip]{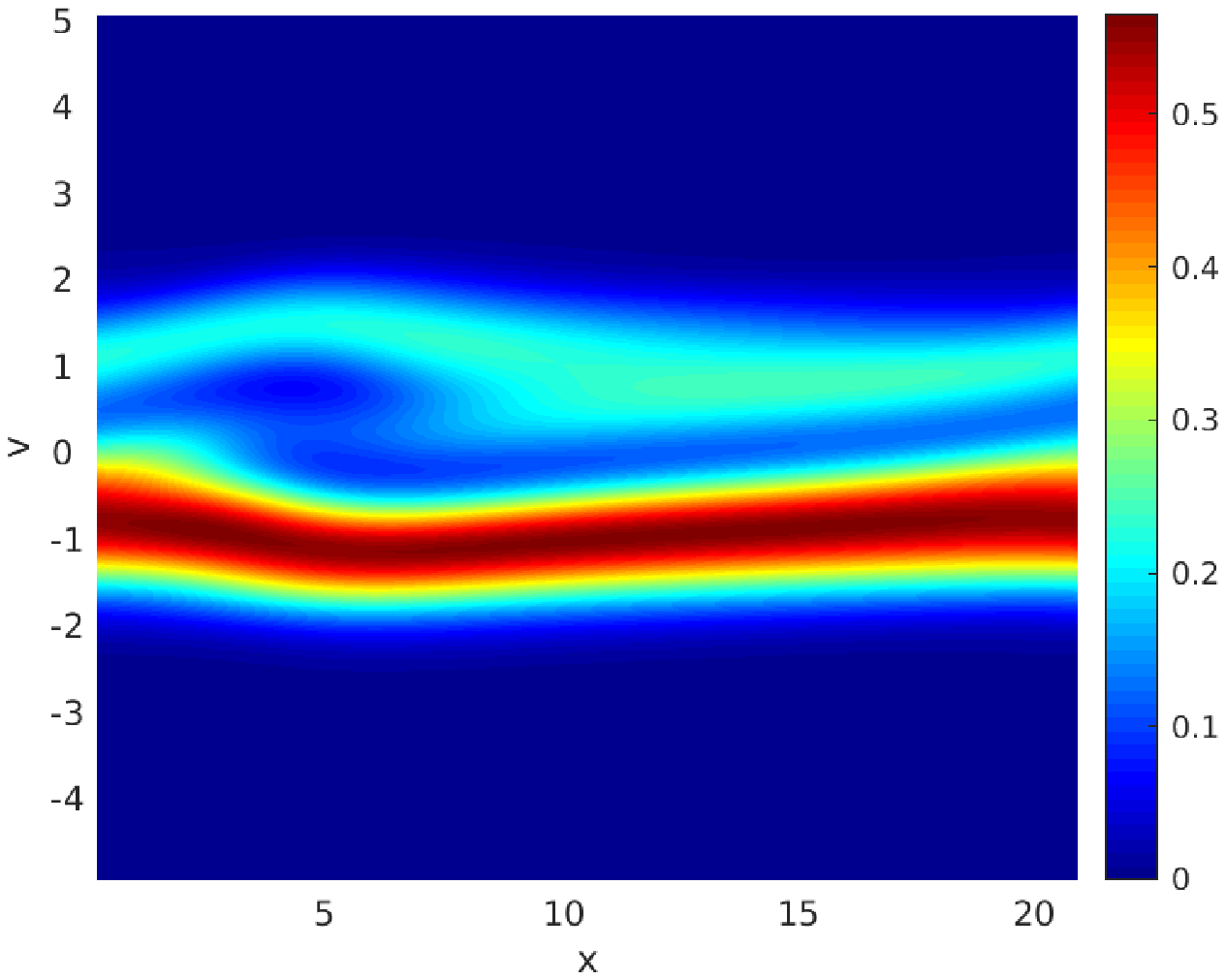}}\hfill
  \subfloat[$t = 30, \nu = 0.001$]
  {\includegraphics[width=0.32\textwidth,
    height=0.21\textwidth,clip]{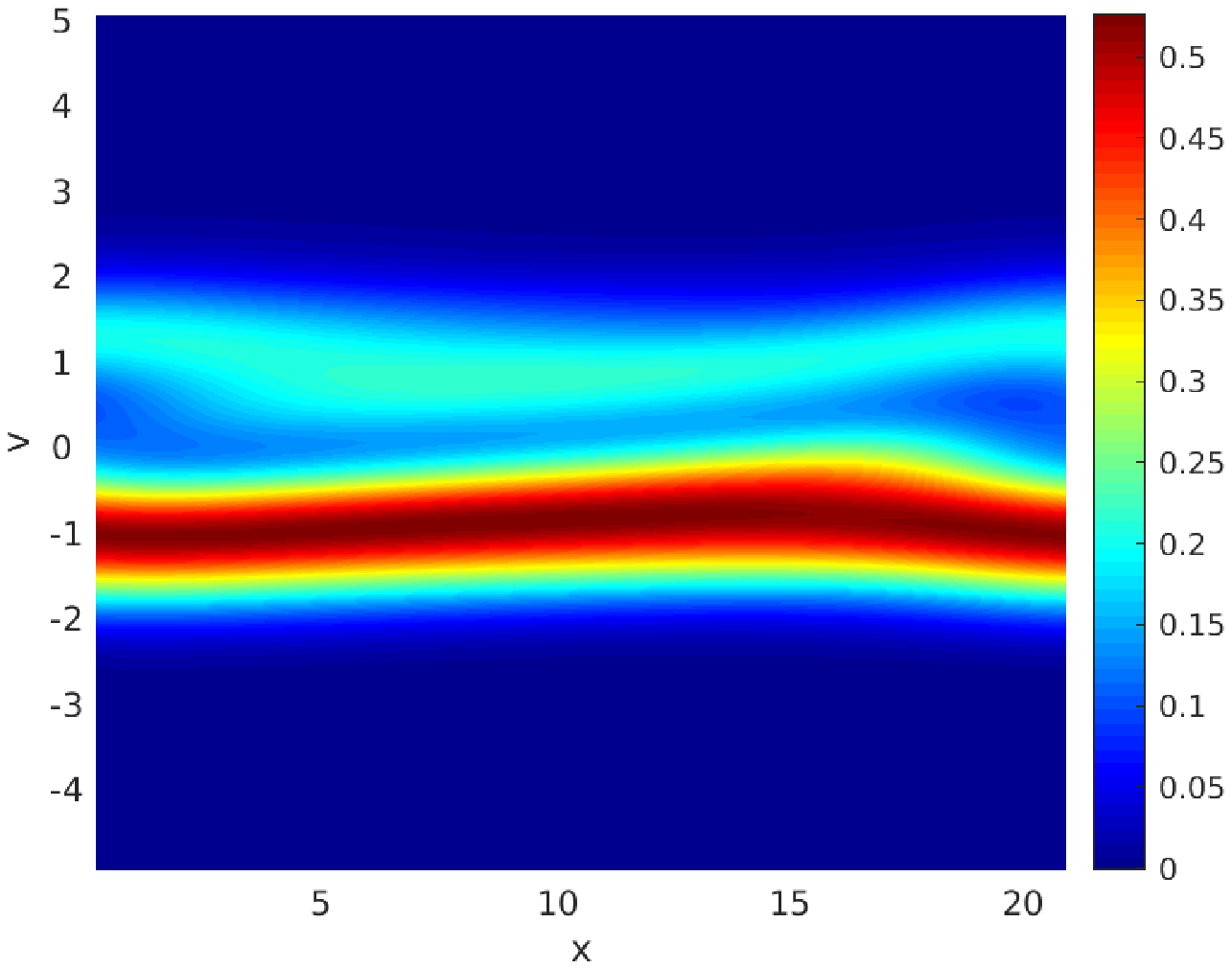}} \hfill
  \subfloat[$t = 30, \nu = 0.01$]
  {\includegraphics[width=0.32\textwidth,
    height=0.21\textwidth,clip]{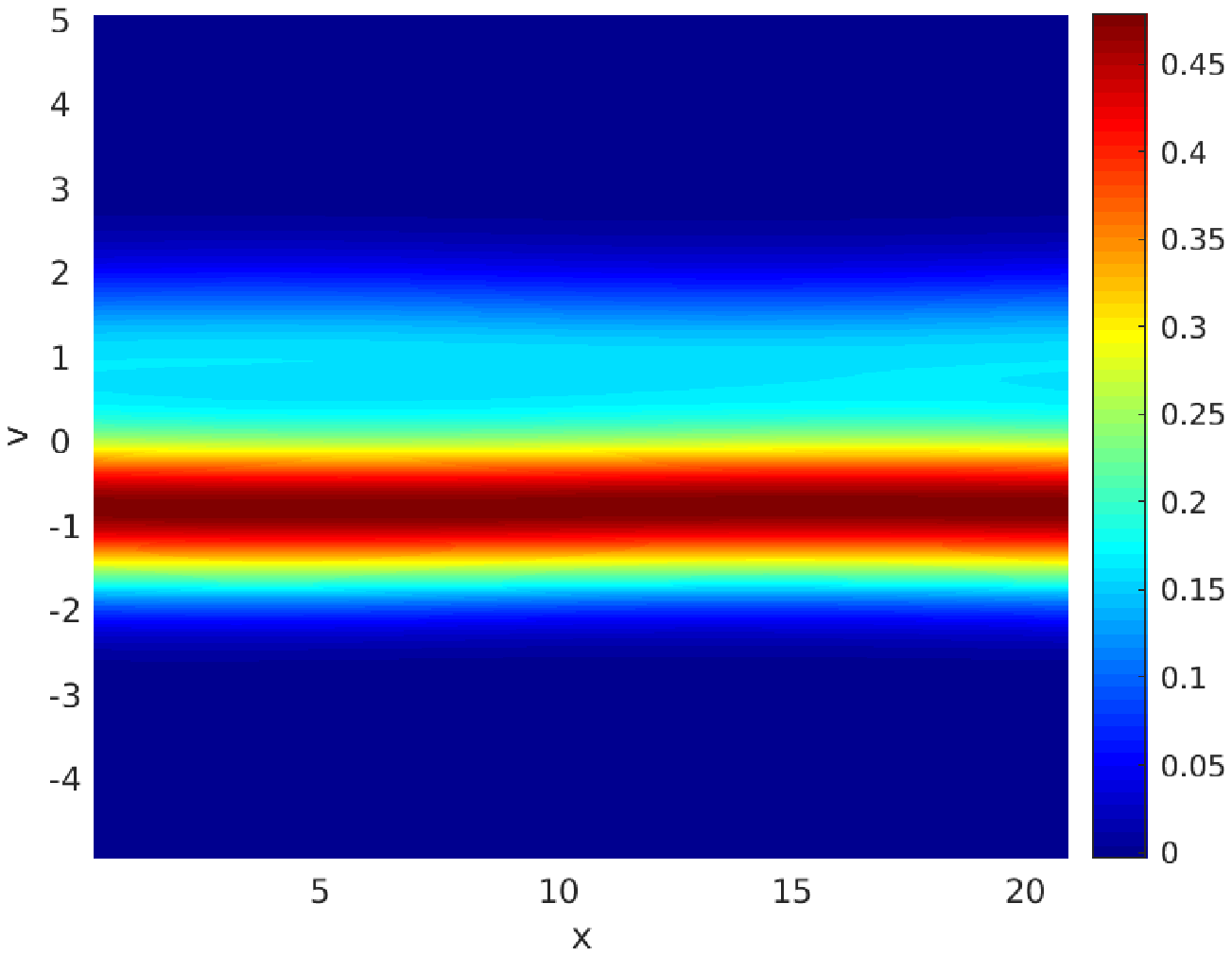}} \\
  \subfloat[$t = 40, \nu = 0.0$]
  {\includegraphics[width=0.32\textwidth, height=0.21\textwidth,
    clip]{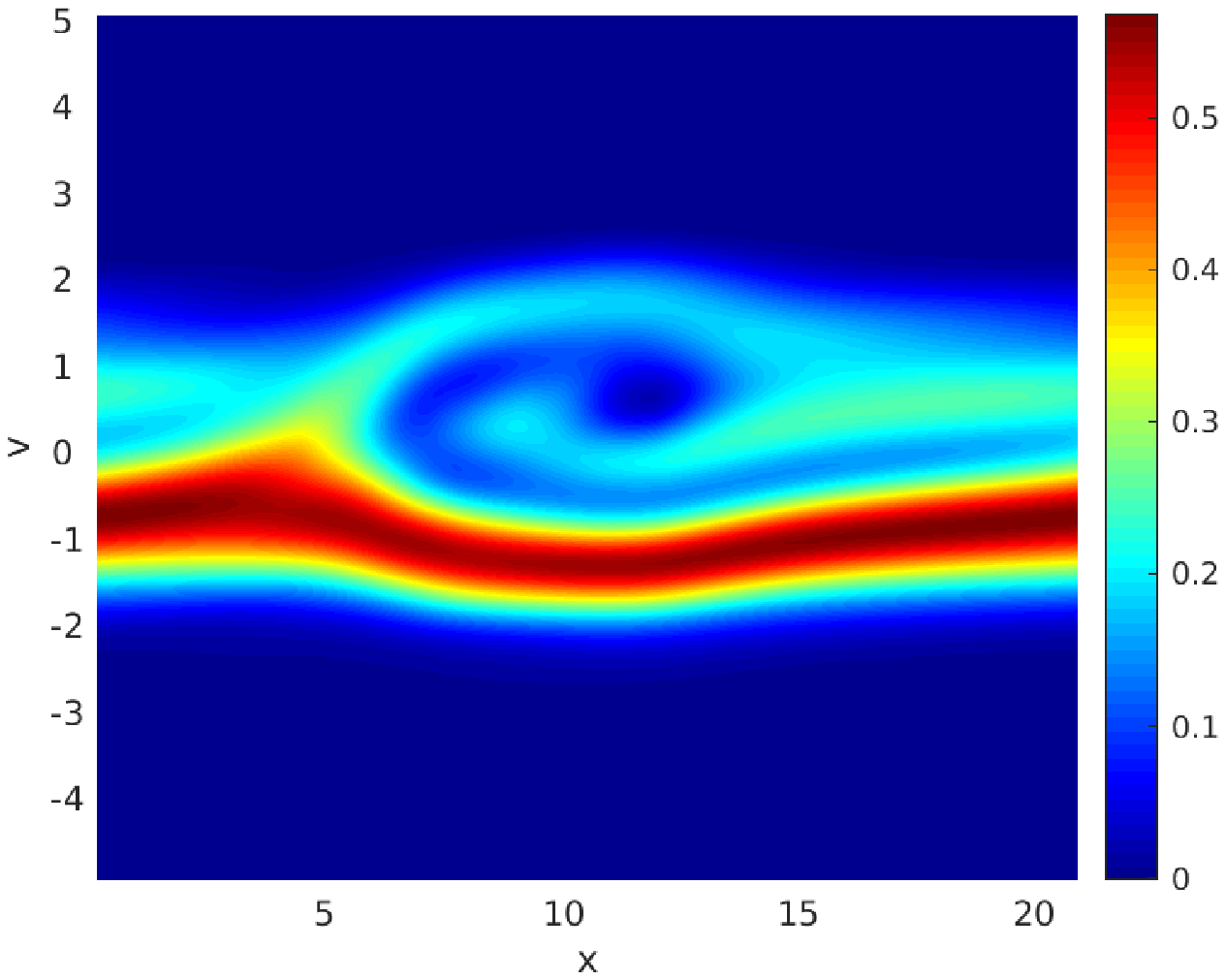}}\hfill
  \subfloat[$t = 40, \nu = 0.001$]
  {\includegraphics[width=0.32\textwidth,
    height=0.21\textwidth,clip]{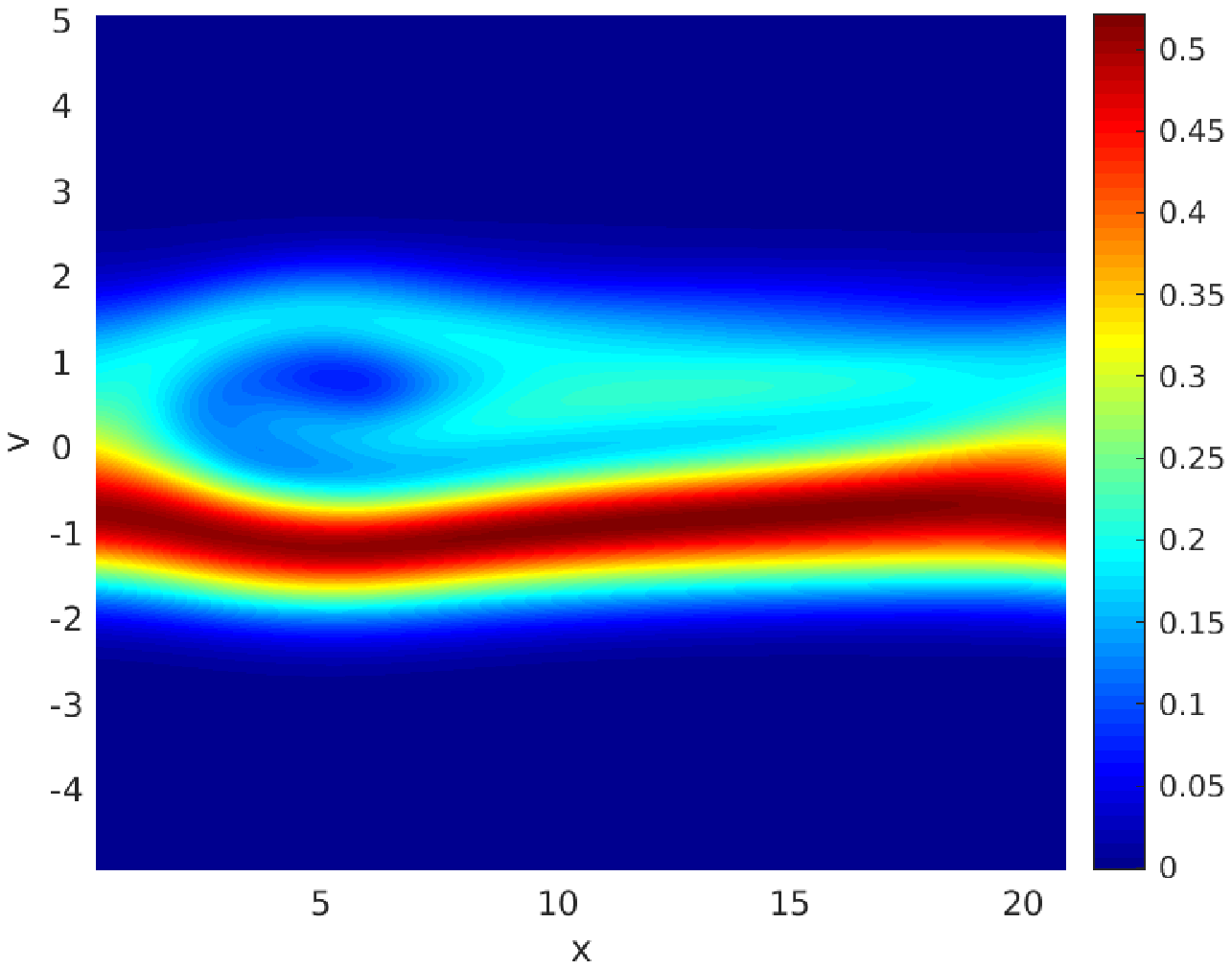}} \hfill
  \subfloat[$t = 40, \nu = 0.01$]
  {\includegraphics[width=0.32\textwidth,
    height=0.21\textwidth,clip]{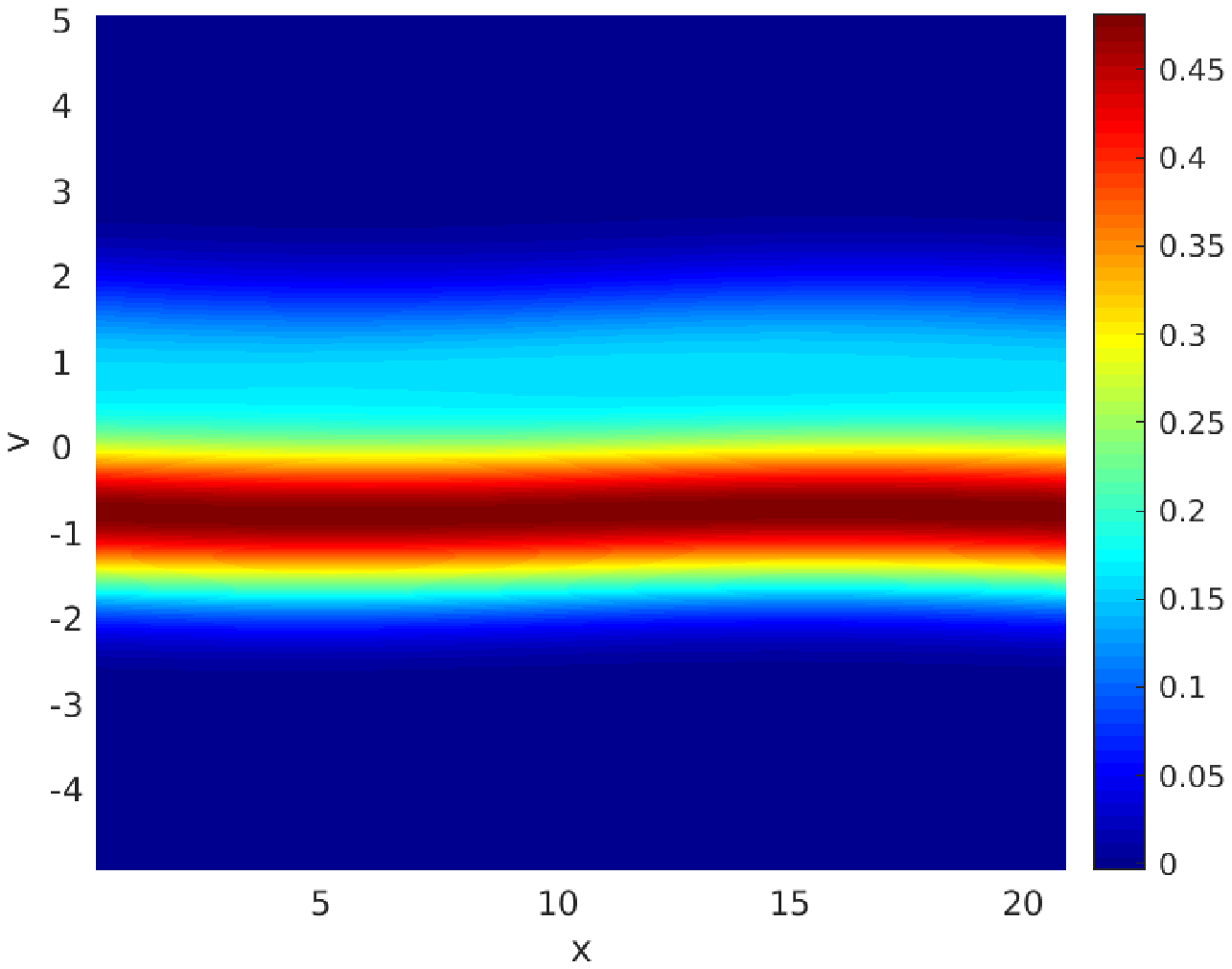}} 
  \caption{Evolution of the marginal distribution function
    $g(t, x, v_1)$ under different collisional frequencies $\nu$
    in the bump-on-tail instability problem. The
    left column corresponds to $\nu =0$
    , the middle column corresponds to $\nu = 0.001$,
    and the right column corresponds to $\nu = 0.01$.}
  \label{fig:ex4_bump_nu}
\end{figure}

Figure \ref{fig:ex4_bump_nu} shows the time evolution of the marginal
distribution function \eqref{eq:marginal} in the $x-v_1$ plane. We can
observe that for the collisionless case, the bump is trapped by the
electric field and gradually forms a crawling vortex-like
structure. For the collisional case, the trapping of the bump is much
weaker, and the distribution of the ``mainstream'' is less affected. In
the case of the collisional frequency $\nu = 0.1$, no vortex-like
structure is perceptible.

\begin{figure}[!htb]
  \centering \includegraphics[width=0.49\textwidth,
  height=0.35\textwidth, clip]{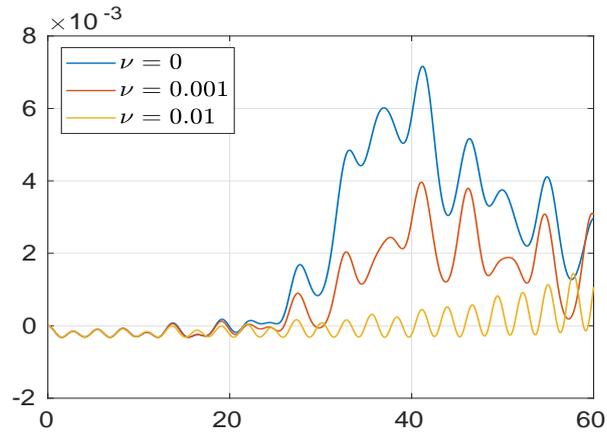}
  \caption{Time evolution of the variation in the total energy
    $ \mE_t(t)$ for different collisional frequencies
    in the bump-on-tail instability problem. The
    variation is defined as
    $(\mE_{t}(t) - \mE_{t}(0)) / \mE_t(0)$. }
  \label{fig:ex4_energy}
\end{figure}

The evolution of the total energy defined in \eqref{eq:total_energy}
is also studied.  Figure \ref{fig:ex4_energy} shows the evolution of
the total energy for different collisional frequencies.  Although the
total energy is not perfectly preserved, the variation in the total
energy is small, especially at the beginning of the evolution and
decreases with the increase in the collisional
frequency.

%% file: article_conclusion.tex
\section{Conclusion}
\label{sec:conclusion}

In this paper, we developed a numerical algorithm for
the FPL equation based on the Hermite spectral method.  Both
collisions within the same species and between different species
were considered to simulate the time evolution of
plasma. A reduced collision model was built by
combining the quadratic FPL collision operator and the diffusive FP
collision operator. A fast algorithm to project between distribution
functions with different expansion centers 
was adopted. Several numerical experiments showed that
our numerical algorithm can capture the time evolution of the
particles accurately and efficiently compared to the fully quadratic
collision model.

The effect of the new reduced collision operator makes this algorithm
promising when dealing with more complicated problems. However, this
method is not capable of dealing with the problems that the state of
the plasma diverges
greatly from equilibrium, which we will work on in the future.
Research on multidimensional problems with the magnetic field is also
ongoing.

\section*{Acknowledgements}
Ruo Li is supported by the National Natural Science Foundation of
China (Grant No. 11971041) and Science Challenge Project
(No. TZ2016002). Yinuo Ren is partially supported by the elite
undergraduate training program of School of Mathematical Sciences in
Peking University. Yanli Wang is supported by Science Challenge
Project (No. TZ2016002) and the National Natural Science Foundation of
China (Grant No. U1930402 and 12031013).

%% file: article_appendix.tex
\section{Appendix}
\label{app}
In this section, we introduce a detailed calculation of the expansion
coefficients for the collision operator \eqref{eq:col_beta} in Section
\ref{app:linear_coe} and the derivation of the governing equations
\eqref{eq:force} for the acceleration step in Section \ref{app:acc}.
\subsection{Calculation of the expansion coefficient
  \eqref{eq:detail_expan_FPL_beta_coe}}
\label{app:linear_coe}
In this section, we introduce a detailed calculation of the expansion
coefficients in \eqref{eq:detail_expan_FPL_beta_coe}. Bringing the
explicit form of the collision operator $\mQ_{\beta}[f]$ in
\eqref{eq:col_beta} into \eqref{eq:expan_FPL_beta} and integrating by
parts, the expansion coefficients
$\mQ_{\beta, \bi}^{[\bu_{\beta}, 1]}$ are
calculated as
\begin{equation}
\label{eq:col_beta_q}
  \begin{aligned}
    \mQ_{\beta, \bi}^{[\bu_{\beta}, 1]}(t, \bx) &=- \frac{1}{\bi!}
    \int_{\bbR^3} \left[ {\bf A}(\bv-\bu_\beta)\nabla_{\bv} f
    \right]\cdot \nabla_{\bv} H_{\bi}\left(\bv - \bu_{\beta}\right)
    \dd \bv.
\end{aligned}
\end{equation}
Expanding the distribution function $f$ as
\begin{equation}
  \label{eq:expan_f}
  f(t, \bx, \bv) \approx\sum_{ \bj\in  \bbN^3}
  f_{\bj}^{[\bu_{\beta}, 
    1]}(t, \bx) \mH_{\bj}^{[\bu_{\beta}, 1]},
\end{equation}
we can calculate \eqref{eq:col_beta_q} as
\begin{equation}
	\label{eq:7.2}
	\begin{aligned}
      \mQ_{\beta, \bi}^{[\bu_{\beta}, 1]}(t, \bx)
      &=- \frac{1}{\bi!}\sum_{\bj\in \bbN^3}
      f_{\bj}^{[\bu_\beta,
        1]}(t, \bx)   C_{\bi}^{\bj}, 
	\end{aligned}
\end{equation}
with
\begin{equation}
  \label{eq:Coe_C}
	\begin{aligned}
      C_{\bi}^{\bj} &= \int_{\bbR^3} \left[ {\bf
          A}(\bv-\bu_\beta)\nabla_{\bv} \mH_{\bj}^{[\bu_\beta,
          1]}(\bv) \right] \cdot \nabla_{\bv} H_{\bi}\left(\bv -
        \bu_{\beta}\right) \dd \bv.
	\end{aligned}
\end{equation}
By changing variables and utilizing the
differentiation property of the basis function
\eqref{eq:Hermite_recur}, we can derive 
\begin{equation}
  \label{eq:Coe_C_detail}
	\begin{aligned}
      C_{\bi}^{\bj} =- \sum_{m,n=1}^3\int_{\bbR^3}
      \left[{\bf A}(\bv)\right]_{mn} \mH_{\bj+\be_n}^{[\bz, 1]}(\bv) i_m
      H_{\bi-\be_m}\left(\bv \right) \dd \bv.
	\end{aligned}
\end{equation}
Recalling the definition of ${\bf A}(\bv-\bu_\beta)$ in \eqref{eq:A},
we expand \eqref{eq:Coe_C_detail} as
\begin{equation}
  \label{eq:aligned}
    C_{\bi}^{\bj} = \Lambda \sum_{m,n=1}^3 i_m
    \Big[\int_{\bbR^3} \delta_{mn}
    \sum_{s=1}^3|\bv|^{\gamma}v_s^2\mH_{\bj+\be_n}^{[\bz,1]}(\bv)H_{\bi-\be_m}\dd \bv  -\int_{\bbR^3}
    |\bv|^{\gamma} v_mv_n \mH_{\bj+\be_n}^{[\bz, 1]}(\bv)i_m
    H_{\bi-\be_m}\left(\bv \right) \dd \bv\Big].
\end{equation}
Finally, with the definition in \cite[Eq.(3.14)]{FPL2018}, i.e.
\begin{equation}
  G_{mn}\big(\gamma,\bi, \bj\big) =\int_{\bbR^3}  |\bv|^{\gamma}
  v_mv_n H_{\bi}\left(\bv \right)
  H_{\bj}(\bv)\dfrac{1}{(2\pi)^{3/2}}\exp{\left(-\dfrac{|\bv|^2}{2}\right)}\dd \bv,
\end{equation}
we can derive the final expression of the expansion coefficients
\eqref{eq:detail_expan_FPL_beta_coe}.

\subsection{Deduction of the governing equation in the acceleration step} 
\label{app:acc}
In this section, we present the deduction of the governing equation
\eqref{eq:force} in the acceleration step with the expansion center
$\tilde{\bu} = \bu$ and $\tilde{T} = T$. A similar deduction can be
found in \cite{Wang}, to which we refer readers for more details. In
this case, the distribution function $f$ is expanded as
\begin{equation}
  \label{eq:expansion_center}
  f(t, \bx, \bv) = \sum_{\bi \in \bbN^3}
  f_{\bi}^{[\bu, 
    T]}(t, \bx) \mH_{\bi}^{[\bu,
    T]}(\bv). 
\end{equation}
Substituting \eqref{eq:expansion_center} into the FPL equation
\eqref{eq:FPL}, we can derive the moment equations with some
rearrangement as
\begin{equation} \label{eq:mnt_eq}
\begin{split}
  & \frac{\partial f_{\bi}}{\partial t} + \sum_{d=1}^3 \left(
      \frac{\partial u_{d}}{\partial t} + \sum_{j=1}^3 u_j
      \frac{\partial u_{d}}{\partial x_j} - F_d \right)
      f_{\bi-\be_d} \\
      & 
       + \sum_{j,d=1}^3 \left[ \frac{\partial u_d}{\partial
          x_j} \left( T f_{\bi-\be_d-\be_j} + (i_j + 1)
          f_{\bi-\be_d+\be_j} \right) + \frac{1}{2} \frac{\partial
          T}{\partial x_j} \left( T f_{\bi-2\be_d-\be_j} +
          (i_j + 1) f_{\bi-2\be_d+\be_j} \right)
      \right] \\
      &+ \frac{1}{2} \left( \frac{\partial
          T}{\partial t} + \sum_{j=1}^3 u_j \frac{\partial
          T}{\partial x_j}
      \right) \sum_{d=1}^3 f_{\bi-2\be_d}+ \sum_{j=1}^3
      \left( T \frac{\partial f_{\bi -\be_j}}{\partial x_j} + u_j \frac{\partial 
          f_{\bi}}{\partial x_j} + (i_j + 1) \frac{\partial
          f_{\bi+\be_j}}{\partial x_j} \right) = Q_{\bi},
\end{split}
\end{equation}
where $[\bu, T]$ are omitted and $ Q_{\bi}$ is the expansion for the
collision term.  Following the method in \cite{Li}, we deduce the mass
conservation in the case of $\bi = \bz$ as
 \begin{equation}
\label{mass_con}
  \frac{\partial f_{\bz}}{\partial x_j} +
  \sum_{j=1}^3 \left(
    u_j \frac{\partial f_{\bz}}{\partial x_j} +
    f_{\bz} \frac{\partial u_j}{\partial x_j}
  \right) = 0.
 \end{equation}
 If we set $\bi = \be_d$, with $ d = 1,2,3$,
 \eqref{eq:mnt_eq} reduces to
\begin{equation}
\label{eq:alpha = e_d}
f_{\bz} \left(
  \frac{\partial u_d}{\partial t} +
  \sum_{j=1}^3 u_j \frac{\partial u_d}{\partial x_j} - F_d
\right) + f_{\bz} \frac{\partial T}{\partial x_d} 
+ T \frac{\partial f_{\bz}}{\partial x_d}
+ \sum_{j=1}^3 (\delta_{jd} + 1)
\frac{\partial f_{\be_d + \be_j}}{\partial x_j} = 0.
\end{equation}
With the splitting method stated in Section \ref{sec:model},
\eqref{eq:alpha = e_d} is split into the convection step 
\begin{equation}
  \label{eq:convection_step}
  f_{\bz} \left(
    \frac{\partial u_d}{\partial t} +
    \sum_{j=1}^3 u_j \frac{\partial u_d}{\partial x_j} 
  \right) + f_{\bz} \frac{\partial T}{\partial x_d} 
  + T \frac{\partial f_{\bz}}{\partial x_d}
  + \sum_{j=1}^3 (\delta_{jd} + 1)
  \frac{\partial f_{\be_d + \be_j}}{\partial x_j} = 0,
\end{equation}
and the force step 
\begin{equation}
  \label{eq:force_step}
   \pd{u_d}{t} - F_d = 0, \qquad d = 1, 2, 3. 
\end{equation}
Then, we obtain the governing equations \eqref{eq:force} for the
acceleration step.